\documentclass{aastex63}
\usepackage[utf8]{inputenc}
\usepackage{hyperref}
\usepackage{algorithm}
\usepackage{verbatim}
\usepackage{algpseudocode}
\usepackage{booktabs}
\usepackage{enumitem}
\usepackage{graphicx}
\usepackage{xcolor}

\usepackage{rotating}

\newcommand\newEBs{7,936~}
\newcommand\knownEBs{2,065~}

\begin{document}

\title{The TESS Ten Thousand Catalog: 10,001 uniformly-vetted and -validated Eclipsing Binary Stars detected in Full-Frame Image data by machine learning and analyzed by citizen scientists}

\author[0000-0001-9786-1031]{Veselin B. Kostov}
\affiliation{NASA Goddard Space Flight Center, Greenbelt, MD 20771, USA}
\affiliation{SETI Institute, 189 Bernardo Ave, Suite 200, Mountain View, CA 94043, USA}

\author[0000-0003-0501-2636]{Brian P. Powell}
\affiliation{NASA Goddard Space Flight Center, Greenbelt, MD 20771, USA}

\author[0000-0001-7062-7632]{Aline U. Fornear}
\affiliation{Citizen Scientist, Eclipsing Binary Patrol Collaboration}

\author[0000-0001-7701-6818]{Marco Z. Di Fraia}
\affiliation{Citizen Scientist, Eclipsing Binary Patrol Collaboration}

\author[0000-0002-5665-1879]{Robert Gagliano}
\affiliation{Citizen Scientist, Glendale, AZ 85308}

\author[0000-0003-3988-3245]{Thomas L. Jacobs}
\affiliation{Citizen Scientist, Missouri City, TX 77459}

\author[0000-0003-3464-1554]{Julien S. de Lambilly}
\affiliation{Citizen Scientist, Eclipsing Binary Patrol Collaboration}

\author[0000-0002-4143-2550]{Hugo A. Durantini Luca}
\affiliation{Citizen Scientist, Eclipsing Binary Patrol Collaboration}

\author[0000-0003-2025-3147]{Steven R. Majewski}
\affiliation{Department of Astronomy, University of Virginia,
530 McCormick Rd., Charlottesville, VA 22904, USA}

\author{Mark Omohundro}
\affiliation{Citizen Scientist, c/o Zooniverse, Department of Physics, University of Oxford, Denys Wilkinson Building, Keble Road, Oxford, OX13RH, UK}

\author[0000-0001-9647-2886]{Jerome Orosz}
\affiliation{Department of Astronomy, San Diego State University, 5500 Campanile Drive, San Diego, CA 92182, USA}

\author[0000-0003-3182-5569]{Saul A. Rappaport}
\affiliation{Department of Physics, Kavli Institute for Astrophysics and Space Research, M.I.T., Cambridge, MA 02139, USA}

\author{Ryan Salik}
\affiliation{Department of Computer Science, Princeton University}
\affiliation{Citizen Scientist, Eclipsing Binary Patrol Collaboration}

\author[0000-0001-5504-9512]{Donald Short}
\affiliation{Department of Astronomy, San Diego State University, 5500 Campanile Drive, San Diego, CA 92182, USA}

\author[0000-0003-2381-5301]{William Welsh}
\affiliation{Department of Astronomy, San Diego State University, 5500 Campanile Drive, San Diego, CA 92182, USA}

\author[0000-0002-4959-8598]{Svetoslav Alexandrov}
\affiliation{Citizen Scientist, Eclipsing Binary Patrol Collaboration}
\affiliation{Institute of Plant Physiology and Genetics, Bulgarian Academy of Sciences, Acad. G. Bontchev Str., BI.23, Sofia 113, Bulgaria}

\author[0000-0001-8953-9149]{Cledison Marcos da Silva}
\affiliation{Citizen Scientist, Eclipsing Binary Patrol Collaboration}

\author[0009-0001-2901-8875]{Erika Dunning}
\affiliation{Department of Astronomy, San Diego State University, 5500 Campanile Drive, San Diego, CA 92182, USA}

\author{Gerd G\"uhne}
\affiliation{Citizen Scientist, Eclipsing Binary Patrol Collaboration}

\author{Marc Huten}
\affiliation{Citizen Scientist, Eclipsing Binary Patrol Collaboration}

\author[0000-0001-8343-0820]{Michiharu Hyogo}
\affiliation{Citizen Scientist, Eclipsing Binary Patrol Collaboration}

\author{Davide Iannone}
\affiliation{Citizen Scientist, Eclipsing Binary Patrol Collaboration}

\author{Sam Lee}
\affiliation{Citizen Scientist, Eclipsing Binary Patrol Collaboration}

\author[0000-0001-6343-4744]{Christian Magliano}
\affiliation{Dipartimento di Fisica ``Ettore Pancini'', Università di Napoli Federico II, 80126 Napoli, Italy}

\author{Manya Sharma}
\affiliation{Citizen Scientist, Eclipsing Binary Patrol Collaboration}

\author{Allan Tarr}
\affiliation{Citizen Scientist, Eclipsing Binary Patrol Collaboration}

\author[0000-0002-0248-4817]{John Yablonsky}
\affiliation{Citizen Scientist, Eclipsing Binary Patrol Collaboration}

\author[0000-0002-7759-0569]{Sovan Acharya}
\affiliation{Citizen Scientist, Eclipsing Binary Patrol Collaboration}

\author[0000-0002-8167-1767]{Fred Adams}
\affiliation{Department of Physics, University of Michigan, Ann Arbor, MI, USA}

\author[0000-0001-7139-2724]{Thomas Barclay}
\affiliation{NASA Goddard Space Flight Center, Greenbelt, MD 20771, USA}

\author[0000-0001-7516-8308]{Benjamin~T.~Montet}
\affiliation{School of Physics, University of New South Wales, Sydney, NSW 2052, Australia}

\author[0000-0001-7106-4683]{Susan Mullally}
\affiliation{Space Telescope Science Institute, 3700 San Martin Dr., Baltimore MD 21212}

\author[0000-0001-8472-2219]{Greg Olmschenk}
\affiliation{NASA Goddard Space Flight Center, Greenbelt, MD 20771, USA}

\author[0000-0002-1913-0281]{Andrej Pr\v{s}a}
\affiliation{Astrophysics and Planetary Science, Villanova University, 800 Lancaster Avenue Villanova, PA 19085}

\author[0000-0003-1309-2904]{Elisa Quintana}
\affiliation{NASA Goddard Space Flight Center, Greenbelt, MD 20771, USA}

\author[0000-0002-4235-6369]{Robert Wilson}
\affiliation{NASA Goddard Space Flight Center, Greenbelt, MD 20771, USA}

\author{Hasret Balcioglu}
\affiliation{Citizen Scientist, Eclipsing Binary Patrol Collaboration}

\author[0000-0002-0493-1342]{Ethan Kruse}
\affiliation{NASA Goddard Space Flight Center, Greenbelt, MD 20771, USA}

\author{the Eclipsing Binary Patrol Collaboration}

\begin{abstract}
    
The Transiting Exoplanet Survey Satellite ({\em TESS}) has surveyed nearly the entire sky in Full-Frame Image mode with a time resolution of 200 seconds to 30 minutes and a temporal baseline of at least 27 days. In addition to the primary goal of discovering new exoplanets, {\em TESS} is exceptionally capable at detecting variable stars, and in particular short-period eclipsing binaries which are relatively common, making up a few percent of all stars, and represent powerful astrophysical laboratories for deep investigations of stellar formation and evolution. We combed Sectors 1-82 of {\em TESS} Full-Frame Image data searching for eclipsing binary stars using a neural network that identified $\sim$1.2 million stars with eclipse-like features. Of these, we have performed an in-depth analysis on $\sim$60,000 targets using automated methods and manual inspection by citizen scientists. Here we present a catalog of 10,001 uniformly-vetted and -validated eclipsing binary stars that passed all our ephemeris and photocenter tests, as well as complementary visual inspection. Of these, \newEBs are new eclipsing binaries while the remaining \knownEBs are known systems for which we update the published ephemerides. We outline the detection and analysis of the targets, discuss the properties of the sample, and highlight potentially interesting systems. Finally, we also provide a list of $\sim$900,000 unvetted and unvalidated targets for which the neural network found eclipse-like features with a score higher than 0.9, and for which there are no known eclipsing binaries within a sky-projected separation of a {\em TESS} pixel ($\approx21$ arcsec).
\end{abstract}

\accepted{ApJS June 2025}

\keywords{Eclipsing Binary Stars --- Transit photometry --- Astronomy data analysis}

\section{Introduction}\label{sec:intro}

Binary stars make up a large fraction of the Galactic stellar population \citep[e.g.,][]{Raghavan2010, Tokovinin2021,Offner2023}. Of these, perhaps the most important subsets are those that produce eclipses due to a favorable geometric configuration with respect to the observer. These eclipsing binary stars (EBs) pave the ``royal road'' to stellar astrophysics \citep[][]{1948HarMo...7..181R} and have long served as a fundamental pillar upon which our understanding of how stars form and evolve stands \citep[e.g.,][]{Osterbrock1953,Andersen1991,Torres2010}. Spectroscopic double-lined EBs enable direct and accurate measurements of the masses, radii, and temperatures of their components, and provide critical calibrators for theoretical models \citep[e.g.,][]{Torres2010}. 

Despite the ubiquitous distribution of binary stars throughout the Solar neighborhood and over two centuries of study \citep[e.g.,][and references therein, including Sewell's letter to S. Vince]{Goodricke1783,Kopal1956,Eggen1957,2001RMxAC..11...23N} pressing questions about these systems remain. For example, it is unclear whether the multiplicity properties of stellar systems are universal or depend on the formation environment and/or stellar mass, what is the origin of the brown dwarf ``desert'' scarcity, and how stellar multiplicity affects planet formation \citep[e.g.,][and references therein]{2017ApJS..230...15M}. These uncertainties are due in large part to the enormous size of the parameter space, since binary stars have extensive distributions of stellar masses and mass ratios, orbital periods, eccentricities, etc., all of which can vary with the environment (e.g., cluster membership).

% Note: probably cite this: https://arxiv.org/abs/2107.10005
% and other works

Large-scale photometric surveys are well-suited for monitoring a large number of binary stars through the detection of eclipses, and have detected hundreds of thousands of eclipsing binary stars. For example, millions of EBs have been observed by Gaia \citep[][]{2023A&A...674A..16M}\footnote{Farewell, Gaia! Thank you for all the amazing science!}, hundreds of thousands by OGLE \citep{2017AcA....67..297S}, ASAS-SN \citep{2022MNRAS.517.2190R}, ATLAS \citep{Heinze2018} and WISE \citep{2021MNRAS.503.3975P}, as well as tens of thousands from primarily exoplanet-focused surveys such as Kepler \citep{2011AJ....142..160S,2011AJ....141...83P,2014AJ....147...45C}, {\em TESS} \citep{2015ApJ...809...77S}, SuperWASP \citep{2021MNRAS.502.1299T}, etc. With its extremely wide sky coverage (${\sim98\%}$) and long dwell time ($\sim$27+ days of nearly-continuous observations), NASA's {\em TESS} mission is an excellent example of the power of all-sky surveys for studying EBs. While the primary science objective of {\em TESS} is finding transiting rocky exoplanets around nearby stars \citep{Ricker2015}, it presents an ideal platform for the detection of thousands of eclipsing binary stars covering a wide range of physical and orbital parameter space \citep[e.g.,][]{Prsa2022,2022RNAAS...6...96H,2023MNRAS.522...29G,Eisner2021,Montalto2023,2024AJ....167..203M,2022MNRAS.513..102C,2023MNRAS.521.3749M,2024A&A...691A.242I,2025arXiv250415875S,2025ApJS..276...57G}

The {\em TESS} mission is also well-suited for exploring the variability of many different classes of stars, and searching for rare systems that may be studied with extensive follow-up observations from space and the ground. The large EB population monitored by {\em TESS} enables statistical studies of the effects of mass, mass ratio, and composition on binary fraction, eccentricity, and orbital period distributions. Such a large sample of EBs covering all stellar types and Galactic environments also helps advance our knowledge of the physics of binary interactions, such as tidal forces, migration, spin-orbit coupling, and mass transfer \citep[e.g.,][and references therein]{1910AN....183..345V,1962P&SS....9..719L,1962AJ.....67..591K,2013MNRAS.435..943P,2018MNRAS.476.4234F,2021MNRAS.502.4479H,2019MNRAS.486.4781F,2019MNRAS.483.4060L,2022MNRAS.511.1362T,2022ApJ...926..195V,2021MNRAS.507.5832K,2021MNRAS.507..560S}. Last, but not least, by better understanding the distribution and properties of eclipsing binaries in the Galaxy, we can improve our priors on background contamination for {\em TESS}'s core mission of exoplanet transit observations.

Given the enormous amount of data produced by the {\em TESS} mission---for example, there are, on average, ${\sim}$3 million stars brighter than T$_{\text{mag}}=15$ observed per sector---we need to develop sophisticated yet efficient analysis techniques to extract the relevant astrophysical information from this unique data set. At the time of writing, several projects have already developed pipelines for the extraction of Full-Frame Image (FFI) lightcurves from {\em TESS} \citep[e.g.,][]{2018AJ....156..132O,eleanor2019,qlp2022,2023AJ....165...71H,2020RNAAS...4..201C,2025PASP..137b4501H}, and have released tools and data products to the public.  
To study binary stars from {\em TESS}, we have developed a local implementation of the \textsc{eleanor} pipeline \citep{eleanor2019} and used it to extract FFI lightcurves for Sectors 1-82 for all targets brighter than T$_{\text{mag}}=15$. To detect EB candidates, we have created and trained a machine learning identification scheme. Here we describe the development and implementation of our extraction and detection pipeline, the processing and analysis of the data by automated methods and human inspection, and present the {\em TESS} Ten-Thousand catalog containing 10,001 uniformly-vetted and -validated EBs. Of these, \newEBs are new EBs and \knownEBs are known EBs for which we update the ephemeris provided in one or more catalogs. We describe the general properties of the population and touch upon individual systems of interests. The catalog provides general target information (TIC ID, sky coordinates, {\em TESS} magnitude, number of sectors observed, effective temperature, Gaia astrometric measurements), ephemerides, eclipse depths and durations, secondary phase, as well as relevant notes and comments. We envision this catalog as a community-facing product to serve as a platform for subsequent studies and analysis of both the populations as a whole and of individual targets of interest, including but not limited to confirmation and modeling efforts, cross-matching against catalogs of {\em TESS} planet candidates, etc. All our data products and results are publicly available as machine-readable online supplements. 

This paper is organized as follows. In Section \ref{sec:lightcurves} we describe the construction of the FFI lightcurves; Section \ref{sec:ml} outlines the identification of EB candidates by a machine learning pipeline while the vetting and validation of the candidates is presented in Section \ref{sec:vetting}. Section \ref{sec:catalog} outlines the catalog of uniformly-vetted and -validated EBs, and the results are summarized in Section \ref{sec:summary}.

\section{Construction of FFI lightcurves}
\label{sec:lightcurves}

While other lightcurve data sets were available to us, such as the MIT Quick Look Pipeline \citep[QLP;][]{2020RNAAS...4..204H} or the {\em TESS} Science Processing Operations Center \citep[SPOC;][]{2020RNAAS...4..201C}, we wanted to pursue potentially unknown systems beyond the scope of available lightcurves. For example, the QLP lightcurves are limited to those stars brighter than $T_{\text{mag}}=13.5$.  As such, we undertook an effort to construct all available {\em TESS} lightcurves to a limit of $T_{\text{mag}}=15.0$ using {\sc eleanor} \citep{2019PASP..131i4502F}.\footnote{Later, we rebuilt these for public release using the {\sc eleanor-lite} pipeline for Sectors 1-26 to a limit of $T_{\text{mag}}=16.0$ \citep{2022RNAAS...6..111P}.  These are available at \url{https://archive.stsci.edu/hlsp/gsfc-eleanor-lite}.}

We started by downloading the full {\em TESS} Input Catalog \citep[TIC;][]{2019AJ....158..138S}, available as a set of CSV files in increments of two degrees of declination, from MAST\footnote{\url{https://archive.stsci.edu/tess/tic_ctl.html}}. Each target in the TIC was queried in parallel using the {\sc tess-point} python package \citep{2020ascl.soft03001B} to determine the sectors of {\em TESS} data in which they were present, effectively translating the overall TIC into a per-sector TIC. 

In preparation for building the lightcurves for a given sector, we then downloaded the {\em TESS} FFIs from MAST and used {\sc eleanor} to create the necessary `postcards' and `backgrounds' required for local construction of the lightcurves (described further in \citealt{2019PASP..131i4502F}).  The per-sector TIC was then used as input to a parallelized implementation of {\sc eleanor} on the NASA Center for Climate Simulation (NCCS) {\em Discover} supercomputer.\footnote{\url{https://www.nccs.nasa.gov/systems/discover}}  The outputs of our parallelized lightcurve construction code were minimized to limit the need for memory storage, and contained only basic metadata along with the times and fluxes.

\section{Machine Learning Identification of EB Candidates}
\label{sec:ml}

Eclipses are an ideal shape for machine learning classification in lightcurves.  They are usually a prominent feature in the lightcurve, with common spatial interrelationships between the eclipse and the baseline, as well as the characteristic point at the eclipse minimum.  These features are uniquely identifiable in lightcurves and lend themselves toward processing with a Convolutional Neural Network \citep[CNN;][]{LeCun1989}.  Rather than limit ourselves to only those lightcurves that demonstrated periodicity with eclipses, we chose to pursue a strategy of training the neural network to find the {\em feature of the eclipse}.  In this manner, we could also treat the lightcurve purely as a 1D shape rather than having to consider time-dependencies, allowing a simpler methodology.  Our intent was to build the neural network for classification purposes, i.e., to produce a single sigmoid-activated output where unity is a positive (indicating that the lightcurve contains an eclipse) and zero is a negative (indicating that the lightcurve does not contain an eclipse).

\begin{figure}
    \centering
    \includegraphics[width=1.\textwidth]{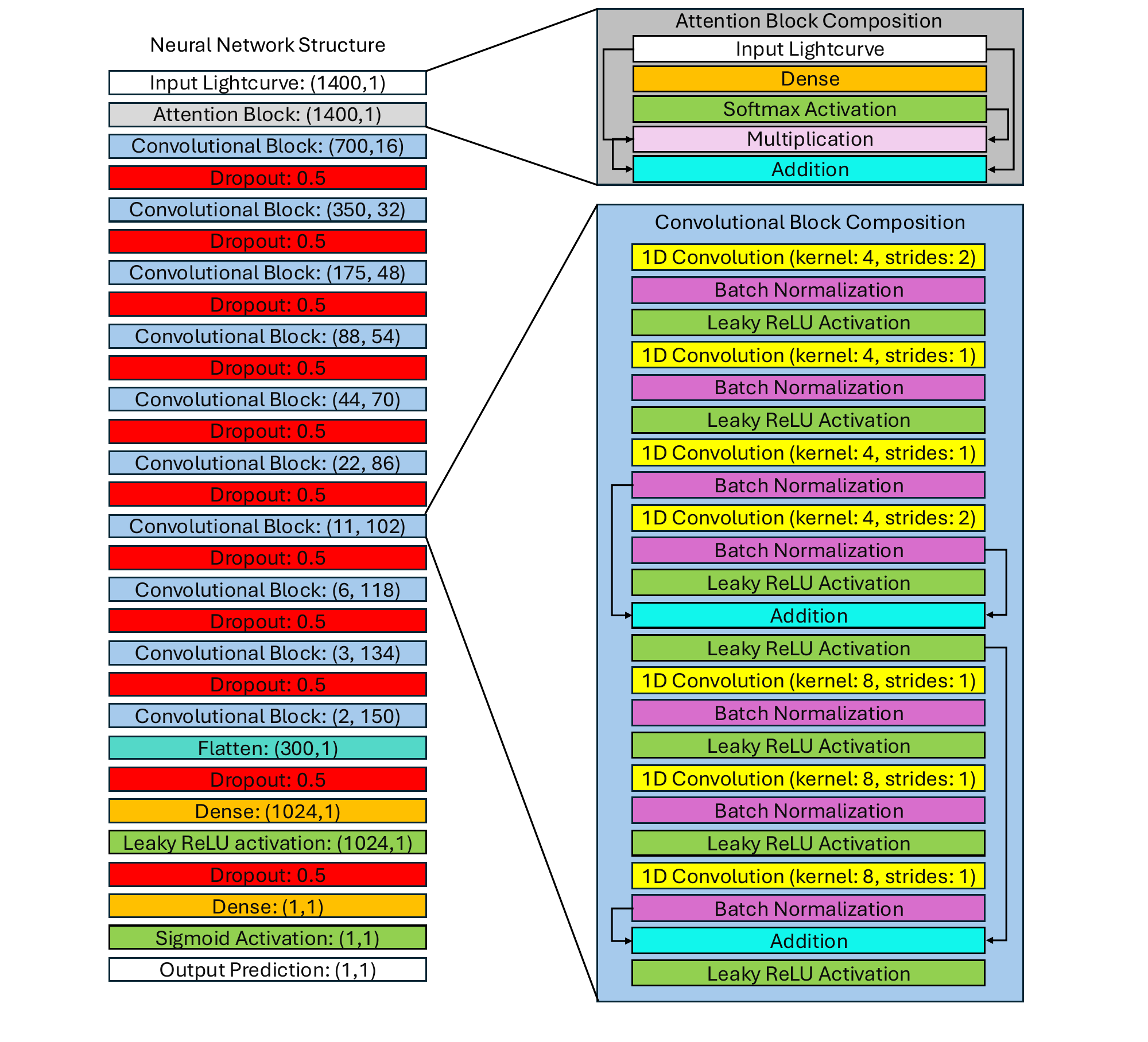}
    \caption{The structure of our neural network, with like layers grouped by color.  The full network summary is shown on the left.  The attention block has a structure as shown on the upper right, while each convolutional block has a structure as shown on the lower right. Arrows into the addition layers indicate the flow of the residual.}
    \label{fig:nn}
\end{figure}

The performance of a CNN as a classifier is broadly tied to depth of the network \citep{Simonyan2014}.  While vanishing or exploding gradients generally limit depth, the concept of residual blocks \citep{He2016} has allowed for depth to be limited only by hardware (in terms of physical memory available) and by training data shape and batch size (in terms of the tradeoff between depth and data size within the physical memory). As such, we designed the general structure of our neural network as a 1D adaptation of ResNet \citep{He2016}, which was originally designed to process 2D images.  Of course, too much depth in conjunction with very little data or overly simplistic data can also prevent convergence.  Since a lightcurve is not a particularly complex data representation requiring extreme depth, we started our development process with a relatively shallow network. We developed the neural network iteratively (using Tensorflow/Keras, \citealt{tensorflow,keras}), and made it deeper as we augmented our training data and ensured that additional data and a deeper network offered continued reduction of the model loss, as given by binary crossentropy \citep{Goodfellow2016} using the RMSprop optimizer \citep{Tieleman2012}. We also found that an additive and multiplicative attention mechanism \citep{Bahdanau2014, Luong2015} at the beginning of the network was beneficial to performance.  Apart from the sigmoid activation on the output layer, we used leaky rectified linear unit \citep[ReLU;][]{Nair2010} activation throughout the network to prevent the problem of vanishing gradients.  The structure of our neural network, shown in Figure \ref{fig:nn}, is rather simple.  In total, our network has 241 layers with $\sim$5.5 million trainable parameters.  

\subsection{Lightcurve Pre-processing}

Concurrent with the development of the neural network, we also needed to refine our method of pre-processing the lightcurves, which, as in many machine learning applications, was critical to the performance of the neural network. 
Machine learning methods require data to be of the same shape.  This of course presents a problem with {\em TESS} FFI lightcurves, especially after the per-timestep quality flags are masked from the lightcurve, resulting in a wide variety of one-dimensional array sizes.  The temporal discontinuities caused by the data downlink gap and the quality mask also create a temporally irregular data set.  We chose to ignore the temporal component and treat the lightcurve as a 1D shape rather than a time-dependent signal.  This approach is consistent with our selection of a CNN rather than a recurrent neural network \citep{Rumelhart1986} or other time-dependent methodology such as, e.g., Long Short-Term Memory \citep[LSTM;][]{Hochreiter1997}, Convolutional LSTMs \citep[ConvLSTM;][]{Shi2015}, or Temporal Convolutional Networks \citep[TCN;][]{lea2017temporal}, among others.

To create homogeneous shapes from the irregular lightcurves, our options were to either truncate longer lightcurves or pad shorter lightcurves.  Truncation, of course, risks missing a lightcurve where a single eclipse occurs in the truncated section.  Padding provides its own risks in providing artificial information in the discontinuity of the data shape.  We decided to pad the lightcurves to a maximum length of 1400 elements, with the padding containing a mirror of the data.  We emphasize again that the neural network has no time-dependency, therefore no understanding of periodicity, and it was determined that the neural network could learn to ignore the discontinuity in the padding in the same manner as it would in the collapsed time gaps.  We also note that we developed the neural network during Year 2 of the {\em TESS} mission, where 30-minute cadence data provided relatively short lightcurves.  With the continued shortening of the {\em TESS} cadence over subsequent years, we have not retrained the network.  Rather, we downsampled longer lightcurves to fit our required data shape.

\begin{figure}
    \centering
    \includegraphics[width=1.\textwidth]{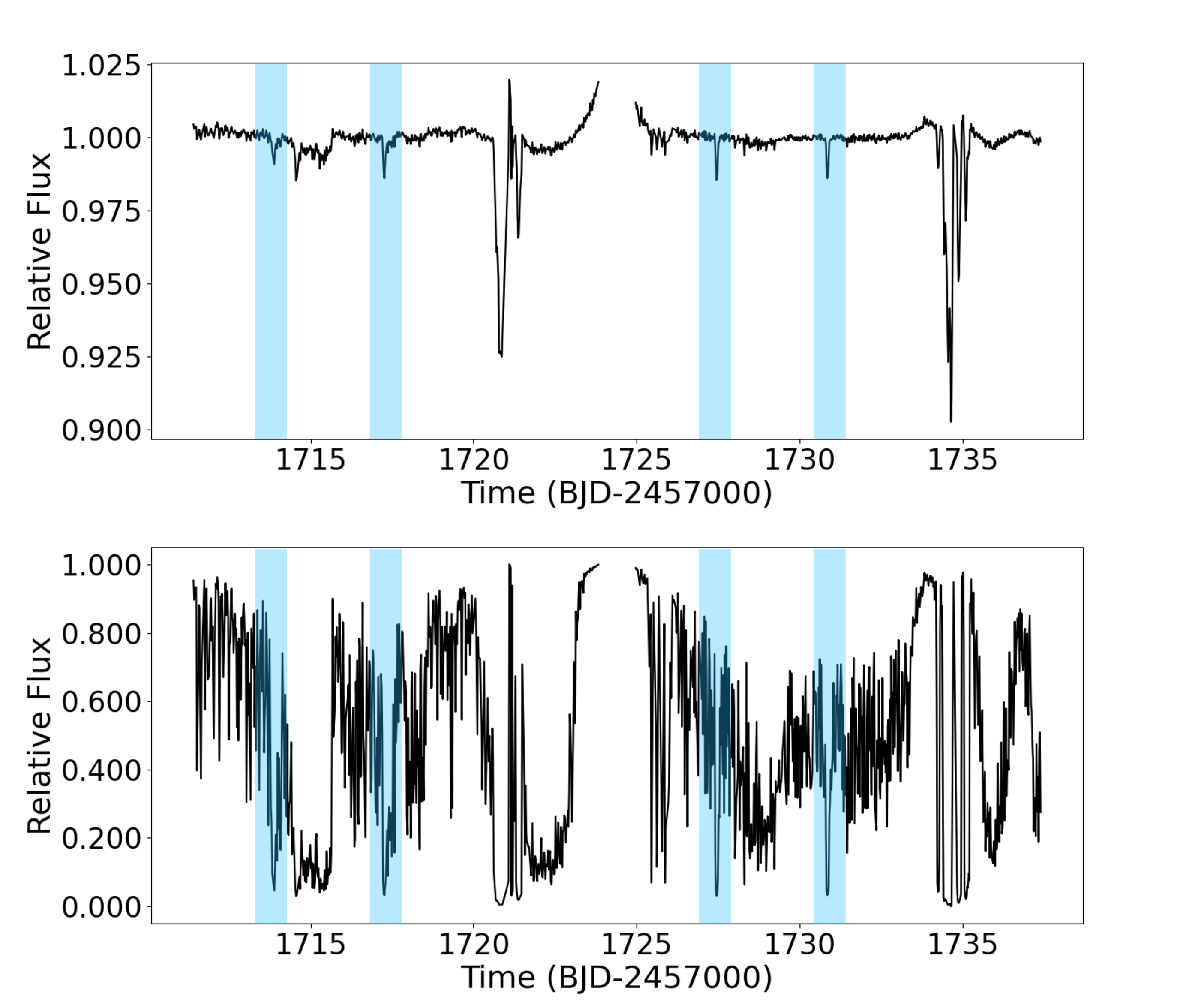}
    \caption{Lightcurve of TIC 139079180 for sector 15 ({\em top panel}), vs. the same lightcurve scaled with a quantile transform ({\em bottom panel}).  Eclipses are highlighted in blue.  The scattered light near the ends of the two segments dominates the lightcurve and also resembles an eclipse, making this type of feature difficult to overcome as a source of false positives.  The quantile transform has the effect of making these events less prominent and emphasizing the actual eclipses.  The inputs to the neural network are the quantile-scaled lightcurves.  Although perhaps more difficult to the human eye, the quantile transform represents the lightcurve to the neural network in a manner that allows it to successfully find and classify the eclipse feature.}
    \label{fig:quantile}
\end{figure}

{\em TESS} systematics presented a different set of challenges.  In many {\em TESS} FFI lightcurves, indeed, the dominant signal is scattered light systematics, which often produce features in the lightcurves that resemble eclipses.  In years 1 and 2 of {\em TESS} data, this problem is particularly pronounced in Sectors 1, 12, 13, 14, 15, 23, 24 and 26.  In neural networks, and machine learning in general, the magnitude of a value is a representation of its importance.  As such, strong eclipse-like systematic signals have the potential to dominate a machine learning method if not properly diminished in importance, hence the need for data scaling.  We specifically selected the quantile transform as our method of scaling, using the built-in {\sc scikit-learn} \citep{scikit-learn} package functionality.  With this method, scattered light systematics are reduced to effectively the same size as the eclipses, forcing the neural network learn the shape of the eclipse signal in contrast to the systematics.  Figure \ref{fig:quantile} demonstrates the outcome of this scaling on a lightcurve dominated by a scattered light feature.  The top panel shows the unscaled lightcurve, while the bottom panel shows the quantile-scaled lightcurve processed as input to the neural network.  Although clearly more difficult for the human eye to distinguish eclipses in the scaled form, this method proved to be superior to the neural network for understanding subtle differences between eclipses and eclipse-shaped noise or systematics. This is not to say that we were able to avoid the neural network classifying such features as eclipses entirely, but this method did substantially diminish the problem.

While quantile scaling underemphasizes large features, it also has the effect of overemphasizing small features.  To our benefit, this helped in the identification of shallow eclipses.  However, we also found that our network will identify planet transits as well as small eclipse-like shapes in noise patterns, which became a substantial source of error (discussed further in Section \ref{sec:performance}).  Thus, we effectively made the decision to trade large noise effects for small noise effects.  We make no assertion that this trade was ideal, nor our method of classifying the eclipse shape vs. an EB directly.

\subsection{Training Data Collection}

A particular challenge of this effort was the collection and augmentation of the training data set.  Generally, the performance of a classifier will track directly with the quantity of training data samples to an asymptotic limit \citep[e.g.][]{Sun2017}.  Additional difficulties arise for our particular application in that eclipses can be vastly different in appearance, thus requiring a substantial amount of training data for a neural network to effectively generalize the features of an eclipse.  

At the time of development of our neural network ({\em TESS} Year 2), we had tens of millions of lightcurves, but only a handful of manually selected eclipsing binary lightcurves.  To gather a sufficiently-sized data set to effectively train our neural network by manually sorting through individual lightcurves would have been an intractable task.  As such, we progressively augmented our data set by iteratively using a weakly-trained neural network to find new training samples from among the full data set of lightcurves.  After each iteration of training and inference, we would select (i) lightcurves given a score near unity which clearly did not show an eclipse, or (ii) lightcurves given a score near zero which clearly showed an eclipse.  The former represented false positives and the latter false negatives.  We used these as properly labeled training samples in the next iteration of training, effectively filling gaps in the understanding of the neural network.  With each of these iterations, the neural network became progressively more capable.  By the time we were satisfied with the performance of the neural network, we had built a training set of $\sim$40k samples.

\subsection{Model Performance}
\label{sec:performance}

\begin{figure}
\centering    
\includegraphics[width=1.0\columnwidth]{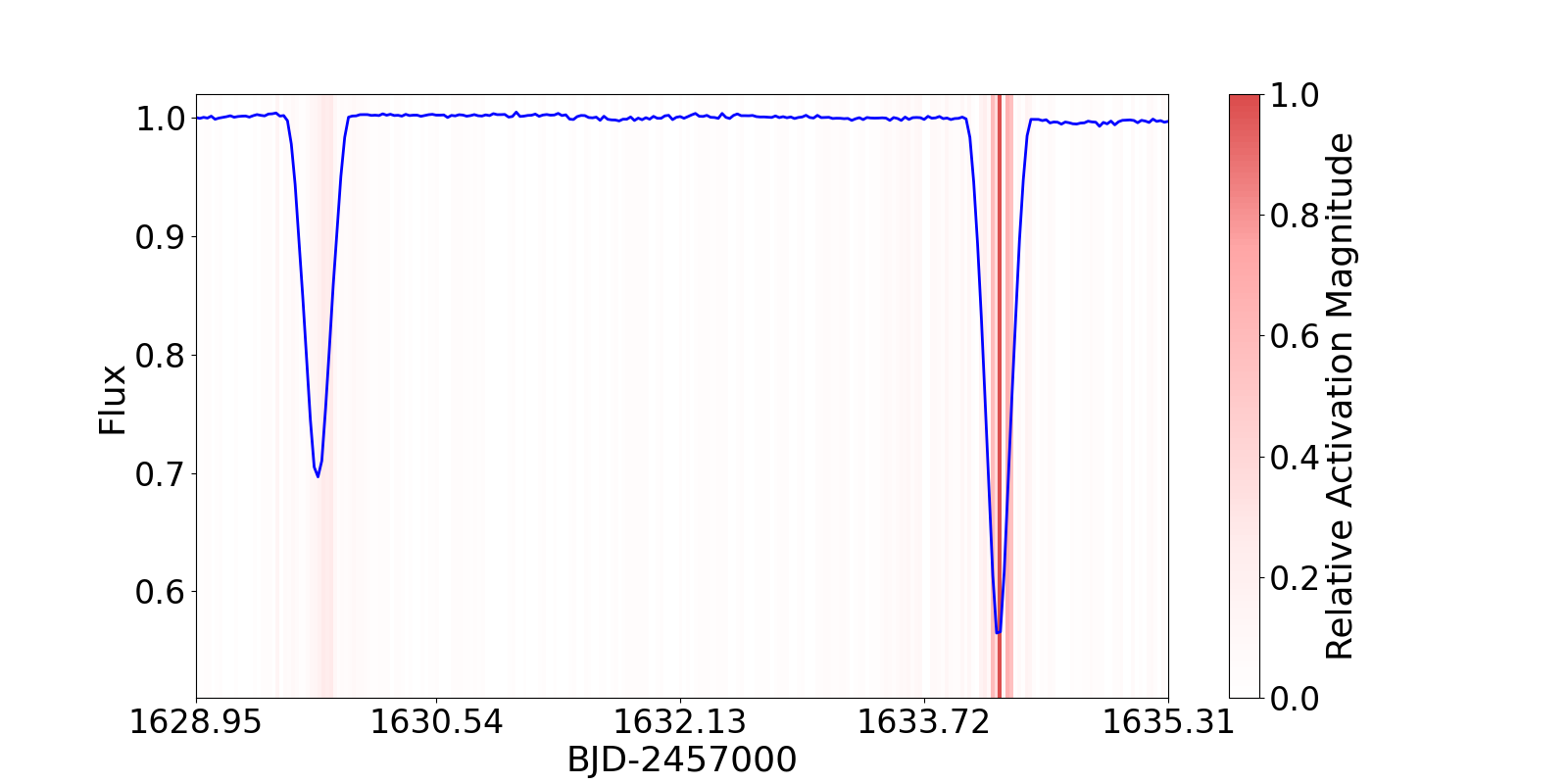}
    \caption{Neural network saliency map (shades of red according to activation magnitude) for a segment of the TIC 214716930 lightcurve in {\em TESS} Sector 12 (blue), demonstrating the activation of the neural network in the penultimate convolutional layer on the feature of the eclipse, made using {\sc Keras-vis} \citep{raghakotkerasvis}.}
    \label{fig:actmap}
\end{figure}

We emphasize that our neural network was {\em not} trained to find EBs.  It has no concept of repeated features or periodicity.  Rather, it was trained to find eclipses, or, more broadly, features resembling eclipses. We show how the neural network activates on the shape of eclipse in the saliency map of the activation weights of the penultimate layer in Figure \ref{fig:actmap}.  Note that the neural network will emphasize a single eclipse in determining the output score of the lightcurve.

Knowing that the neural network would be providing candidates for manual review rather than directly populating a list of near-certain EBs, we wanted to allow for interesting results that would not fit the conventional shape of an eclipse, e.g., a complex syzygy or lopsided eclipse.  This process has allowed us to find multiple star systems with complex outer orbital eclipses, among other interesting phenomena, a body of work on which establishes the effectiveness of our methods given that this neural network has contributed to many discoveries  (e.g. \citealt{2021AJ....161..162P,2021AJ....162..299P,2022ApJ...938..133P,2023MNRAS.524.4296P,2025ApJ...985..213P,2021ApJ...917...93K,2021AJ....162..234K,Kostov2022_quadcat1,2023MNRAS.522...90K,Kostov_quadcat2,2024ApJ...974...25K,2022MNRAS.513.4341R,2023MNRAS.521..558R,2024A&A...686A..27R,2022MNRAS.510.1352B,2024ApJ...975..121J,2024A&A...686C...3M,2025arXiv250415389O,2022ApJS..263...14C}).

Our results have been qualitative, with manual review of outputs in multiple stages (described further in Section \ref{sec:vetting}).  As such, an assessment of performance of the neural network against a data set of known EBs is hardly direct.  However, we provide such a comparison here to provide the reader with the context of our process as well as the contents of our catalog.

An evaluation of the model would be most complete with a section of the sky where we could consider all EBs within {\em TESS}' limiting magnitude to be known. As such, the {\em Kepler} \citep{2010Sci...327..977B} field provided an ideal testing ground, with the full data set having been thoroughly evaluated for the presence of EBs, resulting in the production of an EB catalog\citep{2011AJ....141...83P,2011AJ....142..160S,2012AJ....143..123M,2014AJ....147...45C,2014PASP..126..914C,2015MNRAS.452.3561L,2016AJ....151...68K,2016AJ....151..101A}, hereafter referred to as the ``{\em Kepler} EB catalog.''  By comparing the number and characteristics of the lightcurves identified in our catalog against the 2,920 EBs of the {\em Kepler} EB catalog, we could make an estimate of the performance of our catalog. 

We also considered the catalog of 4,584 {\em TESS} EBs from short cadence data in Sectors 1-26 \cite{2022ApJS..258...16P}, hereafter referred to as the ``{\em TESS} EB catalog.''  Although less comprehensive in terms of a full survey of a section of the sky, a direct comparison to {\em TESS} EBs rather than {\em Kepler} allows for fewer independent sources of error such as, e.g., different noise amplitudes or photometric capabilities.

We cross-referenced our catalog with the the {\em Kepler} field boundaries\footnote{Available from MAST at \url{https://archive.stsci.edu/missions/kepler/ffi_footprints/morc_2_ra_dec_4_seasons.txt}}, finding each object from our catalog that would have been observed by the original {\em Kepler} mission, resulting in 9,768 unique TIC IDs.  We then reduced the {\em Kepler} EB catalog to those objects with $T_{\text{mag}}<15$, as this was the limit of our lightcurve construction, resulting in 2,458 of the original 2,920 EBs.  Our catalog contains 1,371 of these 2,458 EBs, or $\sim$55.8\%.  To compare our catalog to the {\em TESS} EB catalog, we cross-referenced our catalog with lists of the two-minute targets for {\em TESS} sectors 1-26,\footnote{\url{https://tess.mit.edu/public/target_lists/target_lists.html}} from which the {\em TESS} EB catalog is derived.  In total, there were 507,898 unique two-minute targets in these sectors, 8,910 of which were identified by our neural network. Comparing directly to the {\em TESS} EB catalog revealed that our neural network found 3,884 of the 4,584 EBs therein, or  $\sim$84.7\%.  In this comparison of true positives, it is clear that our neural network performed far better against the {\em TESS} EB catalog, which we assess to be likely due to the systematic differences between the {\em TESS} and {\em Kepler} data.

\begin{figure}
    \centering
    \includegraphics[width=.49\textwidth]{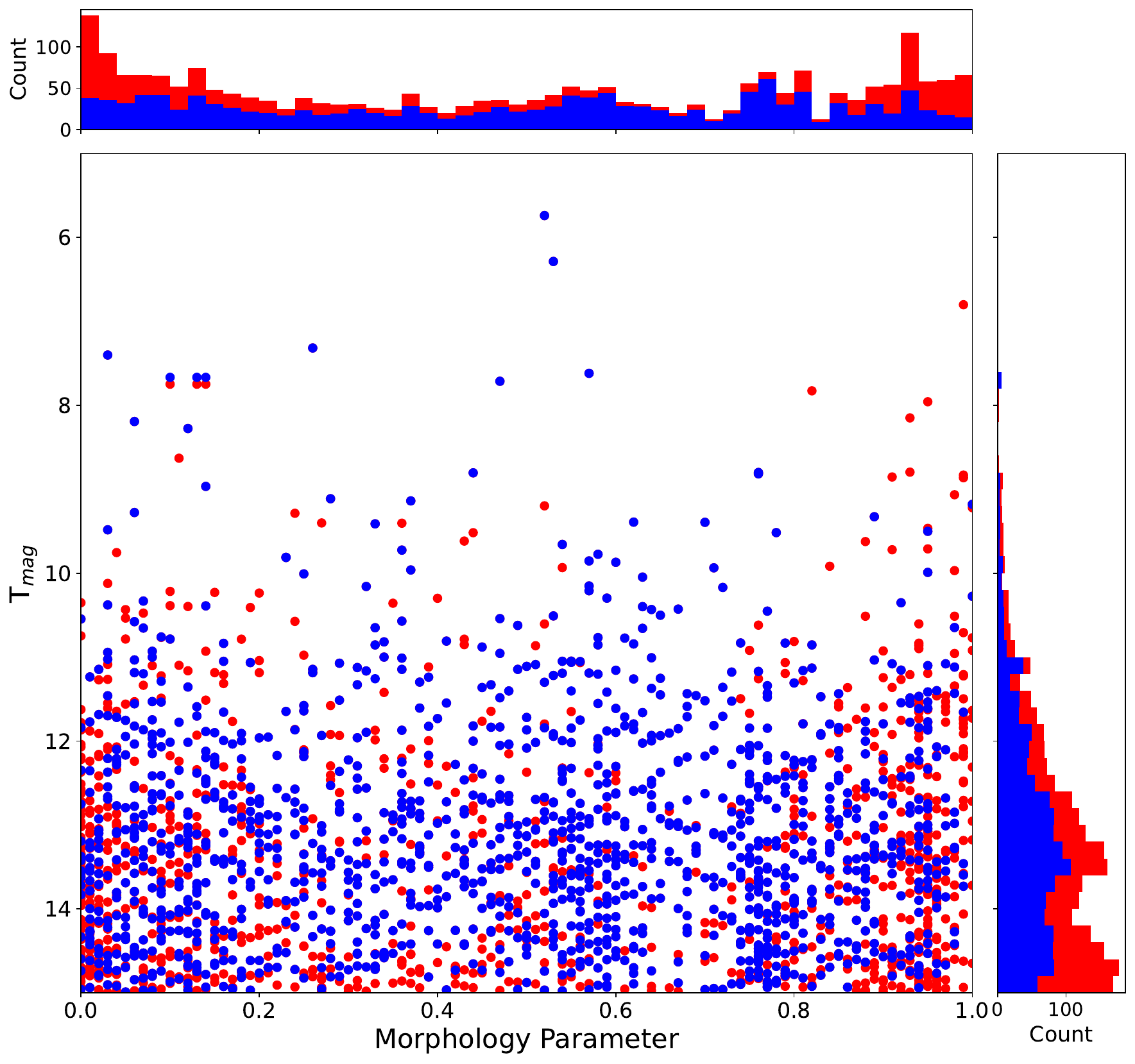}
    \includegraphics[width=.49\textwidth]{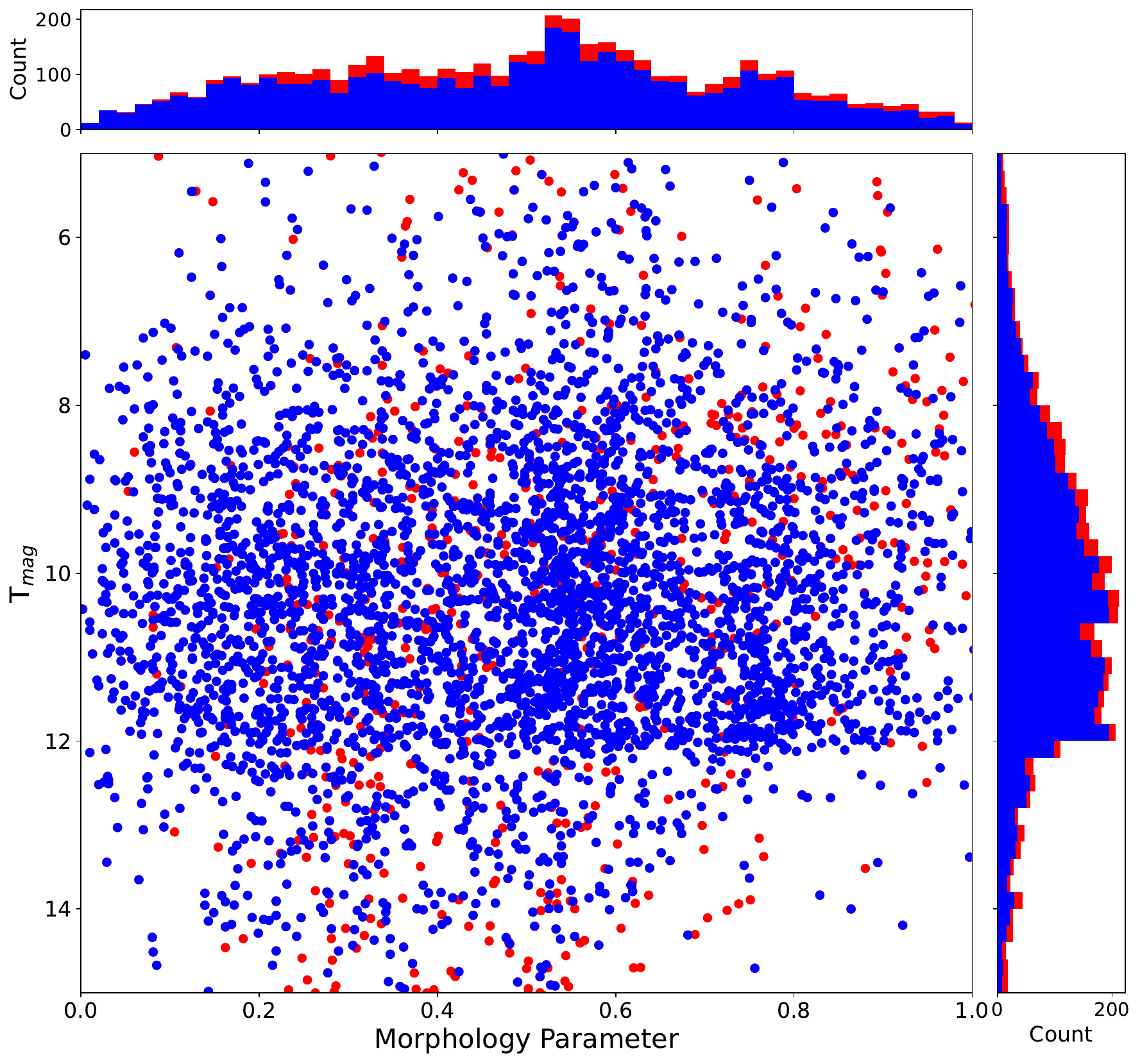}
    \caption{(left panel) Scatter plot of the morphology parameter vs. $T_{\text{mag}}$ for the {\em Kepler} EBs found (blue) and not found (red) by our neural network.  (right panel) The same for the {\em TESS} EB catalog.  For both panels, the top histogram shows the distribution over morphology, while the right histogram shows the distribution by T$_{\text{mag}}$. Our performance against the {\em Kepler} EBs shows a clear preference for the central morphology range as well as diminished performance as magnitude decreases.  Note that the $T_{\text{mag}}$ distribution for the {\em TESS} EBs is limited by their selection as two-minute cadence targets, hence the apparent cutoff at $T_{\text{mag}}\approx12$.}
    \label{fig:morph_v_tmag}
\end{figure}

\begin{figure}
    \centering
    \includegraphics[width=.49\textwidth]{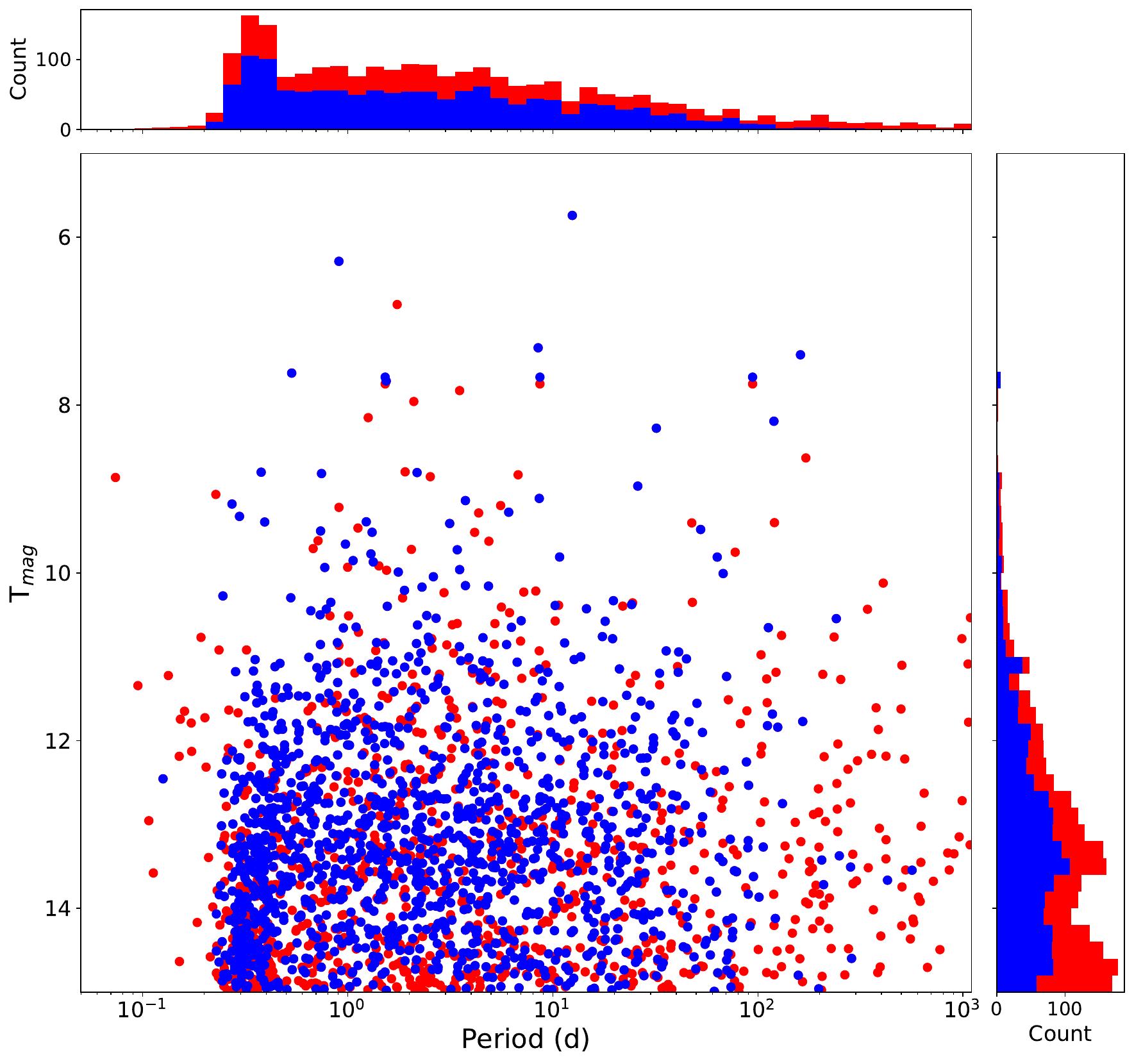}
    \includegraphics[width=.49\textwidth]{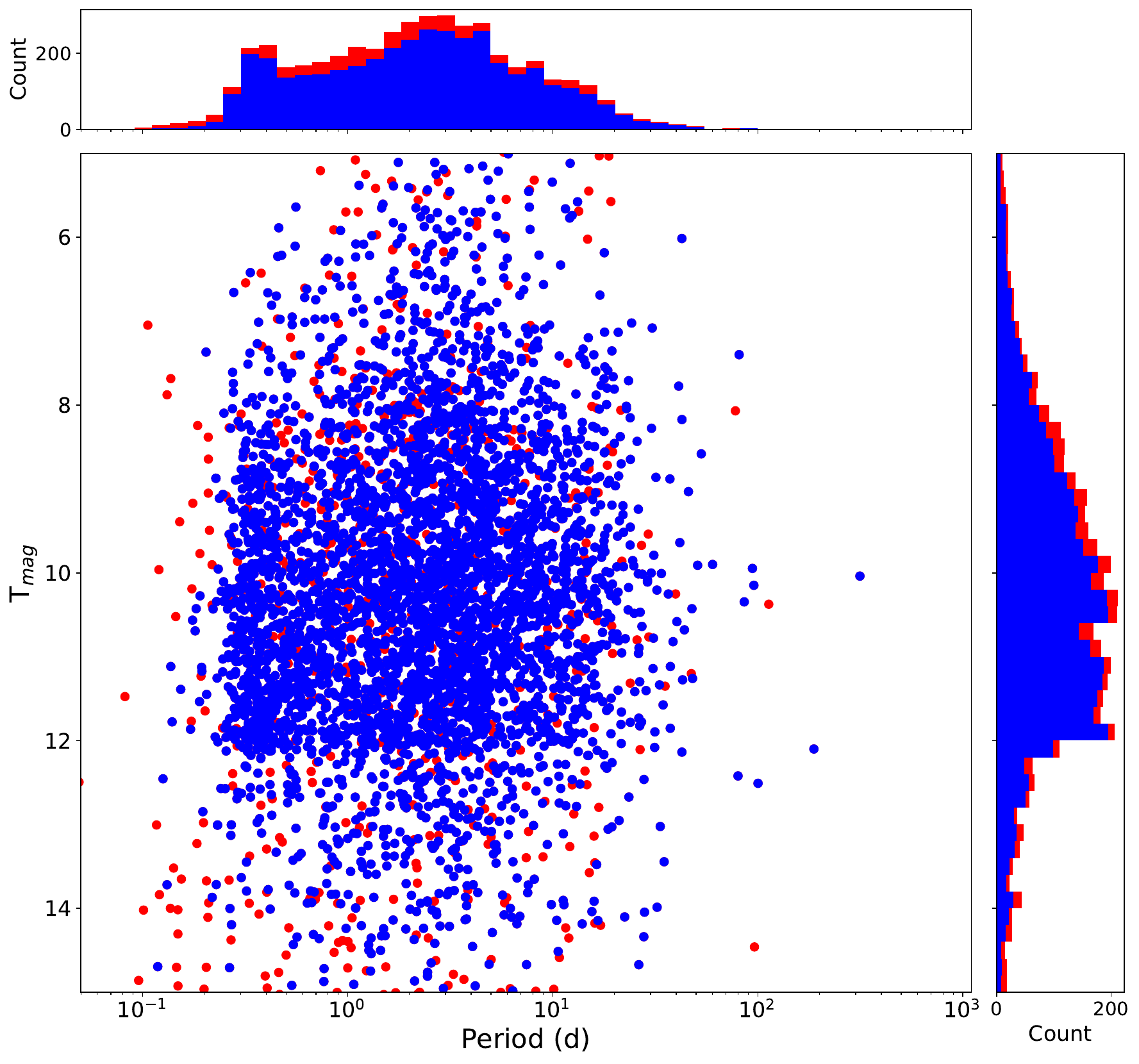}
    \caption{(left panel) Scatter plot of the EB period vs. $T_{\text{mag}}$ for the {\em Kepler} EBs found (blue) and not found (red) by our neural network.  (right panel) The same for the {\em TESS} EB catalog.  For both panels, the top histogram shows the distribution over EB period, while the right histogram shows the distribution by $T_{\text{mag}}$. Again, our performance against the {\em Kepler} EBs shows a  diminished performance as magnitude decreases.  As with Figure \ref{fig:morph_v_tmag}, note that the $T_{\text{mag}}$ distribution for the {\em TESS} EBs is limited by their selection as two-minute cadence targets, hence the apparent cutoff at $T_{\text{mag}}\approx12$.}
    \label{fig:period_v_tmag}
\end{figure}

\begin{figure}
    \centering
    \includegraphics[width=.49\textwidth]{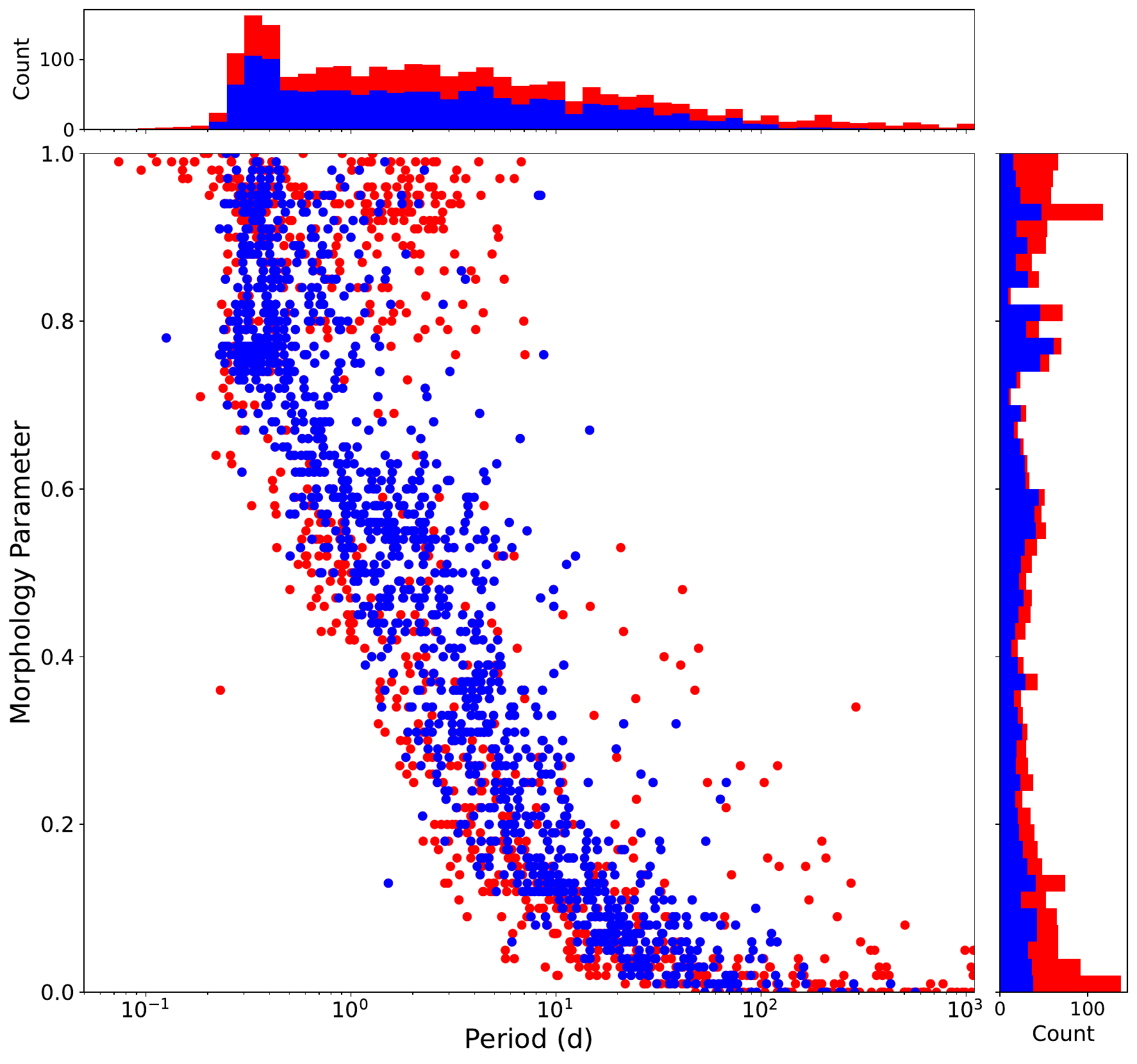}
    \includegraphics[width=.49\textwidth]{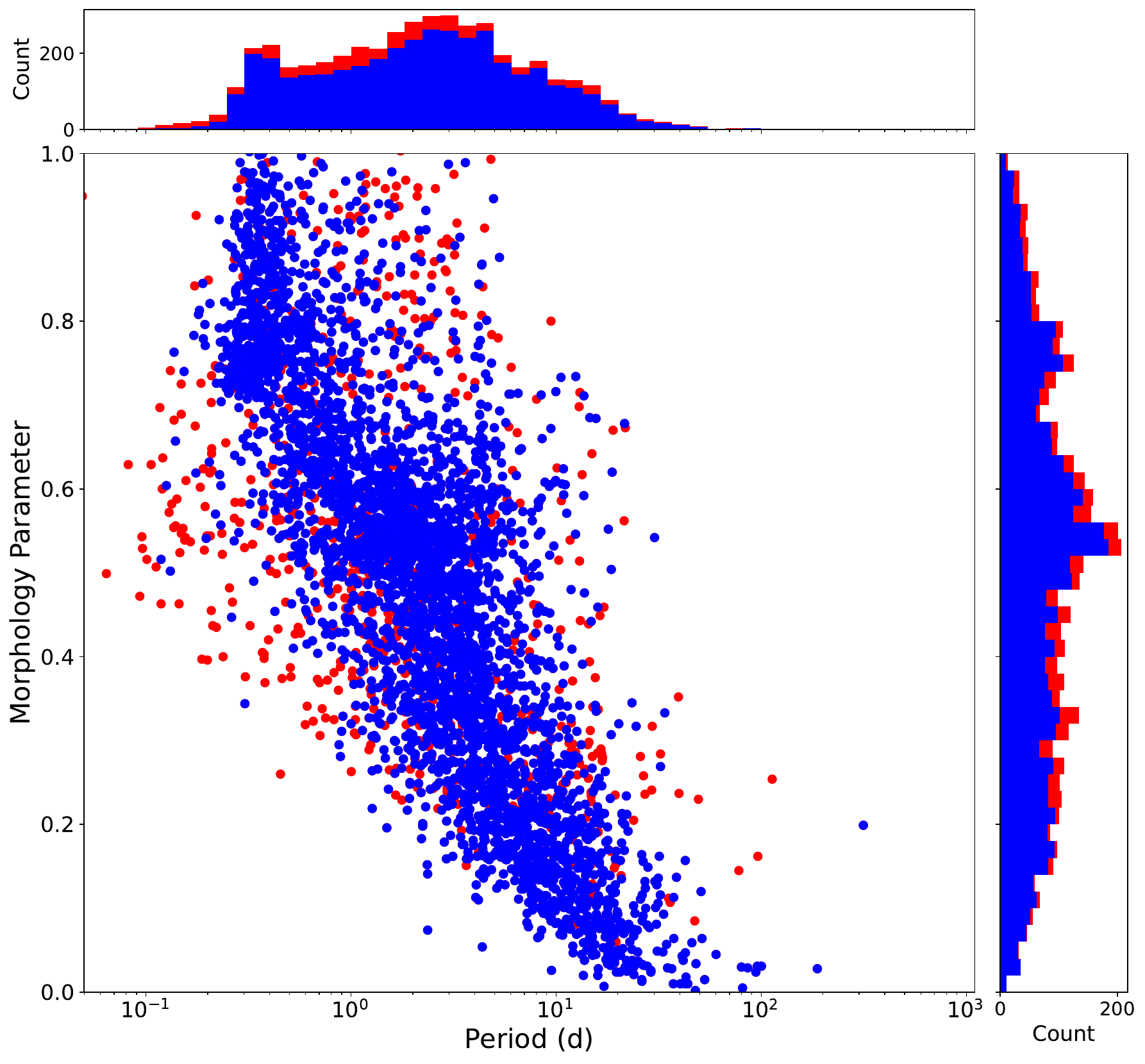}
    \caption{(left panel) Scatter plot of the EB period vs. morphology parameter for the {\em Kepler} EBs found (blue) and not found (red) by our neural network.  (right panel) The same for the {\em TESS} EB catalog.  For both panels, the top histogram shows the distribution over EB period, while the right histogram shows the distribution by morphology parameter.  Besides revealing the relationship between EB period and morphology as described by \citet{2011AJ....141...83P}, we can see a pattern of weakness of our neural network for EBs at the extremes of the period range for any given morphology range.}
    \label{fig:period_v_morph}
\end{figure}

In Figure \ref{fig:morph_v_tmag}, we show a scatter plot of the morphology parameter vs. T$_{\text{mag}}$ of the {\em Kepler} EBs found (blue) and not found (red) by our neural network.  The morphology parameter (described further in \citealt{2011AJ....141...83P}) is a measure of the EB type, with values less than 0.5 corresponding to detached EBs, values in the range 0.5-0.7 being semi-detached EBs, 0.7-0.8 being overcontact EBs, and greater than 0.8 corresponding to ellipsoidal or unknown classifications.  We exclude analysis where the morphology parameter was given a value of -1, indicating the lightcurve was unclassifiable by the methods of \citealt{2011AJ....141...83P}. Furthermore, we make the same comparison using the EB period vs. T$_{\text{mag}}$ in Figure \ref{fig:period_v_tmag} and EB period vs. morphology parameter in Figure \ref{fig:period_v_morph}.  We can make two conclusions from these figures:

(i) The T$_{\text{mag}}$ histograms in the left panels ({\em Kepler} comparisons) of Figures \ref{fig:morph_v_tmag} and \ref{fig:period_v_tmag} show a clear decrease in performance with decreasing T$_{\text{mag}}$, while we do not see the same decrease in the right panels ({\em TESS} comparisons) of the same figures.  Although there is a somewhat artificial lower limit on magnitude in the {\em TESS} 2-minute cadence targets due to the selection bias for bright stars, we can still see a generally uniform trend of the fraction identified in the T$_{\text{mag}}$ histograms of the {\em TESS} comparisons.  However, these targets are still somewhat idealized in comparison to a full sample over a section of the sky.  As such, we consider our neural network's performance against the {\em TESS} EB catalog 2-minute cadence targets to be an upper limit, while the performance against the {\em Kepler} EB catalog should be considered a lower limit.

(ii) For the {\em Kepler} (left) panels in Figures \ref{fig:morph_v_tmag} and \ref{fig:period_v_morph}, the morphology histograms show a clear weakness of our neural network for the extremes of the morphology range.  Other than showing the general relationship between EB period and morphology, Figure \ref{fig:period_v_morph} demonstrates a clear trend of weakness for our neural network in identifying EBs at the extremes of the period range for any given morphology range in both the {\em Kepler} (left) and {\em TESS} (right) comparisons.  That is, the red points seem to dominate the blue both to the far left and far right of the general trendline.  We assess that both of these trends demonstrate a weakness in generalization of the neural network to less common types of eclipse patterns.

Having examined the nature of our true positives and false negatives, we turned to the much larger set of false positives. As previously discussed, 1,371 of the 9,768 unique TIC IDs from the {\em Kepler} field and 3,884 of the 8,910 in the {\em TESS} sector 1-26 short cadence targets found by our neural network were true positives, leaving the remaining 8,397 ($\sim$86\%) of the {\em Kepler} sample and 5,026 ($\sim$56\%) of the {\em TESS} sample as false positives.  Naturally, the questions arise as to what are these false positives and why are they so numerous. To determine their nature, we manually examined a subset of bright false positives in the {\em Kepler} field with T$_{\text{mag}}<10$, totaling 126 unique TIC IDs.  

\begin{figure}
    \centering
    \includegraphics[width=1.0\columnwidth]{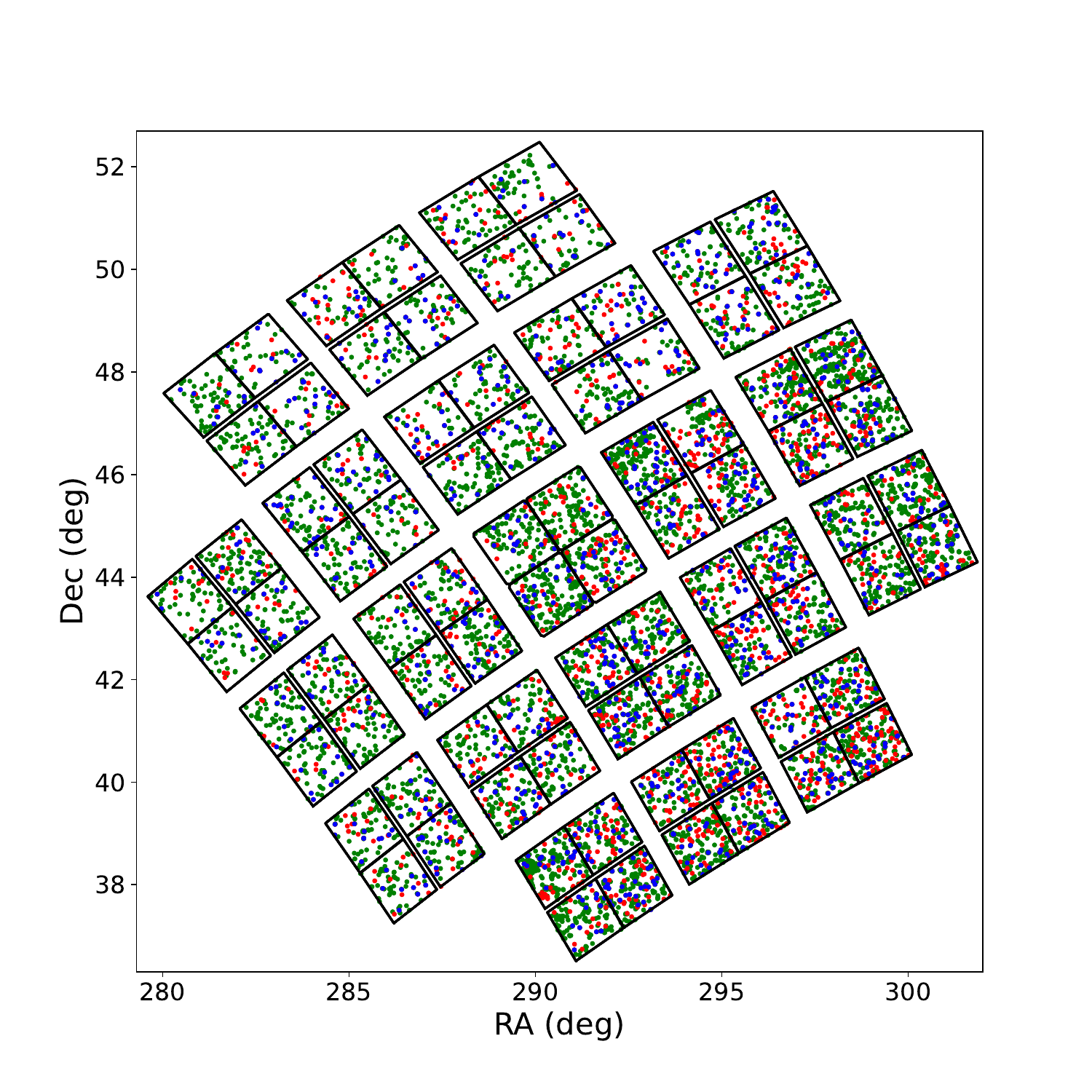}
    \caption{As compared to the {\em Kepler} EB catalog, true positives (blue), false negatives (red), and false positives (green) identified by our neural network, shown here in the {\em Kepler} field. Local groupings of false positives are attributable to very bright EBs contaminating the lightcurves of adjacent stars or systematics such as scattered light, which closely resembled the shape of an eclipse in the overlapping {\em TESS} sector 14.}
    \label{fig:kepfield}
\end{figure}

Of these lightcurves, we found that 16 ($\sim$13\%) showed clear EBs.  Since these were not in the {\em Kepler} EB catalog, we assumed that most of these are likely the result of blending.  That is, the 21'' pixels of {\em TESS} will frequently cause the lightcurves of bright stars to show in the lightcurves of dimmer close neighbors.  To demonstrate this effect, we show the {\em Kepler} field with true positives, false positives, and false negatives in Figure \ref{fig:kepfield} where the clustering of several large groups of false positives can be seen, likely as a result of blending. While this was only a problem in ($\sim$13\%) of our false positives with T$_{\text{mag}}<10$, it can be reasonably expected that lightcurves of dimmer stars will show this type of contamination more frequently. We examined each of the 16 lightcurves showing clear EBs and confirmed that contamination from nearby brighter stars was indeed the source of the signal.  However, we assessed 2 of the 16 lightcurves showing clear eclipses as the true source of an EB, TIC 26542657 (KIC 12013550) and TIC 63454475 (KIC 10342012) .  We confirmed that neither of these targets are present in the {\em Kepler} EB catalog.  Furthermore, we found that there was no {\em Kepler} data available for TIC 63454475, while the {\em Kepler} lightcurves for TIC 26542657 (first identified as an EB in the {\em TESS} EB catalog by \citealt{Prsa2022}) indeed showed no eclipses, confirming that neither of the targets were missed accidentally in the creation of the {\em Kepler} EB catalog.  Given that TIC 26542657 has Kepler lightcurves without eclipses, we assess that it must be a higher order system, likely a triple, with so-called `disappearing eclipses.' We will discuss this particular system further (also confirming it as a triple) as well as identifying other examples of this type of system in Section \ref{sec:interesting}.  Although it is beyond the scope of this effort to analyze these systems, we note briefly that the changes in binary inclination causing periods of eclipsing and non-eclipsing behavior are driven by interaction with the outer body, and we refer the reader to \citet{2022Galax..10....9B} for a detailed discussion of the nature of this type of triple system, among others.  Given we found one such system out of only 126 in our cross-match with the {\em Kepler} EB catalog with T$_{\text{mag}}<10$, we expect there to be several more such systems in our 8,397 false positives from the {\em Kepler} field, which may merit an investigation in its own right.

\begin{figure}
    \centering
    \includegraphics[width=.49\linewidth]{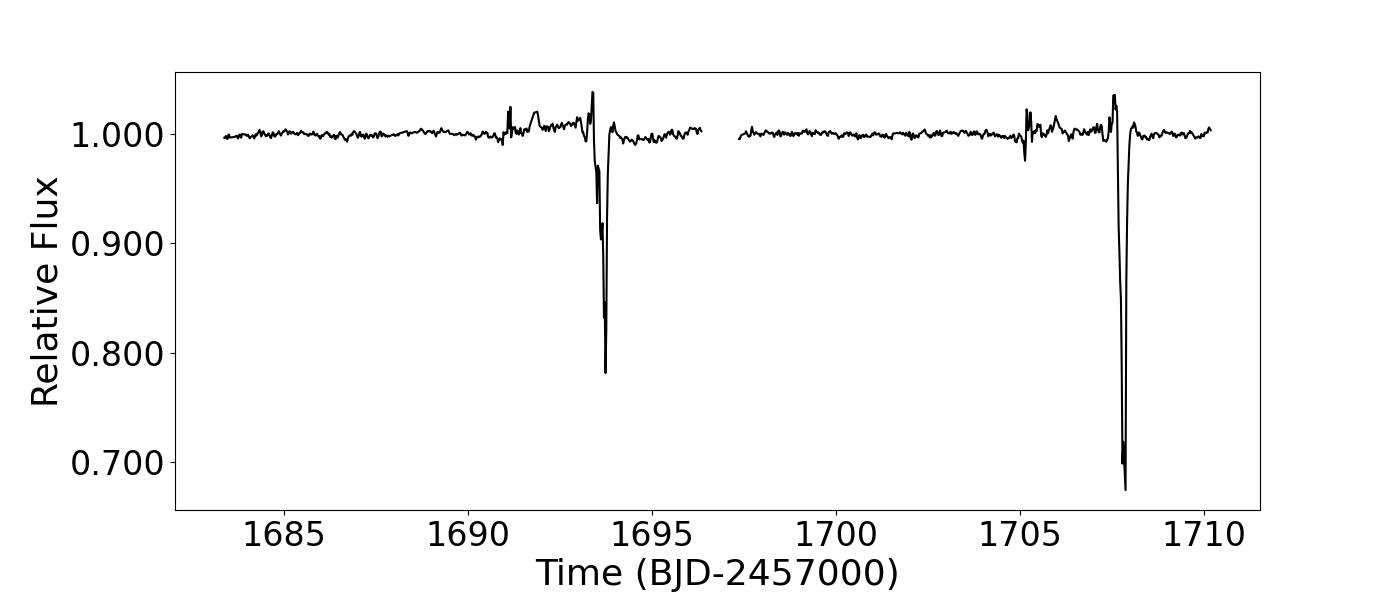}
    \includegraphics[width=.49\linewidth]{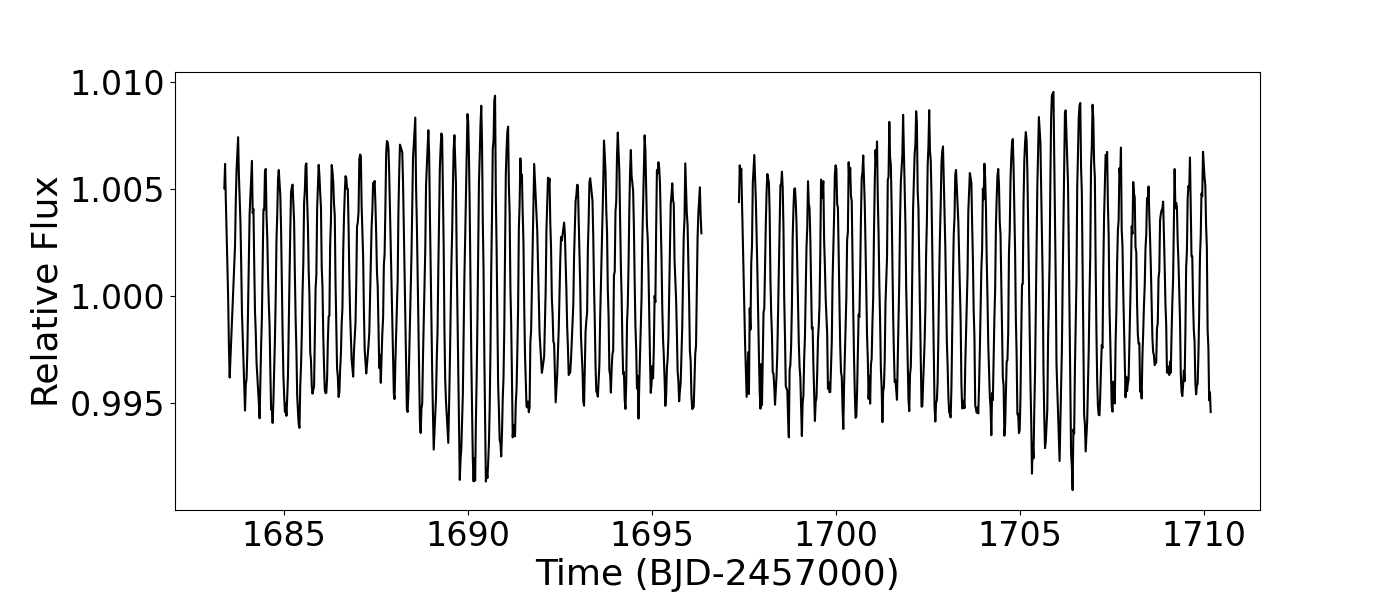}
    \includegraphics[width=.49\linewidth]{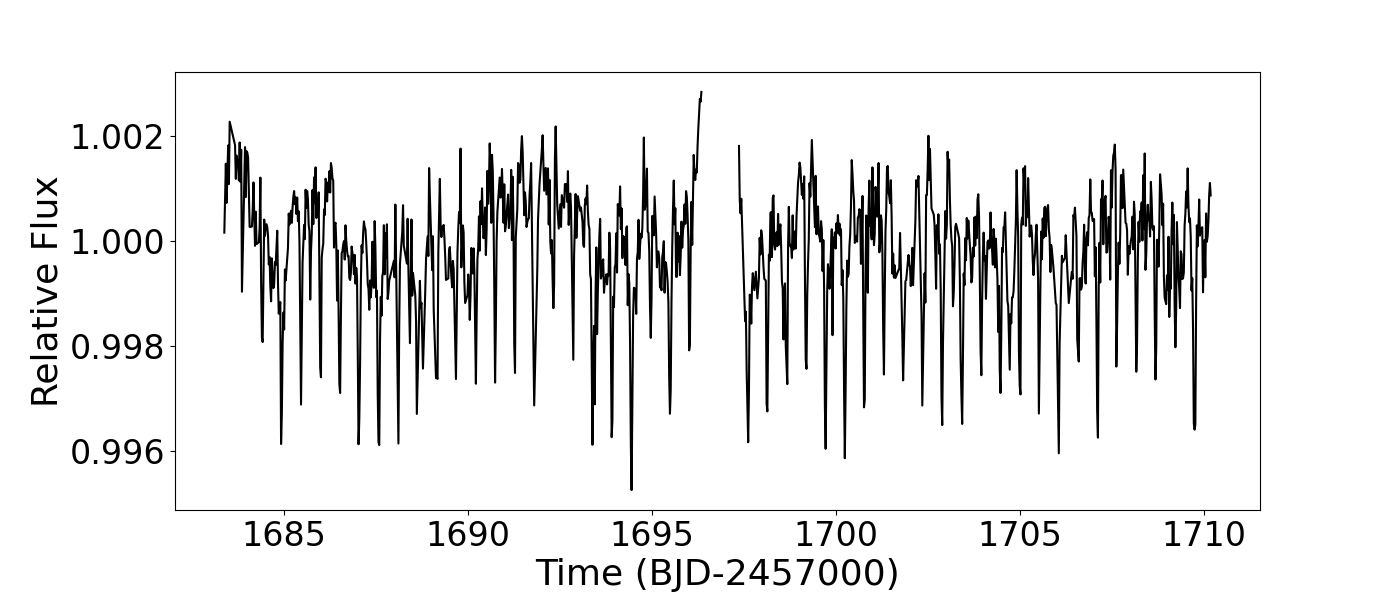}
    \includegraphics[width=.49\linewidth]{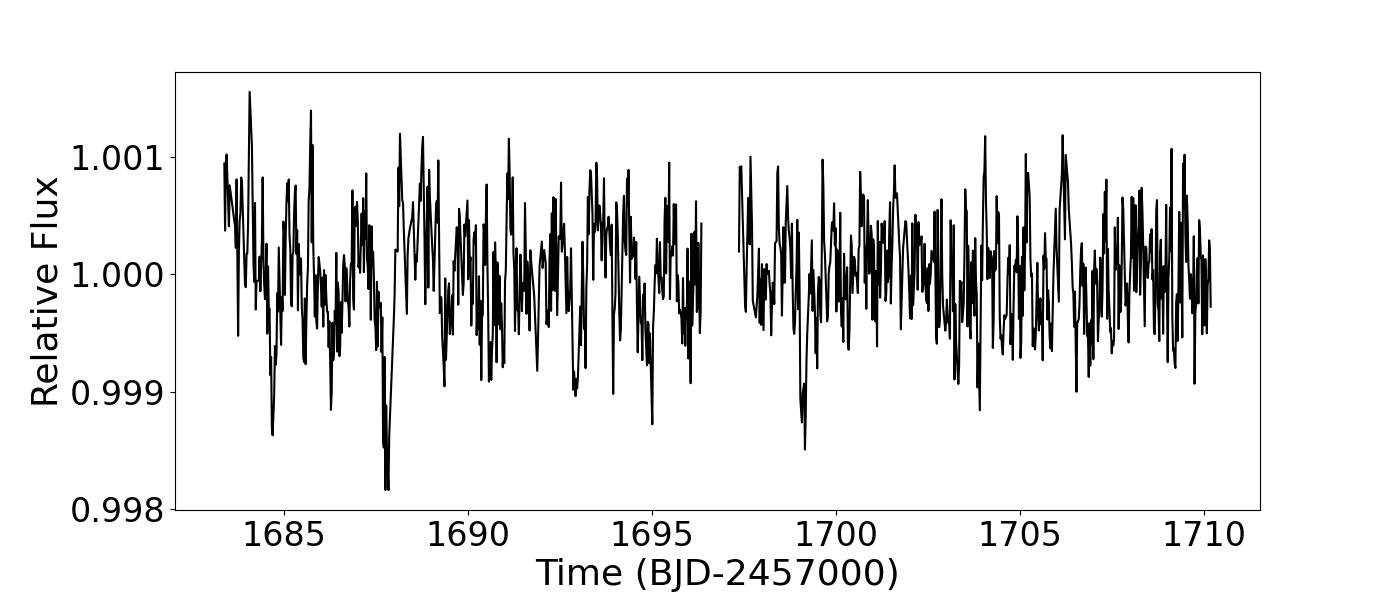}
    \caption{Examples of several different types of false positives returned by our neural network shown by {\em TESS} sector 14 lightcurves. The TIC 120693310 lightcurve (upper left) demonstrates scattered light resembling eclipses, a particularly difficult false positive to train against.  The TIC 120426180 lightcurve (upper right) shows pulsations with local minima broad enough to be classified an an eclipse.  The TIC 26656569 lightcurve (bottom left) shows a clear EB, but this lightcurve is contaminated by the nearby TIC 26656583 (KIC 11560447), which is in the {\em Kepler} EB catalog.  The TIC 123447105 lightcurve (bottom right) shows a noise pattern with several features that could be mistaken for eclipses.}
    \label{fig:fps}
\end{figure}

Returning to our false positives from the $T_{\text{mag}}<10$ sample, oscillations resembling eclipses comprised 16 ($\sim$13\%) of the false positives, while 21 ($\sim$17\%) showed scattered light systematics resembling eclipses, as in Figure \ref{fig:quantile}.  The latter were particularly prominent in {\em TESS} sector 14, which overlapped with the {\em Kepler} field.  The bulk of the false positives, 74 ($\sim$59\%), contained noise patterns that resembled eclipses. We provide examples of a sample of these false positives in Figure \ref{fig:fps}.

Another type of scientifically valuable false positive found by our neural network were planet transits.  We compared our catalog to the NASA Exoplanet Archive {\em Kepler} catalog and found an overlap of 1,502 of the 8,397 TIC IDs.  125 of these were determined to be genuine exoplanet candidates.  The overlap with the NASA Exoplanet Archive {\em TESS} exoplanet candidate catalog\footnote{Both {\em Kepler} and {\em TESS} exoplanet catalogs are available at \url{https://exoplanetarchive.ipac.caltech.edu/index.html}} is even more pronounced.  Of the 7,576 {\em TESS} exoplanet candidates, our catalog contains 2,445 ($\sim$32\%). Given how we scale the lightcurves (see Figure \ref{fig:quantile}), the neural network is generally not able to discern a difference between an eclipse and a clean transit signal, as it was not trained to do so.  As such, {\em we expect that our catalog is also rich with planet candidates}.  Of particular interest, since our neural network has no periodicity requirement, we expect there to be many single transit events that evade discovery through periodicity-based analyses.

\subsection{Limitations and Caveats of Our Results}

We emphasize, again, that our neural network was trained to find features resembling eclipses, not EBs.  Our results were qualitative and manually reviewed.  As such, the comparisons provided in the previous section should not be considered a fully accurate measure of the performance of the neural network as much as context for the reader to understand our process and the contents of our catalog. Our method should be considered as a means of reducing an extremely large data set (hundreds of millions of lightcurves) to a much smaller, manageable data set with a high concentration of scientific value.

\begin{figure}
    \centering
    \includegraphics[width=0.95\linewidth]{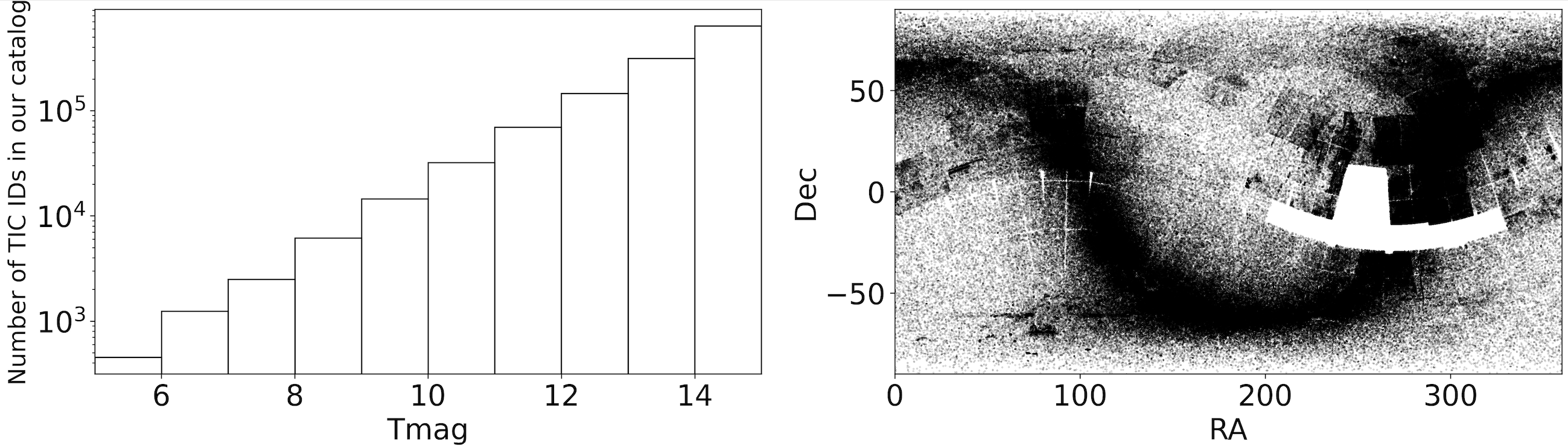}
    \caption{Left: Number of {\em TESS} targets exhibiting eclipse-like features from Sectors 1-82 as a function of {\em TESS} magnitude. Right: Corresponding RA vs Dec.}
    \label{fig:EB_histo}
\end{figure}

Our full catalog consists of 1,223,603 unique TIC IDs with lightcurves that our neural network gave a score of $\geq$ 0.9.  The distribution of these, in terms of {\em TESS} magnitude and ecliptic coordinates, is shown in Fig. \ref{fig:EB_histo}. Most of the targets are on the faint side, with a median {\em TESS} magnitude of $\approx14$, and the majority are near the Galactic plane. It is from these candidates that we distilled the much smaller catalog of vetted EBs to be discussed in the remainder of this paper.  This catalog could be employed by interested researchers with the caveats of the analysis in the preceding section.  We use the {\em Kepler} EB catalog comparison as the lower end of our estimate and the {\em TESS} EB catalog comparison as the upper end of our estimate in the following summary of caveats and contents: 
\begin{enumerate}
\item This is a catalog of lightcurves with eclipse-like features, not an EB catalog.
\item 14\% - 44\% of the catalog should be expected to be EBs.
\item It follows that 56\% - 86\% of the catalog should be expected to {\em not be} EBs.  These will consist of contaminated lightcurves, transiting exoplanets, dippers, systematics, and other sharp ellipsoidal features.
\item The catalog should be expected to contain 55\% - 84\% of all EBs in {\em TESS} lightcurves for Sectors 1-82 with T$_{\text{mag}}<15$.  The fraction of completeness will decrease with brightness.
\item The full catalog is entirely unvetted and we offer no guarantee as to its contents.  We have, however, found the neural network outputs to be scientifically valuable, so we offer it to the community in its entirety for their own purposes.
\end{enumerate}

With these caveats in mind, we transition in the next section to explaining how we analyzed a subset of these, producing 10,001 as genuine EBs for inclusion in a well-vetted catalog.

\section{Vetting and Validation of the EB candidates}
\label{sec:vetting}

Contamination from nearby (in terms of sky projection) eclipsing binary stars can result in a not-insignificant contribution of light to the aperture used to extract the target's lightcurve -- and thus mimic eclipse-like signals that seem to come from the target star. This is a common occurrence in {\em TESS} observations where it is not unusual to see one or more field stars within 2-3 pixels of the target star (where each pixel is $\approx21$ arcsec), often even falling within the same pixel, thus adding their often significant signal to that of the target of interest \citep[e.g.,][and references therein]{2020RNAAS...4..204H,2020RNAAS...4..206H,2024AAS...24323105K,Kostov2022_quadcat1,Kostov_quadcat2}. Thus detection of eclipse-like features from a particular star in {\em TESS} does not immediately tell us their origin and additional investigations are required before an EB candidate is verified. In the absence of radial velocity measurements to confirm or rule out a potential EB (or planet) candidate, one can capitalize on the information-rich content of the available photometric data. 

\subsection{Photocenter Vetting}

A particularly powerful method to constrain the pixel position of the source of detected eclipses (or transits) is the photocenter-based analysis that is routinely used in surveys aimed at finding transiting exoplanets \citep[e.g.,][and references therein]{2014AJ....147..119C,2015ApJ...812...46T,2018ApJS..235...38T,2018PASP..130f4502T,2019AJ....157..124K,2021MNRAS.504.5327A,2025arXiv250209790V}. Briefly, the method uses the target pixel files to measure the in-eclipse center-of-light (``photocenter'') pixel position for each eclipse detected in the difference image\footnote{The difference image is created by subtracting the out-of-eclipse target pixel data from the in-eclipse pixel data, both covering the measured duration of the eclipse.}, and compare it to the out-of-eclipse photocenter and/or to the catalog pixel position of the target. If there is no statistical difference between these and the in-eclipse photocenter, the eclipses are considered to be `on-target'; otherwise the eclipses are 'off-target', likely coming from a nearby field EB, and the candidate is marked as a false positive. 

\begin{figure}
    \centering
    \includegraphics[width=0.95\linewidth]{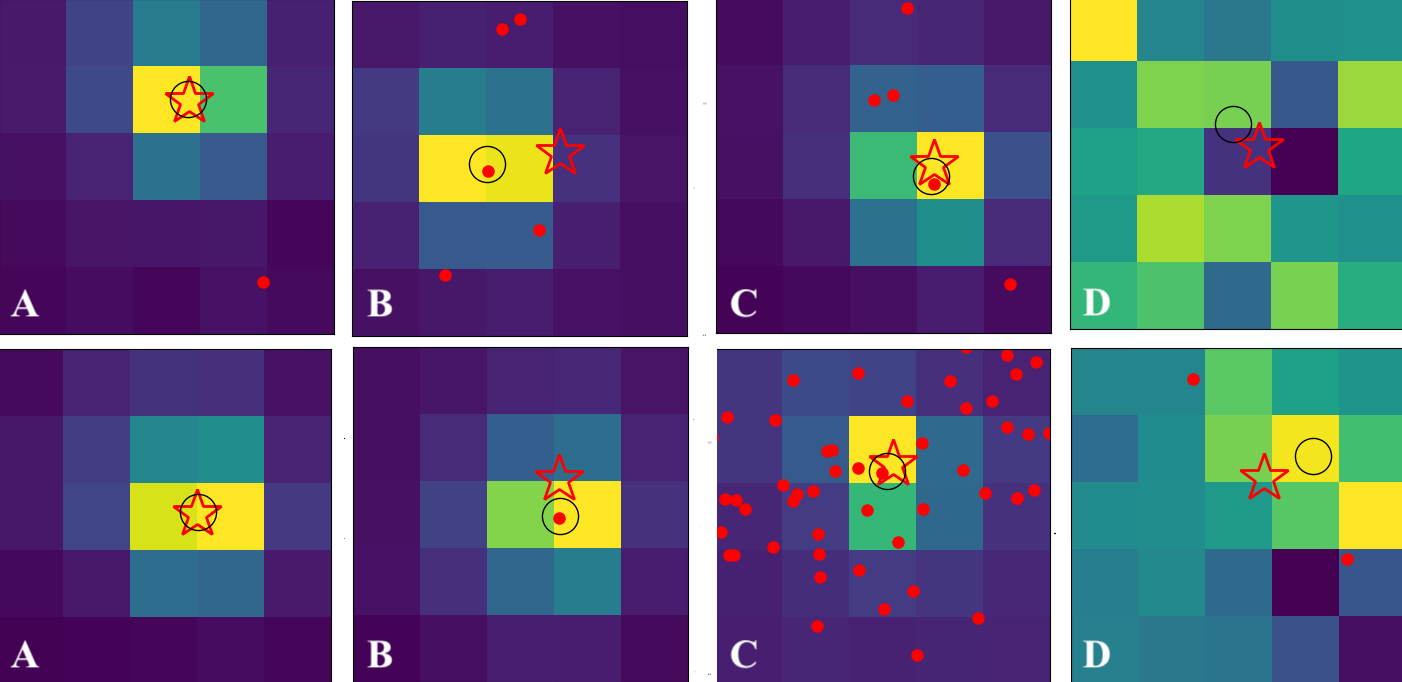}
    \caption{Example 5x5 pixels sector-averaged difference images used for photocenter measurements of {\em TESS} EB candidates. The red star symbols represent the pixel position of the target star, the open black circles represent the sector-averaged photocenter, and the small red symbols represent nearby field stars that are bright enough to produce the detected eclipses as contamination to the target star. The first three columns from the left (labels A, B, and C, respectively) show difference images that are reasonably well-suited for photocenter measurements. The photocenters in the first column (A) indicate that the detected eclipses are on-target, whereas the second (B) and third (C) columns show false positives due to significant photocenter offset. Column B highlights the case for a well-separated target and contaminator, while the two are within the same pixel in column C -- and also reside in crowded fields. Column D shows difference images that are dominated by systematic effects, thus making them inadequate for reliable photocenter measurements. }
    \label{fig:example_centroids}
\end{figure}

Obtaining robust photocenter measurements depends on multiple factors such as the SNR of the target pixel files, depth of the detected eclipses, presence of nearby comparably-bright field stars (worst case scenario much brighter {\it and} variable on timescales comparable to the duration of the eclipses), etc. In theory, the difference images used to measure the per-eclipse photocenters resemble a well-defined, bright pixelated spot superimposed on an otherwise dark background (Fig. \ref{fig:example_centroids}, first three columns from the left). In practice, the difference images are often distorted due to various astrophysical, systematic, or instrumental effects (Fig. \ref{fig:example_centroids}, rightmost column), making the corresponding photocenter measurements unreliable \citep[e.g.,][]{2022MNRAS.513..102C,2023MNRAS.521.3749M,Kostov2022_quadcat1,Kostov_quadcat2}. 

Overall, based on our experience with {\em TESS} data -- and depending on the peculiarities of the specific target -- measurements of genuine photocenter offsets of $\gtrsim0.2-0.3$ pixels (i.e., $\gtrsim4-6$ arcsec) are often trustworthy \citep[][]{2023MNRAS.521.3749M,Kostov2022_quadcat1,Kostov_quadcat2}. However, considering measured offsets of $\lesssim0.1-0.2$ pixels as significant can be extremely challenging to potentially even impossible. Thus, throughout this work we adopt a photocenter offset threshold of 0.2 pixels ($\approx4$ arcsec) such that cases below that are considered as likely `on-target' and those above -- `off-target'.

Ideally, genuine photocenter offsets 3-5 times larger than this threshold (i.e.,$\sim12-21$ arcsec) should be relatively easy to measure (and trust). Thus, to account for potential false positives due to known EBs, we first evaluated whether the {\em TESS} EB candidates our neural network identified are within a 21 arcsec (1 {\em TESS} pixel) sky-projected separation of EBs listed in various catalogs. In particular, we queried EB catalogs from ASAS-SN ($\sim150,000$ EBs), ATLAS ($\sim30,000$ EBs), Gaia ($\sim2,200,000$ EBs), OGLE ($\sim430,000$ EBs), Simbad ($\sim2,400,000$ EBs, $\sim200,000$ spectroscopic binaries), {\em TESS} ($\sim50,000$ EBs, $\sim10,000$ planet candidates from ExoFOP-TESS (\url{https://exofop.ipac.caltech.edu/tess/})), VSX ($\sim900,000$ EBs), and WISE ($\sim50,000$ EBs), taking into account the respective overlaps between the different data sets. Unsurprisingly, given the pixel size of {\em TESS} and the corresponding crowding, about a quarter of our $\sim1.2$ million candidates fulfill the above criteria (see Table \ref{tab:overlaps}). We note that this consideration does not immediately rule these out as false positives, but it marks them as likely suspects. Conversely, those that are not within 1 pixel of known EBs ($\sim900,000$ TICs) can potentially be confirmed as bona-fide new EBs through careful photocenter analysis. For the benefit of the community, we provide these targets as an auxiliary data set (see Table \ref{tab:900k}).

\begin{table}[!ht]
\scriptsize
\centering
\begin{tabular}{l|r}
\hline
\hline
Source & Number of targets \\
\hline
This work & 1223603 \\
ASAS-SN & 119126 \\
ATLAS & 21244 \\
Gaia & 259631 \\
OGLE & 6841 \\
Simbad & 268669 \\
Simbad$^a$ & 14076 \\
TESS$^b$ & 62681 \\
VSX & 136190 \\
WISE & 41085 \\
ZTF & 18336 \\
\hline
Total overlap & 343014 \\
\hline
\hline
\multicolumn{2}{l}{\textbf{Notes:} (a) As spectroscopic binaries;}\\
\multicolumn{2}{l}{\textbf{Notes:} (b) As EBs or planet candidates;}\\
\end{tabular}
\caption{Likely overlap between the 1,223,603 {\em TESS} targets exhibiting eclipse-like features detected by our neural network and known EBs, defined here as a sky-projected separation of less than 1 {\em TESS} pixel ($\approx$ 21 arcsec). Duplicates are removed from the total number of overlaps.}
\label{tab:overlaps}
\end{table}

\begin{table}[]
    \centering
    \begin{tabular}{c|ccc}
    \hline
    \hline
        TIC ID & RA [deg] & Dec [deg] & Tmag \\
        \hline
        1051 & 218.815978 & -28.267080 & 14.77 \\
        4482 & 218.858361 & -25.722840 & 13.42 \\
        8639 & 219.017912 & -27.561259 & 14.75 \\
        17084 & 219.309771 & -25.415941 & 14.68 \\
        17361 & 219.336321 & -24.958481 & 11.34 \\
    \hline
    \end{tabular}
    \caption{Identifying information for $\sim900,000$ unvetted and unvalidated targets for which (i) the neural network identified eclipse-like events with a score greater than 0.9; and (ii) no EBs from the sources listed in Table \ref{tab:overlaps} are within $\approx21$ arcsec. Table available in full as a machine-readable supplement.}
    \label{tab:900k}
\end{table}

To investigate this matter further, we conducted a deep dive into a subset of $\sim60,000$ targets (hereafter 60K), representing $\sim5$\% of the likely known and potentially new EB candidates from our preliminary list. The targets were randomly selected and evenly split on either side of the 21-arcsec demarcation line, and are representative of the {\em TESS} magnitude and sky position distributions shown in Fig. \ref{fig:EB_histo}. These 60K targets were subjected to comprehensive ephemerides determination and photocenter measurements, and analyzed in-depth via the 2-step process outlined below. 

\subsection{In-depth analysis of 60,000 targets}

During the first step, we developed and utilized an automated pipeline to calculate ephemerides and measure photocenter offsets. Specifically, we applied the Box-Least Squares algorithm \citep[BLS][]{BLS} to the available \textsc{eleanor} FFI lightcurves to measure periods and conjunction times, limiting the minimum/maximum period searched for to 0.5/40 days, respectively. The BLS results were further improved by fitting a generalized Gaussian model to each detected eclipse adopting the methodology of \cite{Kostov2022_quadcat1}, and testing for period deviations from linear ephemeris. The latter helps take into account potential eclipse timing variations (ETV) that may decrease the precision of the BLS measurements, and also provides robust measurements for the eclipse depths and durations. Next, we used the refined ephemerides and eclipse durations to construct the appropriate difference images for each EB candidate, following the prescription of \cite{2019AJ....157..124K}. Finally, we obtained photocenter measurements by fitting to each difference image the {\em TESS} Pixel Response Function and a Gaussian Point Spread Function, and adopting the average of the two as the corresponding photocenter of the image. 

\begin{figure}[!ht]
    \centering
    \includegraphics[width=0.98\linewidth]{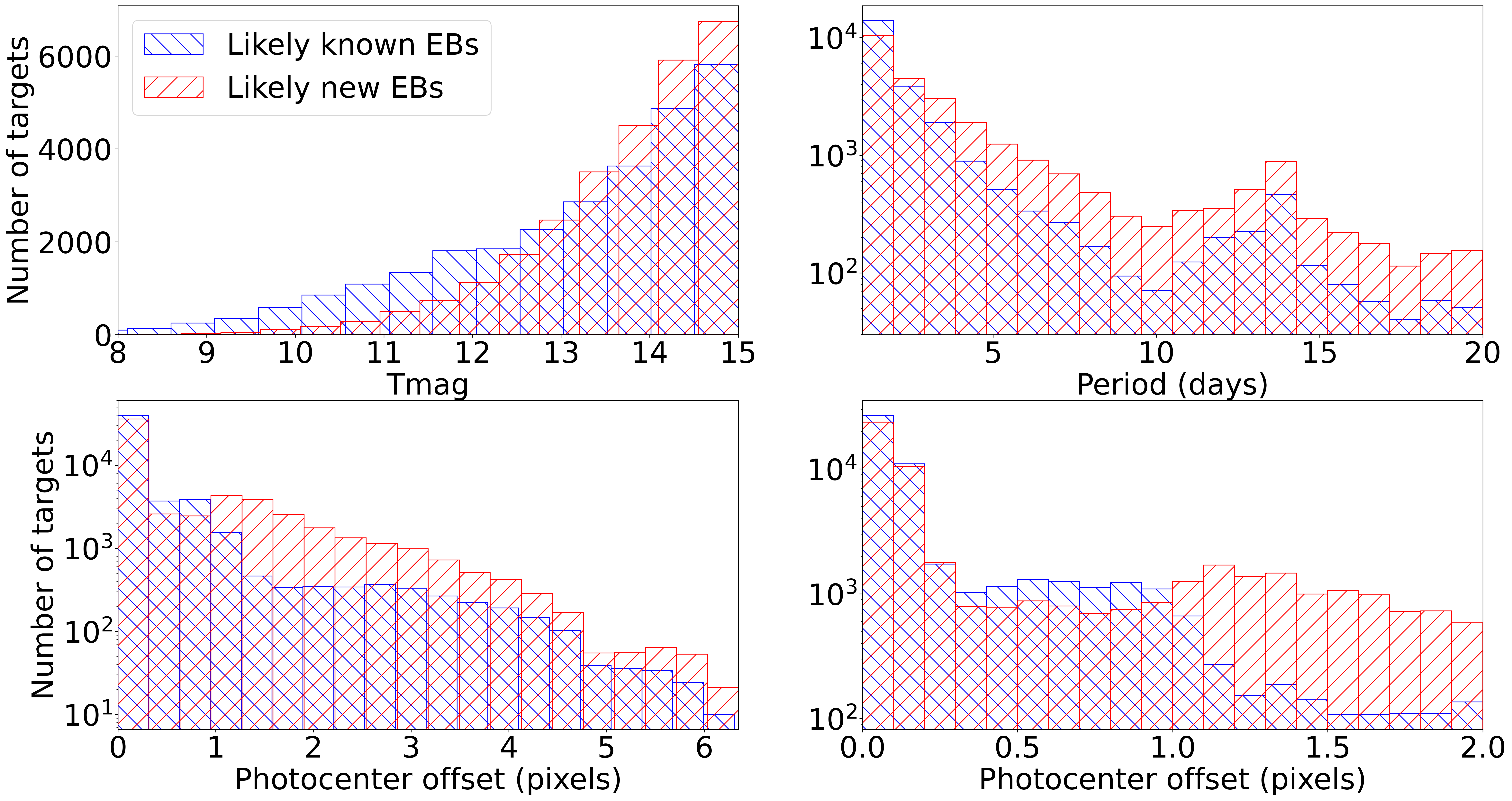}
    \caption{Distributions of the {\em TESS} magnitudes, preliminary measured periods, and respective photocenter offsets for the 60K sample. The lower right panel is a zoomed-in version of the lower left panel, highlighting the distribution of measured photocenters smaller than 2 pixels. See text for details.}
    \label{fig:60k_sample}
\end{figure}

Preliminary results from the first step are highlighted in Fig. \ref{fig:60k_sample}, showing the distributions of the 60K sample in terms of {\em TESS} magnitude, measured period, and photocenter offset. At this stage, the relevant measurements have not yet undergone the rigorous vetting and validation analysis required for promoting a target as a genuine EB, and are thus likely affected by various systematics. For example, the period distribution of the new EBs seen in Fig. \ref{fig:60k_sample} shows a local maximum near 14 days. This is close to half the duration of a {\em TESS} sector, which makes the potential periods suspicious. 

\begin{figure}
    \centering
    \includegraphics[width=0.98\linewidth]{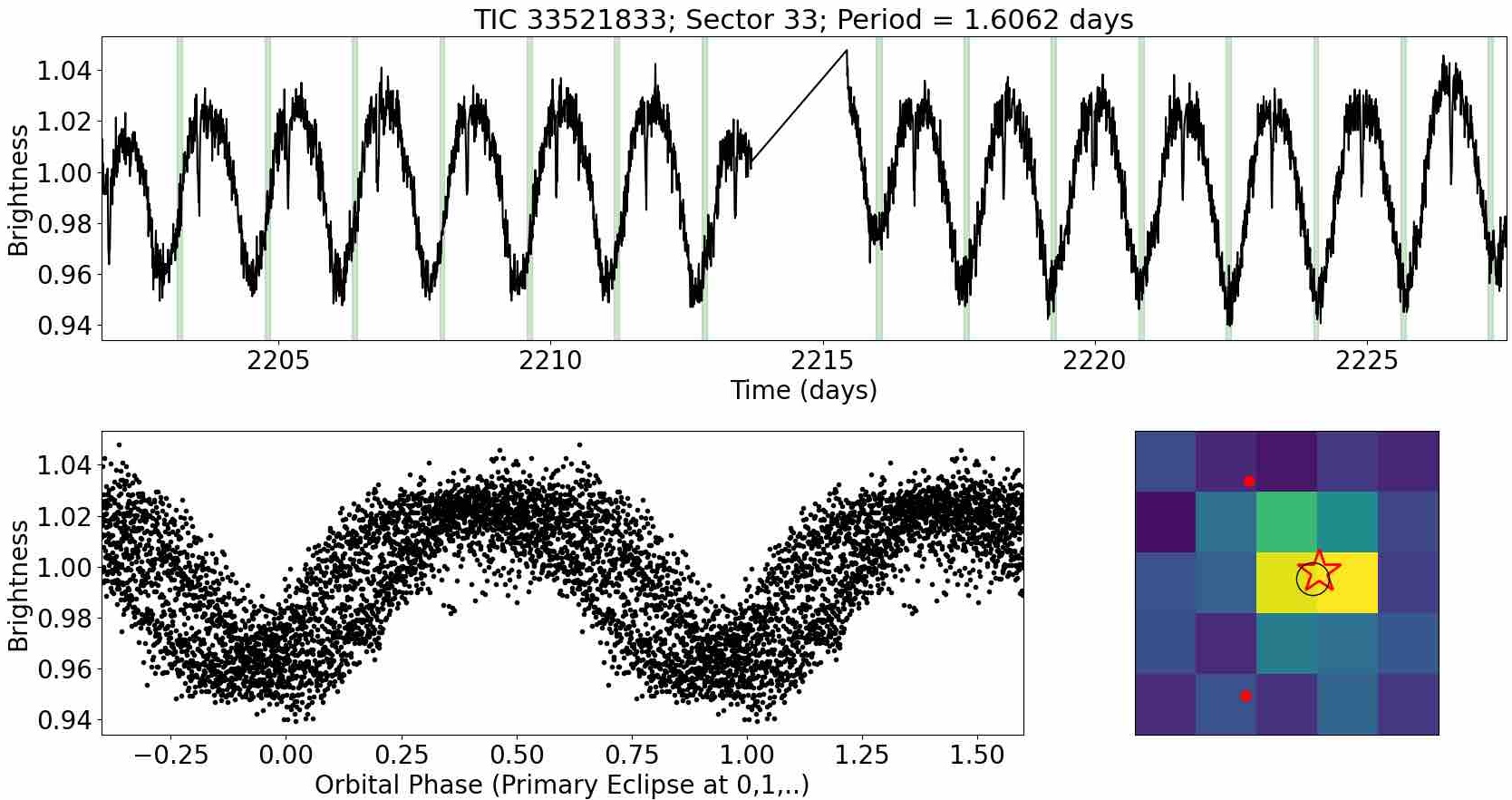}
    \caption{An example issue associated with the automated ephemeris measurements of the {\em TESS} FFI \textsc{eleanor} lightcurve and difference image for TIC 33521833. The upper panel shows one sector of {\em TESS} data with the automatically-detected periodic signal highlighted by the green vertical bands. Lower left panel: corresponding phase-folded data from all available sectors. Lower right panel: corresponding difference image and photocenter measurements. Here, the lightcurve is dominated by stellar variability and the automatically-measured period is incorrect.}
    \label{fig:period_issues_1}
\end{figure}

\begin{figure}
    \centering
    \includegraphics[width=0.98\linewidth]{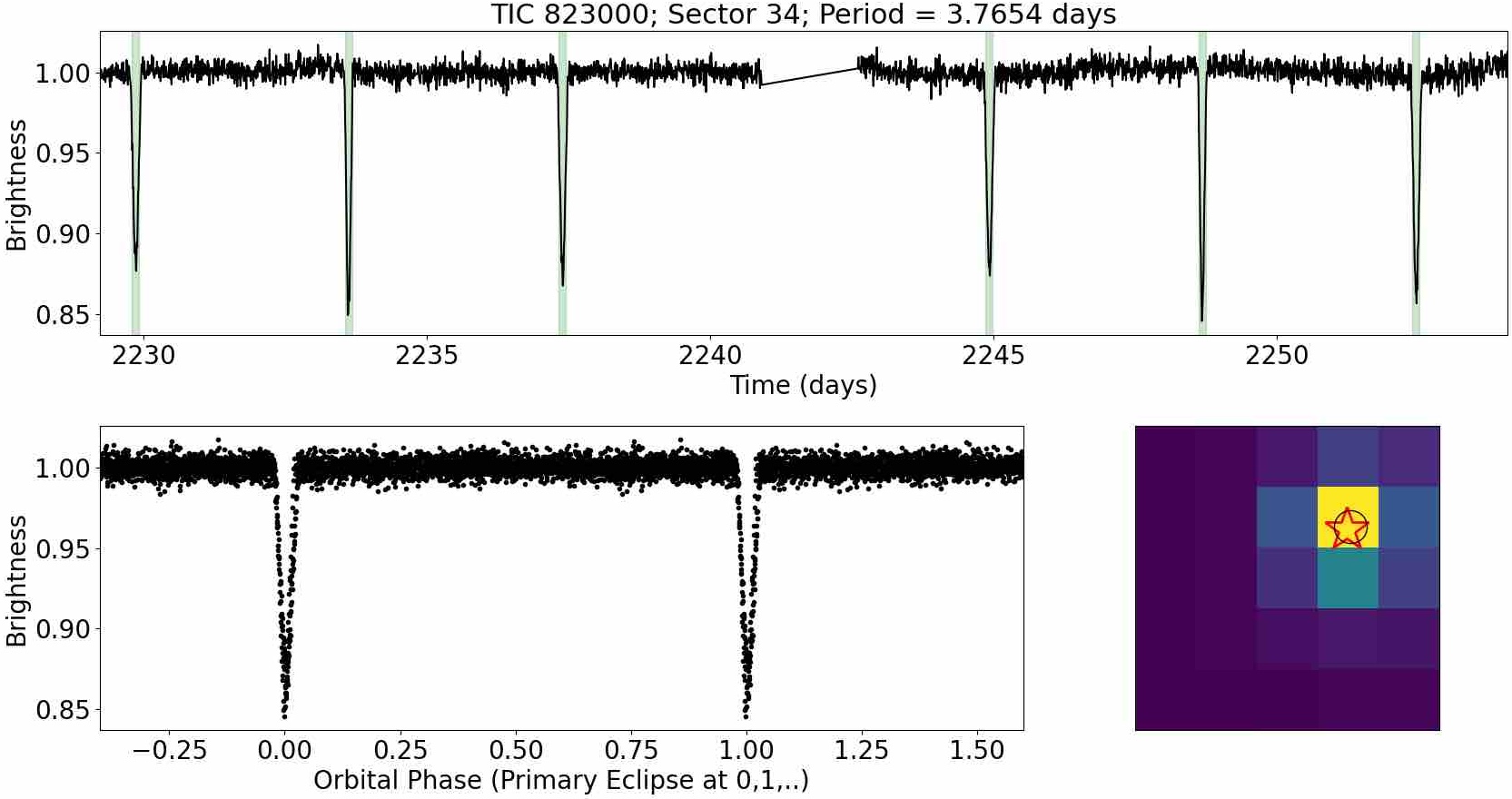}
    \caption{Same as Fig. \ref{fig:period_issues_1} but for TIC 823000. Here, the primary and secondary eclipses have similar depths (130 parts-per-thousand vs 120 parts-per-thousand, respectively) and the automatically-measured period is off by a factor of two.}
    \label{fig:period_issues_2}
\end{figure}

The second step of the process addresses the issue of potentially incorrect ephemeris measurements and, consequently, incorrect photocenter measurements produced by the automated pipeline outline above. This can occur when the period search is misled by, for example, strong systematic effects such as prominent out-of-eclipse lightcurve variations (due to, e.g., starspots) that can dominate or even completely overwhelm the eclipse signal (see Fig. \ref{fig:period_issues_1} and \ref{fig:period_issues_2}, upper panels). Additionally, EBs where the primary and secondary eclipses are similar in depth, duration, and shape can lead to situations where the automatically-measured period is an integer ratio of the true period (see Fig. \ref{fig:period_issues_2}, lower panels). While we try to minimize the impact of such issues on the ephemeris and photocenter pipeline as much as possible, it is challenging to account for all possible complications without negatively affecting the signals of interest. For example, we only use data points with good quality flags as provided by \textsc{eleanor}, remove those that are either near known issues, such as momentum dumps ({\em TESS} Instrument Handbook\footnote{\url{https://archive.stsci.edu/files/live/sites/mast/files/home/missions-and-data/active-missions/tess/_documents/TESS_Instrument_Handbook_v0.1.pdf}}), or we identified as potentially suspected sections of the lightcurve, and, where appropriate, utilize low-order polynomial detrending. 

To account for these and other challenges, we manually inspected the products of the automated pipeline for each of the 60K targets as outlined below. 

\subsection{Citizen Science}
\label{sec:cs}

Analyzing vast amounts of data using automated methods remains a highly non-trivial process as various sources of `noise' -- be it astrophysical, instrumental, systematic, etc -- can introduce subtleties that are challenging to automatically account for. Additionally, while autonomous methods excel at, for example finding a periodic signal in time-series such as the transits of an exoplanet (e.g., using BLS), they often lack the insight to discover unique and unexpected features, or various unusual and uncommon astrophysical objects or phenomena. For example, {\it all} the known transiting circumbinary planets have been identified by visual inspection of Kepler and {\em TESS} lightcurves \citep{2023Univ....9..455K}. Notably, while human intervention can help fill in the voids left by the algorithms, given the complexity and scope of the astronomical data sets it is quite challenging for a traditional research group to manually inspect hundreds of thousands of targets.

Thankfully, to the rescue come volunteers from all walks of life that boost the capacity of bandwidth-limited professional astronomers many-fold and help tackle the ever-increasing volume of publicly available astronomical data. This so-called citizen science approach is not a new concept -- professional and amateur astronomers have a fantastic and strongly intertwined history. One famous example is the ``Harvard Computers'' where some of the participants started the project with no formal astronomy training yet helped revolutionize astronomy and became some of the most successful professional astronomers \citep[e.g.,][]{Nelson2008}. %(e.g., Nelson, Nature volume 455, pages 36–37 (2008)). 
Among some of the more recent examples, many of the transiting planets with orbital periods longer than 1 year have been discovered by citizen scientists \citep[e.g.,][]{Wang2015}, and hundreds of eclipsing triple and quadruple star systems have been spotted by eagle-eyed volunteers \citep[e.g.,][and references therein]{2022Galax..10....9B}.
Citizen scientists have been responsible for many `firsts', e.g., (i) the unusual Boyajian’s Star \citep[][]{Boyajian2016}; (ii) an exocomet transiting its host star \citep[][]{Rappaport2018}; 
and (iii) a newly-discovered class of objects called ``tidally tilted pulsators'' \citep[e.g.,][]{Handler2020}. Time and again, the volunteers have demonstrated they can extract interesting signals from noise in numerous cases. 

The state of citizen science is strong, with multiple projects tackling various astronomical data sets and making important scientific contributions on a regular basis thanks to e.g., Planet Hunters and Planet Hunters {\em TESS}\footnote{\url{https://blog.planethunters.org/2010/12/16/planet-hunters-introduction/}, \citep[][]{Fischer2012}; \url{https://www.zooniverse.org/projects/nora-dot-eisner/planet-hunters-tess}}, Exoplanet Explorers\footnote{\url{exoplanetexplorers.org}}, Citizen ASAS-SN\footnote{\url{ https://www.zooniverse.org/projects/tharinduj/citizen-asas-sn0}}, SuperWASP variable stars\footnote{\url{https://www.zooniverse.org/projects/ajnorton/superwasp-variable-stars}}, Planet Patrol\footnote{\url{https://www.zooniverse.org/projects/marckuchner/planet-patrol/}}, Eclipsing Binary Patrol\footnote{\url{https://www.zooniverse.org/projects/vbkostov/eclipsing-binary-patrol}}, Exoplanet Watch \footnote{\url{https://exoplanets.nasa.gov/exoplanet-watch/about-exoplanet-watch/overview/}}, UNITE: Unistellar Network Investigating TESS Exoplanets \footnote{\url{https://science.unistellar.com/exoplanets/unite/}}, Visual Survey Group \citep[VSG;][]{2022PASP..134g4401K}. The power and dedication of citizen scientists is truly incredible -- for example, members of the VSG group have visually inspected {\it tens of millions} of lightcurves from Kepler and {\em TESS} \citep{2022PASP..134g4401K}, and helped make important new discoveries in multiple branches of astrophysics.

\subsubsection{Exogram}

\begin{figure}
    \centering
    \includegraphics[width=0.95\linewidth]{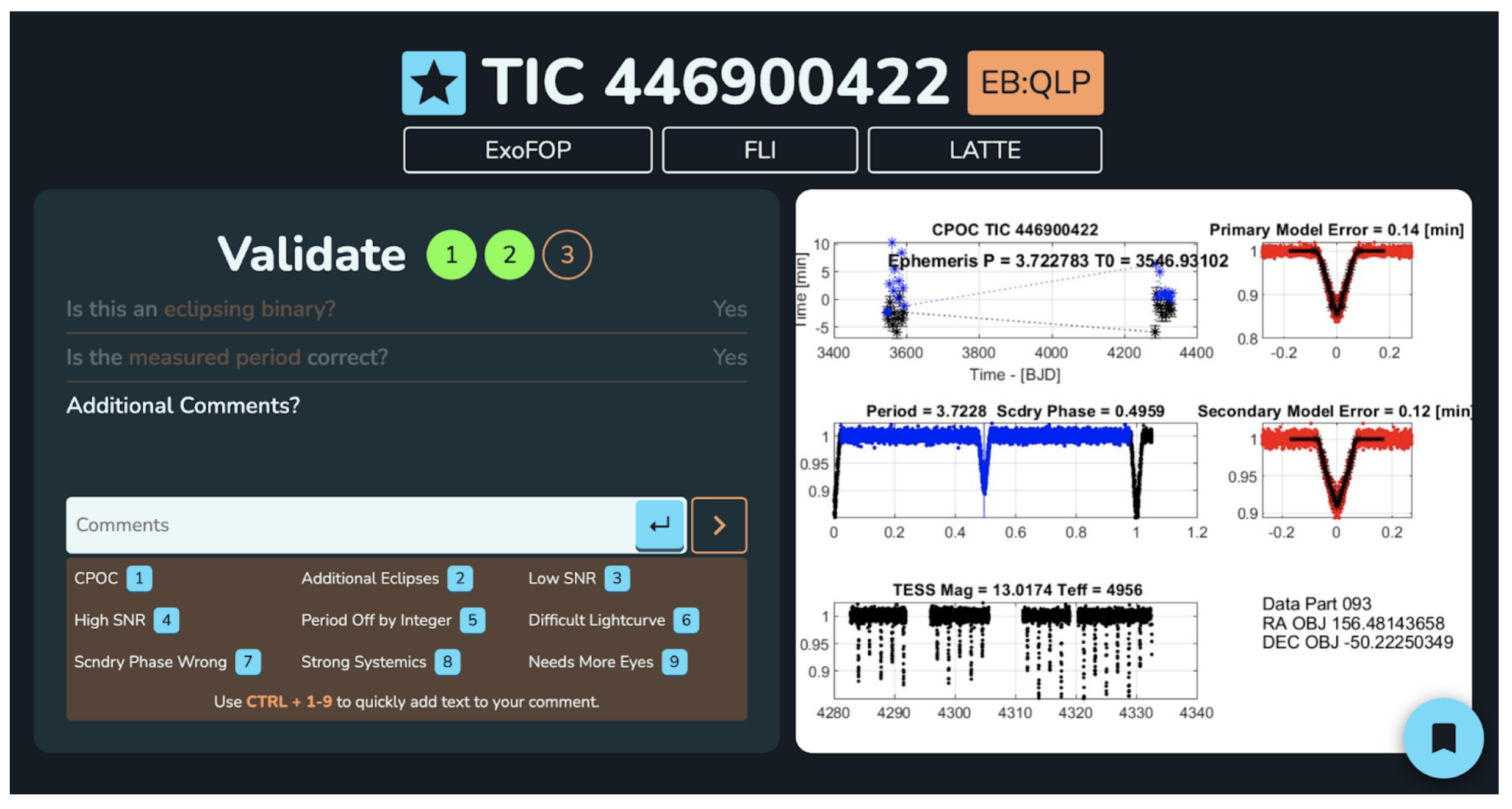}
    \caption{An example screenshot of the workflow and interface of Exogram EB vetting, enabling rapid image classification.}
    \label{fig:exogram}
\end{figure}

The manual inspection of the 60K targets proceeded as follows. First, we adapted our online vetting portal ``Exogram''\footnote{\url{https://exogram.vercel.app/}, developed by one of the citizen scientists on our team (RS).} for rapid visual scrutiny of a randomly-drawn subset of about 10K targets out of the 60K set by our core science team 
composed of professional astronomers and highly-knowledgeable citizen scientist ``superusers''. The team formed during the Planet Patrol project, and we have been working together ever since. The custom interface breaks down the vetting process into three main questions: ``Is this an eclipsing binary?'',  ``Is the measured period correct?'', and a space for additional comments (both pre-defined and free text). An example screenshot from the Exogram EB vetting portal is shown in Fig. \ref{fig:exogram}.

Briefly, the user first evaluates the data for clear evidence of eclipse-like features, paying close attention to the two panels on the right, and to the lower left panel. Next, Exogram automatically proceeds to the second question, where the user scrutinizes the phase-folded plot (middle left panel) and decides whether the measured period is correct. Finally, the portal takes the user to the last question, which provides the opportunity to mark the target as particularly noteworthy. Several predefined options are provided, corresponding to the most commonly observed characteristics. 

Importantly, Exogram EB is designed to enable fast image classification through the use of keyboard shortcuts. We found this to be a critical advantage as it allows an expert vetter to classify images with a typical ``cruising speed'' on the order of seconds, especially when the data clearly indicates a typical EB system\footnote{For the easiest cases, the vetting speed is practically limited by the reaction time for pressing the relevant shortcut.}. Naturally, more interesting cases such as those exhibiting additional eclipse-like features take longer to inspect, as do targets dominated by systematics such as momentum dumps, but even for these the access to keyboard shortcuts significantly decreases the response time. 

The portal also provides links to external tools such as the Fast Lightcurve Inspector (FLI\footnote{\url{https://fast-lightcurve-inspector.osc-fr1.scalingo.io}}) and LATTE (Eisner et al. 2020) that enable deeper investigation of potentially interesting or particularly challenging targets. Both tools allow researchers to interactively examine the entire lightcurve, making it easier to, for example, distinguish between genuine eclipses and momentum dumps or other sources of interference. It is worth noting that FLI was created by one of our superusers (JdL) and is designed as a free, online, user-friendly, interactive tool for visual inspection of {\em TESS} data, including BLS analysis and phase-folding. FLI uses the lightkurve package \citep{lightkurve} to query MAST for all available \textsc{GSFC-ELEANOR-LITE}, \textsc{QLP}, \textsc{SPOC}, and \textsc{TESS-SPOC} lightcurves, along with corresponding diagnostics (e.g., background flux and centroid measurements), and presents them in a Bokeh/Plotly environment. 

Exogram was developed using SvelteKit, a modern web application framework. The database and authentication are handled by Google’s Firebase platform, and the website itself is hosted by Vercel. We store the lightcurve images on Google Drive. Behind the scenes, the Exogram server tallies the amount of responses, labels targets as fully classified if three users had already commented and removes them from the pool of images shown. Additionally, the Exogram platform integrates social media-esque features to encourage collaboration between users. For example, users can ``star'' a target to save it for later inspection. Starred EBs are public, and users can see which targets were saved by others. This makes it easy for vetters to find targets that others deemed interesting or unusual. Users can also share targets with each other, and the integrated notification feature alerts users when something was shared with them. Finally, the platform also shows a vetting leaderboard to encourage friendly competition among the users.

\subsubsection{Eclipsing Binary Patrol}

To further capitalize on the power of citizen science, and inspired by the success of the Planet Patrol project \citep{2022PASP..134d4401K}, we proceeded with the investigation of all 60K targets by developing the Eclipsing Binary Patrol (EBP) project\footnote{The targets inspected through Exogram are included in EBP as an additional layer of scutiny and validation}. EBP is hosted on Zooniverse and provides a streamlined, interactive, and user-friendly platform to visually inspect a summary of the results produced by the automated pipeline described above. EBP launched on Sep 3, 2024 and was completed in March 26, 2025, during which period $\sim1,800$ participants produced $\sim320,000$ classifications. 

\begin{figure}[!ht]
    \centering
    \includegraphics[width=0.98\linewidth]{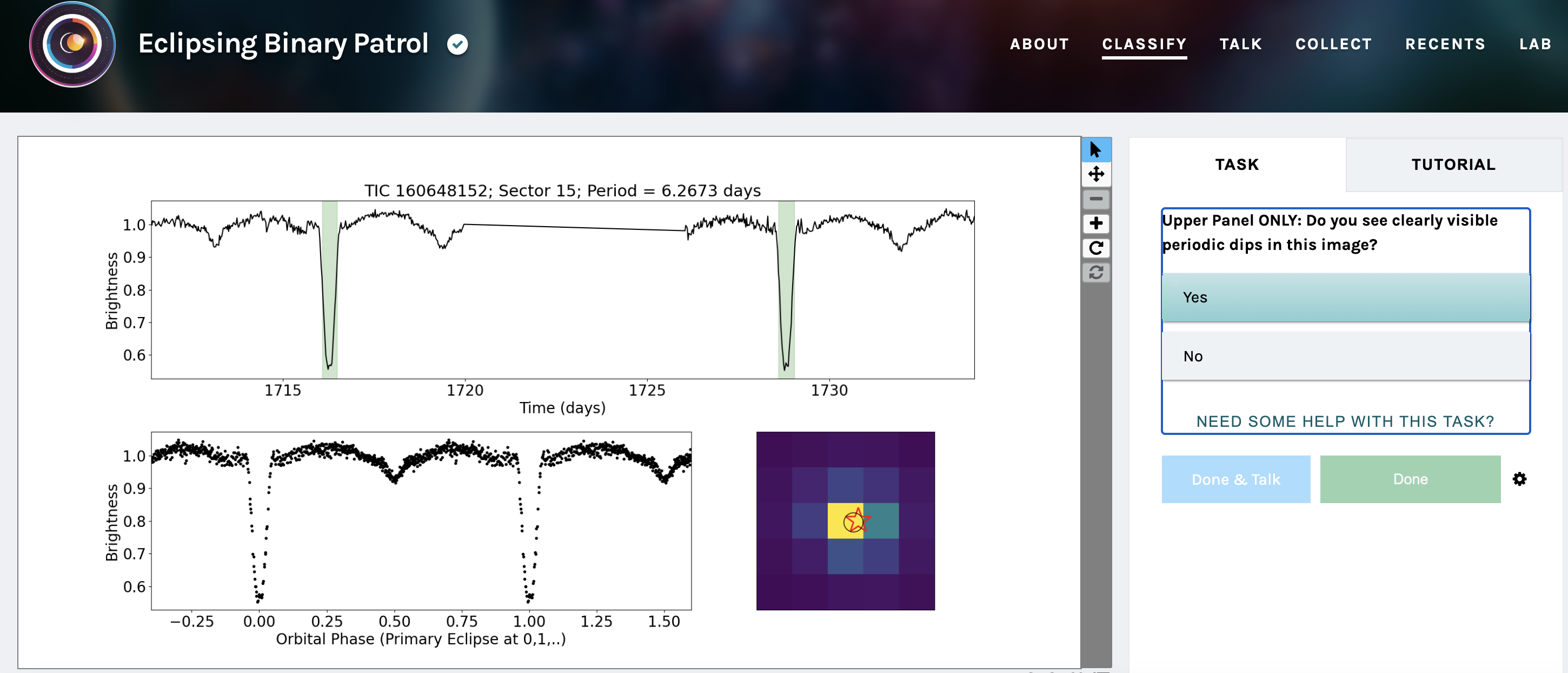}
    \caption{An example screenshot highlighting the first question of the EBP workflow. The upper panel of the figure shows one sector of {\em TESS} data, highlighting the detected eclipses in green, and listing the measured period. The lower left panel shows the corresponding phase-folded lightcurve, and the lower right panel shows the difference image used for the photocenter measurements. The difference image also shows the pixel position of the target (red star) and the sector-averaged measured photocenter (open black circle). The user decides whether the target shows clearly visible periodic dips (indicating an EB), answers ``yes'' or ``no'', and is then taken to the second question.}
    \label{fig:zooniverse_example}
\end{figure}

The EBP workflow consists of four questions aimed at evaluating whether the target is indeed an EB candidate, the calculated period is correct, and the photocenter measurements are reliable. An example screenshot of the classification scheme is shown in Fig. \ref{fig:zooniverse_example}. The volunteers inspect the original and phase-folded lightcurves, and decide whether they see periodic eclipse-like signals, check if the period is correct, and scrutinize the lightcurve for secondary eclipses. Additionally, they evaluate the quality of the difference image and classify it as either appropriate for photocenter measurements -- i.e., the image shows a well-defined bright spot on an otherwise dark background -- or otherwise. 

The EBP portal provides extensive background information on the science of EBs and their astrophysical importance, a comprehensive tutorial with a step by step demonstration of the workflow, a field guide presenting relevant examples, edge cases, etc. The portal also includes guidelines on how to interpret and classify the images, as well as an active talk board where the volunteers can discuss targets of interest and ask for help from the science team. Each image presented to the volunteers also contains additional auxiliary information enabling more detailed investigation of the inspected target, in particular with the help of FLI. We note that classifications on EBP are not strictly blind. Volunteers could freely look up outside information, based on the provided TIC ID, which could potentially affect their evaluation. Finally, volunteers interested in contributing to the vetting process beyond the Zooniverse project are invited to join the science team.

For completeness, we would like to briefly share the experience gained and lessons learned from EBP. First and foremost, frequent interactions between the science team and volunteers on the Talk boards, especially during the initial stages, were critical for the success of the project. These interactions ensured timely resolution of technical issues, addressed vetting and scientific questions, and enabled live updates aimed at improving the overall workflow. For example, prompted by feedback from volunteers we quickly refined the FAQ and tutorial by adding new examples, clarifying existing instructions, etc. Finally, consistent communication, including social media posts highlighting interesting targets and celebrating milestones, helped retain user engagement throughout the duration of the project, averaging about 1,000 classifications per day, even months after launch. 

The workflow of EBP is designed such that each target is considered as fully classified when at least five different volunteers have inspected the corresponding image and answered the provided questions. It is worth pointing out that at the launch of the project, the image `retirement' limit was set to nine. However, that proved to be too high as the rate of completed classifications was rather slow. Thus, in order to speed up the vetting and complete the 60K sample in a timely manner, three weeks after the project was launched we reduced the limit to seven, and shortly after down to five. 

To adopt an aggregate response to each question, we tested three options: (i) a simple majority, i.e., at least 3 out of 5 volunteers select the same answer; (ii) at least 4 out of 5; and (iii) 5 out of 5. To evaluate the reliability of these aggregates, we checked the corresponding classifications for a random sample of 1,000 targets where the measured period was classified as correct. Overall, we found that about 75\%/85\%/90\% of the responses are correct for options (i), (ii), and (iii), respectively. In order to increase the fidelity and maximize the reliability of the classifications, members of the science team performed complementary visual inspection of all potentially new EB candidates where at least 3 out of 5 volunteers indicate that the period as correct. 

\begin{figure}
    \centering
    \includegraphics[width=0.95\linewidth]{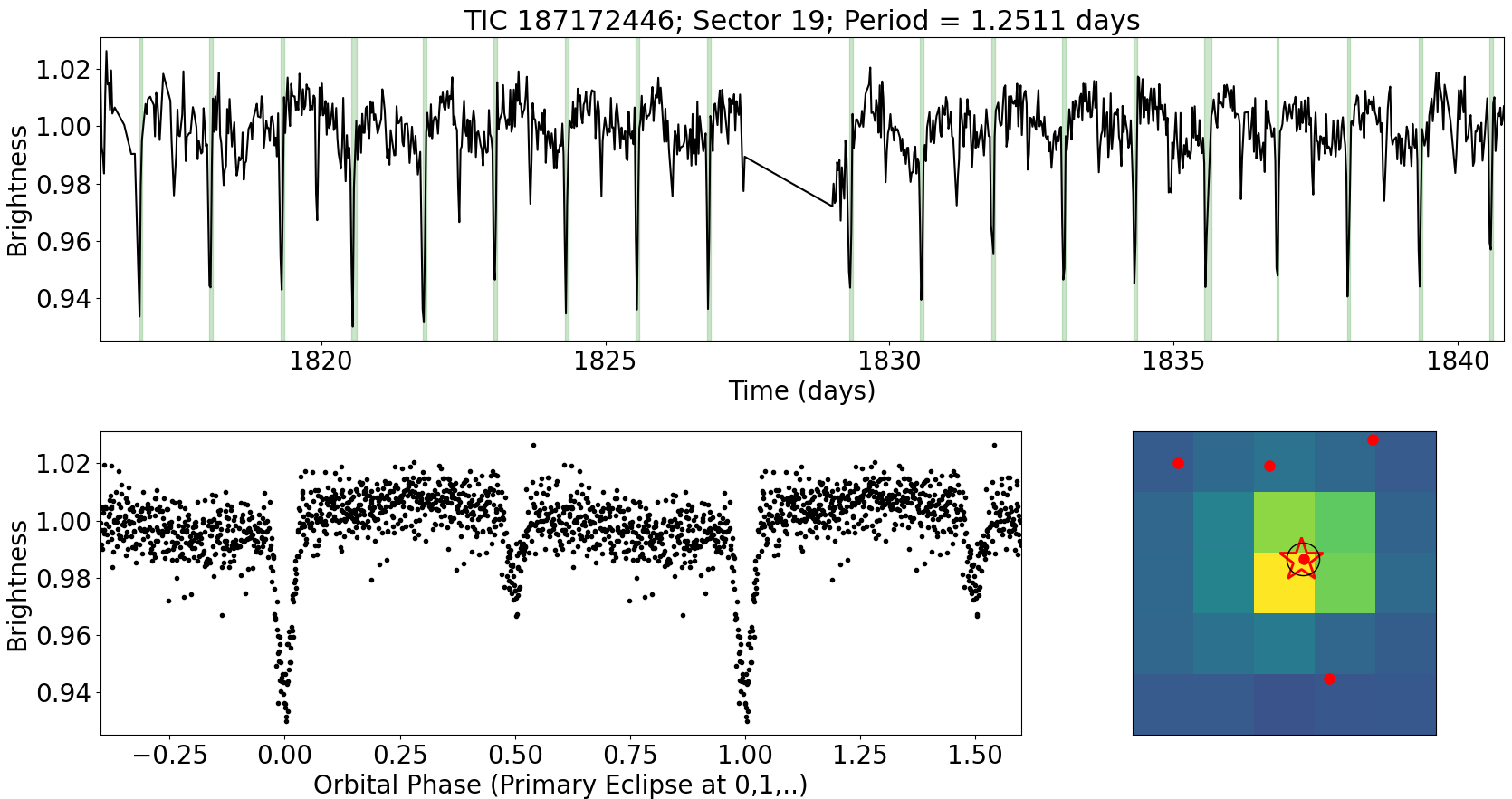}
    \caption{Same as Fig. \ref{fig:period_issues_1} but for TIC 187172446. Here, the lightcurve shows a clear EB signal, the measured period is correct, the difference image is adequate for reliable photocenter analysis, and the measured photocenter offset is $\sim0.14$ pixels. However, there is a nearby field star, TIC 510123334, that is bright enough to be the source of the eclipses and too close to the target (projected separation of about 0.71 arcsec) for the photocenter measurements to pinpoint the source of the detected eclipses. Such targets are not included in our catalog.}
    \label{fig:tic_187172446}
\end{figure}

Altogether, \newEBs targets passed the following vetting and validation tests: (1) the \textsc{eleanor} lightcurve shows clear eclipses; (2) the measured period is correct; (3) the difference images used for photocenter analysis are of sufficiently high quality for reliable measurements; (4) the measured photocenter offsets are smaller than 0.2 pixels; and (5) no field stars from the Gaia and TIC catalogs are within 0.2 pixels of the target, and bright enough to produce the detected eclipses as contamination. An example of a target that passes the first four tests but fails the last is TIC 187172446, shown in Fig. \ref{fig:tic_187172446}. Here, there is a nearby field star, TIC 510123334 that is about 1 {\em TESS} magnitude fainter and at a projected separation of 0.17 arcsec. Thus, while the measured photocenter offset is about 0.14 pixels, it is impossible to tell from the {\em TESS} data which of these two stars is producing the detected eclipses. 

\section{A catalog of uniformly-vetted and -validated EBs from {\em TESS} FFI data}
\label{sec:catalog}

The final product of the process outlined above is a uniformly-vetted and -validated catalog of new EB candidates identified in {\em TESS} FFI data. The catalog contains \newEBs targets with verified ephemerides, eclipse depths and durations, and, where applicable, phase of secondary eclipses. Table \ref{tab:eb_catalog} showcases the content of the catalog, which also includes the TIC ID of the target, sky position, {\em TESS} magnitude, number of sectors observed, Gaia astrometric information, relevant comments, etc. We note that 29 of the \newEBs targets are listed in Gaia as single-lined spectroscopic binaries. In addition, we provide updated ephemerides for \knownEBs known EBs where the period listed in one or more catalogs is incorrect. Most of our corrections are with respect to the Gaia EB catalog -- 1233 out of 1889 targets, followed by 312 out of 986 ASAS-SN EBs, and 308 out of 1015 VSX EBs.

% \begin{table}[!ht]
\begin{sidewaystable}[]
    \centering
    \footnotesize
    \begin{tabular}{l | c c c c c c c c c c c c c c c c c}
        \hline
        \hline
        TIC ID & RA & Dec & Tmag & Period & ${\rm T_{0,prim}}$ & ${\rm Depth}$ & ${\rm Duration}$ & ${\rm T_{0,sec}}$ & ${\rm Phase}$ & ${\rm Depth}$ & ${\rm Duration}$ & Sectors & RUWE & AEN & AENS & ${\rm T_{eff}}$ & Comments \\
         & [deg] & [deg] & & [days] & [BJD$^a$] & [prim$^b$] & [prim, hrs] & [BJD$^a$] & sec & [sec$^b$] & [sec, hrs] & & & & & [K] & \\
        \hline
        8636 & 219.005555 & -27.569441 & 12.36 & 3.886554 & 1603.2923 & 121 & 3.0 & 2355.3404 & 0.5 & 45 & 3.0 & 2 & 1.78 & 0.17 & 39.28 & 4965 & -- \\
        672717 & 73.486075 & -26.691694 & 14.44 & 7.18193 & 1448.2325 & 121 & 2.6 & 1445.1831 & 0.58 & 113 & 2.6 & 2 & 8.61 & 2.71 & 912.95 & -- & -- \\
        737910 & 74.543779 & -28.491438 & 13.32 & 2.297452 & 1447.8866 & 31 & 3.6 & 1453.6230 & 0.5 & 30 & 3.5 & 2 & 1.06 & 0.07 & 1.08 & 4949 & -- \\
        823000 & 127.043916 & -16.737116 & 12.05 & 7.530827 & 2979.1752 & 132 & 3.5 & 2967.8872 & 0.5 & 121 & 4.6 & 4 & 0.99 & 0.00 & 0.00 & 5991 & -- \\
        890432 & 127.043916 & -13.796373 & 14.63 & 14.201932 & 2235.1764 & 143 & 4.5 & 2980.5780 & 0.49 & 58 & 4.9 & 5 & 1.12 & 0.09 & 8.63 & 4841 & -- \\
        891636 & 127.057470 & -12.512750 & 13.61 & 2.023998 & 1496.7517 & 99 & 2.1 & 2973.2540 & 0.5 & 38 & 2.7 & 5 & 39.40 & 4.10 & 11889.59 & -- & -- \\
        1102444 & 155.353410 & 27.595890 & 12.8 & 4.04989 & 2612.6915 & 16 & 4.9 & -- & -- & -- & -- & 2 & 1.01 & 0.00 & 0.00 & 4087 & -- \\
        1124666 & 36.948680 & -7.826921 & 13.78 & 3.524638 & 2148.3515 & 139 & 3.5 & -- & -- & -- & -- & 2 & -- & 6.11 & 9190.83 & -- & -- \\
        1195217 & 136.100062 & -11.945334 & 13.29 & 1.788685 & 1541.2277 & 16 & 1.8 & -- & -- & -- & -- & 4 & 1.02 & 0.00 & 0.00 & 4156 & -- \\
        1195846 & 71.095974 & -32.854422 & 14.69 & 3.141602 & 2145.8888 & 33 & 2.5 & 2191.4521 & 0.5 & 8 & 2.7 & 3 & 1.02 & 0.00 & 0.00 & 4518 & -- \\
        1196032 & 71.069268 & -33.584354 & 14.64 & 15.15001 & 1444.4083 & 173 & 4.8 & 1449.5224 & 0.34 & 124 & 5.0 & 3 & 0.96 & 0.00 & 0.00 & 5516 & -- \\
        1199301 & 71.333957 & -35.113030 & 13.08 & 5.26431 & 2188.7857 & 62 & 6.0 & 1443.8504 & 0.49 & 3 & 6.1 & 3 & 0.98 & 0.00 & 0.00 & 6086 & -- \\
        1309535 & 72.858580 & -32.250238 & 12.12 & 13.531491 & 2187.5640 & 192 & 3.9 & 1452.1739 & 0.65 & 10 & 4.2 & 2 & 0.89 & 0.04 & 1.86 & 5987 & -- \\
        1471956 & 146.911387 & -15.546746 & 13.03 & 9.007584 & 2995.8203 & 135 & 3.3 & 2991.3347 & 0.5 & 81 & 3.3 & 4 & 18.99 & 2.10 & 2515.51 & 4326 & -- \\
        1503836 & 131.959336 & -24.266628 & 13.07 & 1.021917 & 1527.7987 & 21 & 1.6 & 2993.7410 & 0.5 & 12 & 1.7 & 7 & 1.92 & 0.21 & 34.13 & -- & -- \\
        1541478 & 169.541503 & -14.092368 & 13.52 & 14.665098 & 3026.3663 & 417 & 3.1 & 3033.0097 & 0.45 & 359 & 3.0 & 4 & 1.19 & 0.05 & 0.76 & 3729 & -- \\
        1616408 & 132.133251 & -24.902464 & 14.71 & 3.115479 & 1524.3949 & 28 & 3.9 & -- & -- & -- & -- & 6 & 1.10 & 0.06 & 0.87 & 5861 & -- \\
        1755837 & 78.624341 & 32.548174 & 14.28 & 4.28833 & 2502.8285 & 128 & 5.0 & 3645.6576 & 0.5 & 21 & 3.8 & 8 & 1.01 & 0.00 & 0.00 & 5368 & -- \\
        1942820 & 78.947929 & 34.186537 & 14.34 & 3.381414 & 1826.4456 & 141 & 2.8 & -- & -- & -- & -- & 8 & 4.14 & 0.83 & 179.45 & 5780 & -- \\
        2099994 & 79.341909 & 30.521062 & 10.91 & 3.827636 & 2496.3074 & 8 & 4.0 & 2521.1347 & 0.49 & 1 & 3.9 & 6 & 8.15 & 1.26 & 3159.49 & 5923 & -- \\
        2158899 & 79.451105 & 30.893429 & 14.69 & 1.805987 & 2497.1628 & 20 & 3.5 & -- & -- & -- & -- & 8 & 1.06 & 0.04 & 0.31 & 6622 & -- \\
        2438442 & 80.004098 & 35.227821 & 12.81 & 3.147522 & 1826.3919 & 133 & 4.0 & 3647.2077 & 0.49 & 55 & 3.8 & 7 & 18.74 & 2.18 & 3404.77 & 5972 & -- \\
        2508333 & 80.125447 & 32.402184 & 13.85 & 2.183682 & 1840.1360 & 60 & 4.7 & -- & -- & -- & -- & 8 & 1.51 & 0.13 & 9.05 & 7104 & -- \\
        2509102 & 79.993792 & 32.023402 & 11.78 & 2.486683 & 1825.6144 & 20 & 5.1 & -- & -- & -- & -- & 8 & 1.00 & 0.07 & 6.26 & 7141 & -- \\
        2601042 & 80.292516 & 35.601431 & 13.57 & 3.476145 & 2493.7629 & 4 & 2.8 & 2481.5927 & 0.5 & 1 & 3.1 & 6 & 1.13 & 0.09 & 3.01 & 7681 & -- \\
        2678574 & 80.370472 & 32.400301 & 13.59 & 1.329262 & 2475.2094 & 14 & 2.7 & -- & -- & -- & -- & 8 & 0.95 & 0.00 & 0.00 & 6825 & -- \\
        2679754 & 80.290739 & 31.819113 & 12.32 & 3.184818 & 2508.1125 & 9 & 3.4 & -- & -- & -- & -- & 8 & 0.96 & 0.04 & 2.04 & 7065 & -- \\
        2680580 & 80.337560 & 31.421225 & 14.28 & 6.874418 & 2492.9904 & 74 & 5.8 & 1829.4379 & 0.48 & 46 & 6.3 & 8 & 1.08 & 0.00 & 0.00 & 6650 & -- \\
        2762701 & 356.676090 & -16.842597 & 14.08 & 1.216169 & 1366.2573 & 179 & 1.3 & 1359.5735 & 0.5 & 69 & 1.4 & 3 & 1.05 & 0.08 & 0.75 & -- & pETVs \\
        2840082 & 243.496472 & -41.192966 & 11.49 & 2.887738 & 3073.6308 & 42 & 9.0 & 3092.4026 & 0.5 & 7 & 9.7 & 3 & 3.66 & 0.41 & 211.15 & 6456 & -- \\
        \hline
        \multicolumn{18}{l}{\textbf{Notes:} (a) BJD - 2,457,000; (b) in parts per thousand (ppt); pETVs = potential ETVs;}\\
    \end{tabular}
    \caption{Parameters of the \newEBs new EBs\label{tab:EBparameters}. Table available in full as machine-readable online supplement.}
    \label{tab:eb_catalog}
% \end{table}
\end{sidewaystable}

\begin{figure}
    \centering
    \includegraphics[width=0.99\linewidth]{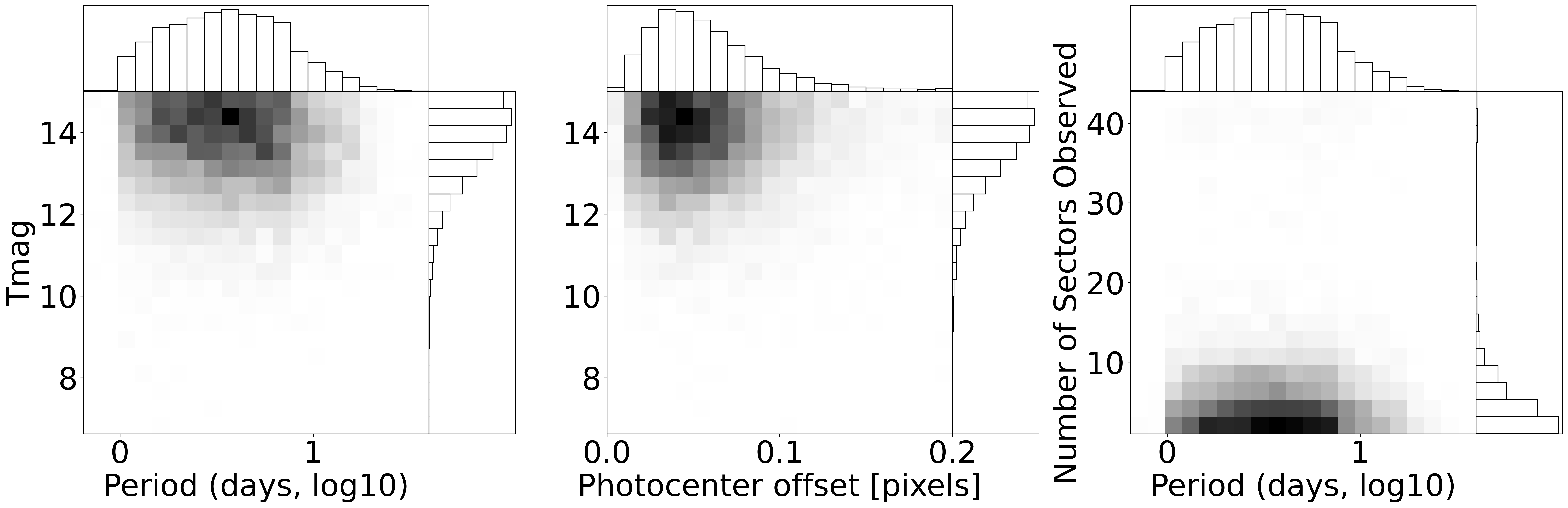}
    \caption{The distributions of the {\em TESS} magnitudes, orbital periods, photocenter offsets, and number of sectors observed for the \newEBs new EBs in our catalog. Darker shade represents higher number of targets. Most of the EBs are faint, have short orbital periods, the detected eclipses originate within $\sim1-2$ arcsec of the respective target star, and have been observed by {\em TESS} in at least three sectors.}
    \label{fig:tmag_vs_period_vs_offset}
\end{figure}

The distributions of the TESS magnitudes, orbital periods, photocenter offsets, and number of sectors observed for the \newEBs new EBs are shown in Fig. \ref{fig:tmag_vs_period_vs_offset}. The period distribution has a mean and median values of $\approx4.5$ and $\approx3.5$ days, respectively, and a 95th-percentile of $\approx11.6$ days. This is comparable to the Kepler EB catalog, where the median period is also about 3.5 days \citep[][]{2011AJ....142..160S,2011AJ....141...83P,2014AJ....147...45C,2018haex.bookE..34W}. The shortest period in our catalog is $\approx0.65$ days, while the longest is $\approx40$ days. As seen from the figure, the measured photocenter offsets are remarkably small, with mean/median/95th-percentile values of 0.053/0.05/0.12 pixels, respectively, confirming that the detected eclipses originate from within $\sim1-2$ arcsec of the target stars. Additionally, granted that most of the EBs are on the fainter end (i.e., mean/median/95th-percentile of 13.6/13.8/14.9 mag) there is no strong correlation between a target's brightness and the magnitude of the corresponding photocenter offset, highlighting the excellent quality of the {\em TESS} FFIs even for the fainter stars. Finally, {\em TESS} observed the majority of the EBs three times or more, such that $\approx88\%/32\%/13\%$ of the targets were covered in at least 3/6/9 sectors, respectively.

\subsection{Depth, Durations, and Secondary Eclipses}

To measure the eclipse times, depths and durations, we adopt the methodology of \cite{Kostov2022_quadcat1} and, for each sector, fit each eclipse with their generalized Gaussian model:

\begin{equation}
{\rm F(t) = A - B e^{-(\frac{|t-t_o|}{\omega})^{\beta}} + C(t-t_o)}
\label{eq:GG}
\end{equation}

\begin{figure}[!ht]
    \centering
    \includegraphics[width=0.99\linewidth]{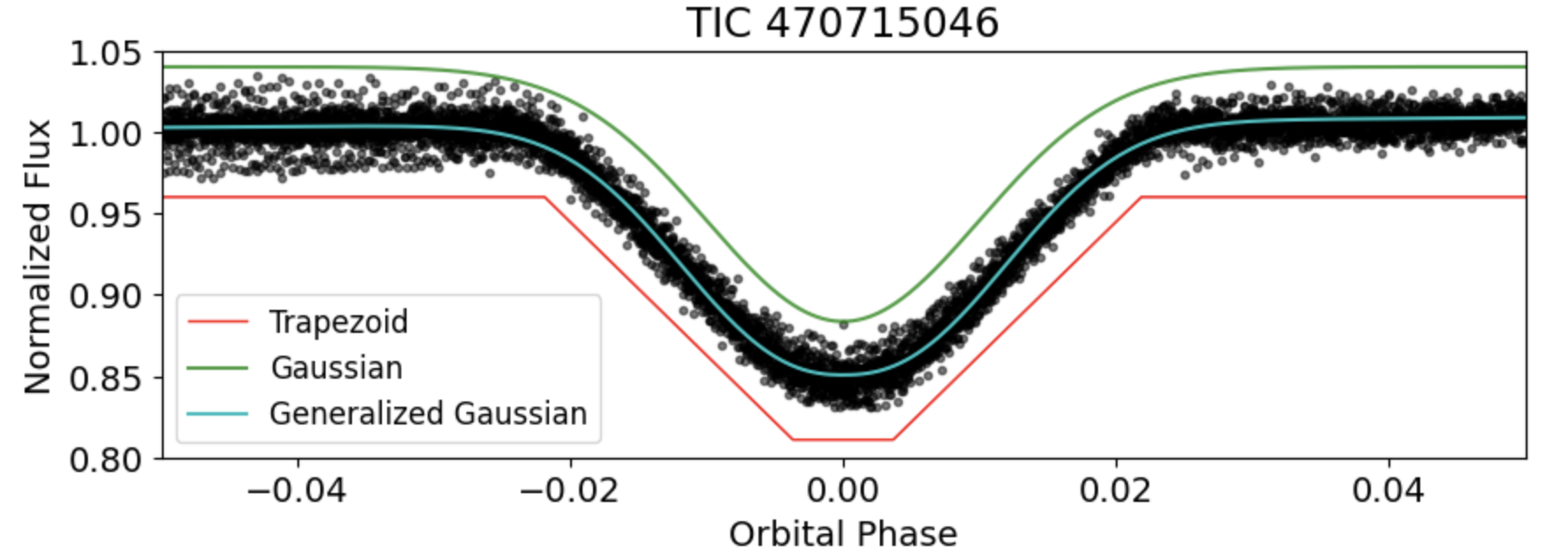}
    \caption{Example fits to all phase-folded primary eclipses of TIC 470715046 for three models: trapezoid (red), Gaussian (green), and generalized Gaussian (cyan). The red and green curves are vertically offset for clarity. As seen from the figure, the generalized Gaussian model provides a much better fit to the data than either of the other models.}
    \label{fig:dep_dur_fits}
\end{figure}

For illustrative purposes, Fig. \ref{fig:dep_dur_fits} shows the model fit to all phase-folded primary eclipses of TIC 470715046, as well as a Gaussian and a trapezoid fit for comparison. As seen from the figure, the generalized Gaussian model provides an excellent fit to the data -- certainly better than both the trapezoid and the narrower Gaussian model -- and we use it for measuring the eclipse depths and durations. 

\begin{figure}[!ht]
    \centering
    \includegraphics[width=0.99\linewidth]{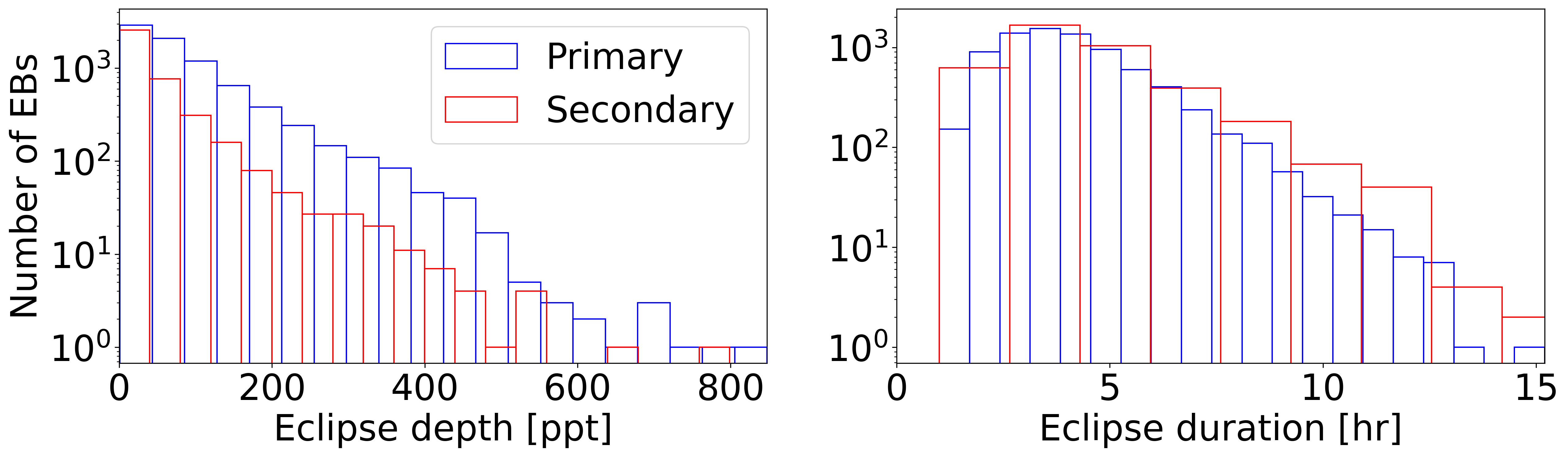}
    \caption{Distributions of the median primary and, where present, secondary eclipse depths (left panel, in parts-per-thousand, ppt), and durations (right panel, in hours) for the \newEBs new EBs presented here.}
    \label{fig:dep_dur}
\end{figure}

\begin{figure}[!ht]
    \centering
    \includegraphics[width=0.98\linewidth]{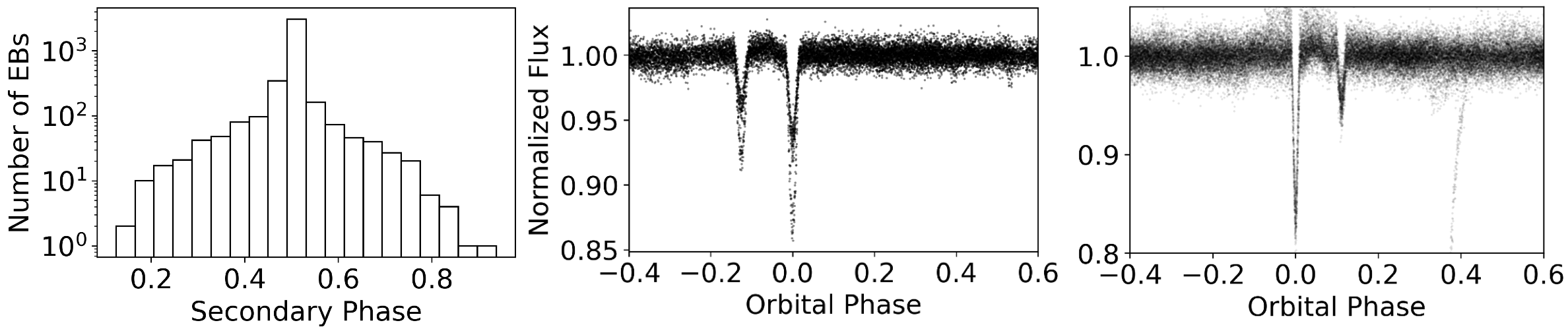}
    \caption{Left panel: Distribution of the detected secondary eclipses as a function of the orbital phase. Middle and right panels: The two EBs with the most extreme secondary phases, TIC 149673382 with phase $\approx0.87$ (middle) and TIC 337097515 with phase $\approx0.11$ (right panel), both exhibiting a heartbeat `bump'.}
    \label{fig:sec_phase}
\end{figure}

Table \ref{tab:eb_catalog} provides the median depths and durations for the new EBs presented here. The corresponding distributions are shown in Fig. \ref{fig:dep_dur}. The primary depth distribution has mean/median/95-th percentile values of 91/62/276 parts-per-thousand, respectively; the mean/median/95-th percentile values for the primary duration distribution are 4.2/3.9/7.4 hours, respectively. Roughly half of the targets exhibit secondary eclipses. As highlighted in Fig. \ref{fig:sec_phase}, most of these occur near orbital phase of 0.5, and about 95/99\% of them reside within a phase range of $\sim0.43-0.58$/$\sim0.3-0.72$, respectively. The two most extreme secondary phases in our catalog are for TIC 149673382, with a secondary phase of $\approx0.87$, and TIC 337097515, with a secondary phase of $\approx0.11$, both exhibiting a pronounced heartbeat `bump' in-between the primary and secondary eclipses. 

\begin{figure}[!ht]
    \centering
    \includegraphics[width=0.48\linewidth]{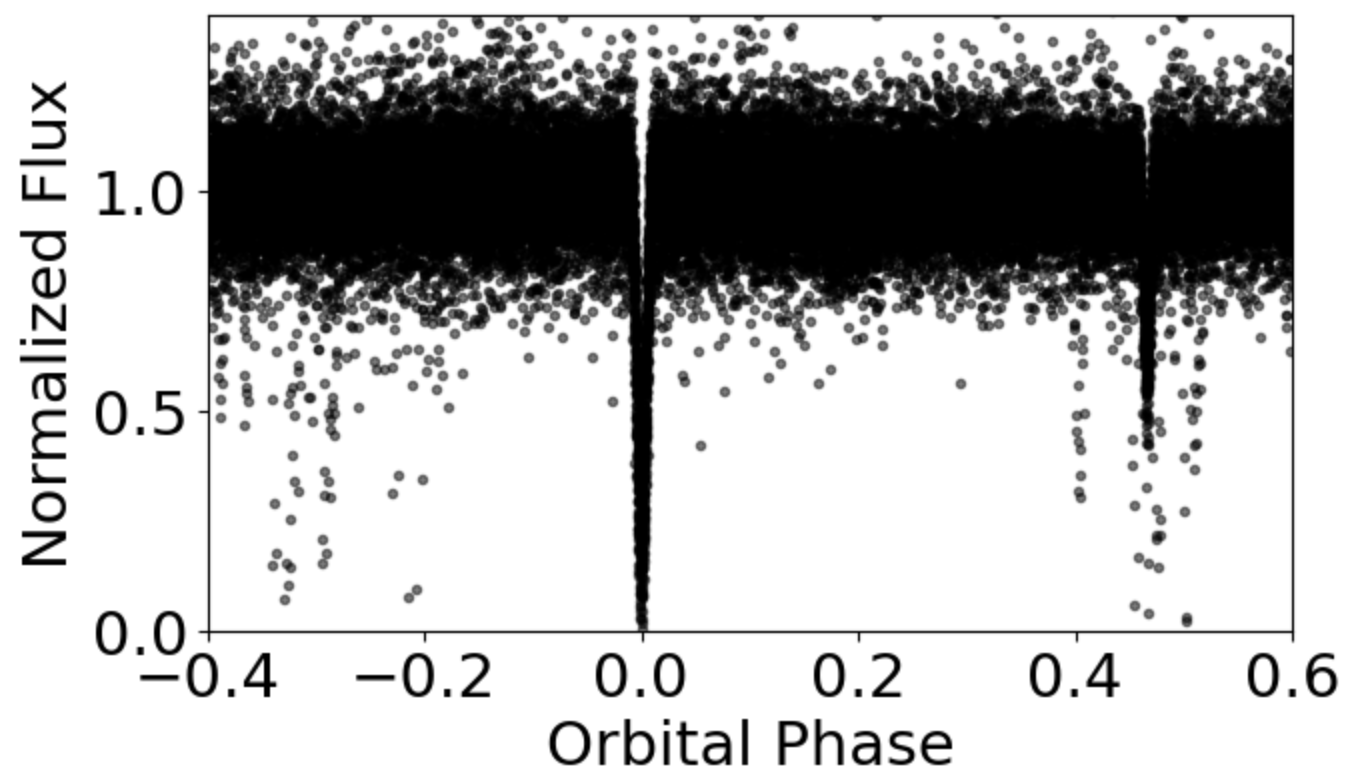}
    \includegraphics[width=0.48\linewidth]{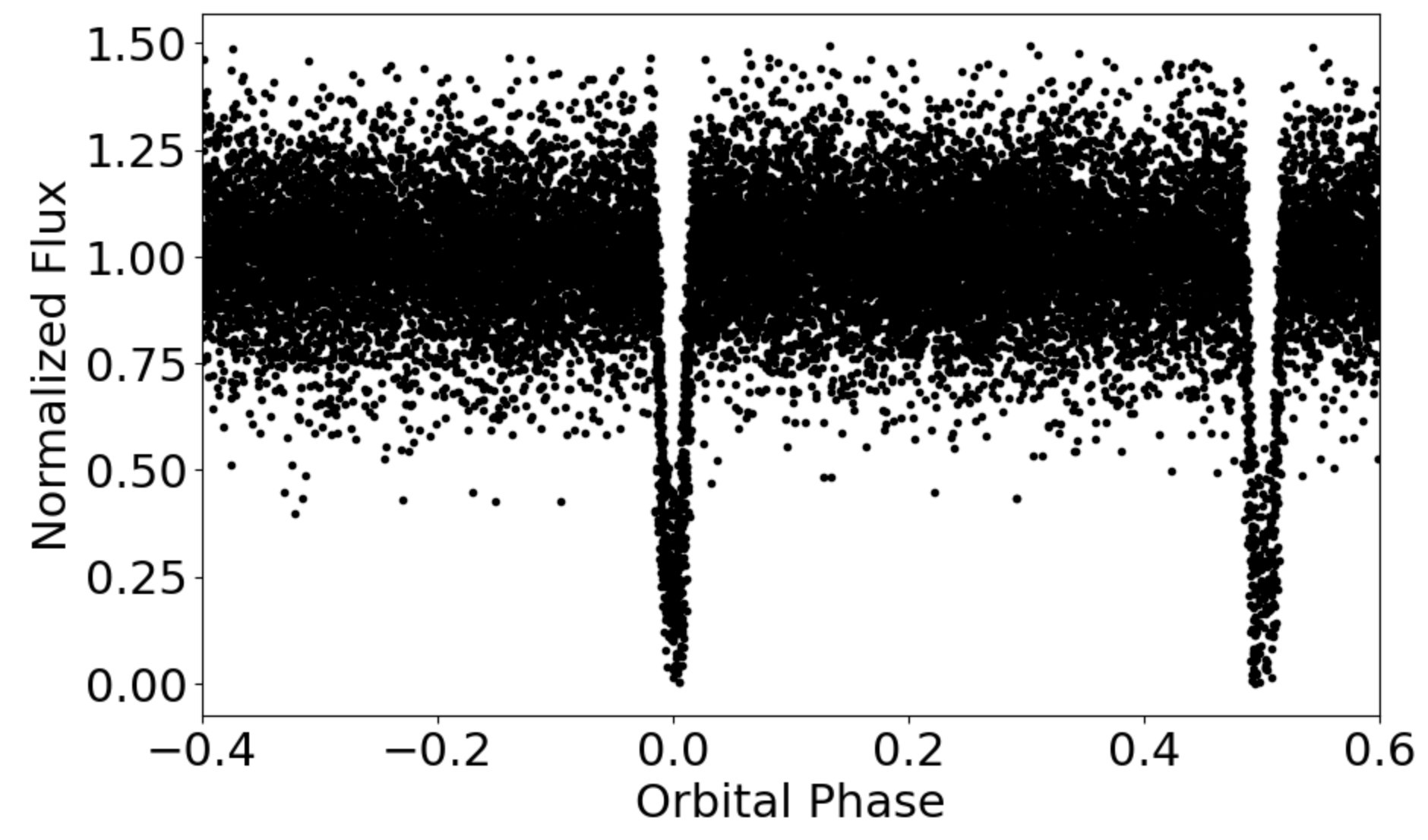}
    \caption{Phase-folded lightcurves for two targets with primary eclipse depths close to unity: TIC 42066695 (sector-averaged depth of $\approx848$ ppt, left panel) and TIC 446208053 (sector-averaged depth of $\approx804$ ppt, right panel).}
    \label{fig:dep_misc}
\end{figure}

We note that several targets in our catalog have primary depths larger than 0.5 according to the current version of \textsc{eleanor} data, likely due to systematics. The three most extreme cases are TIC 42066695 (average depth of $\approx848$ ppt), TIC 446208053 (depth of $\approx804$ ppt), and TIC 192305147 (depth of $\approx751$ ppt); the phase-folded lightcurves for the first two are shown in Fig. \ref{fig:dep_misc}. The average primary depth for 42066695 is much smaller, $\approx434$ ppt; for the other two targets there is no publicly-available QLP data at the time of writing.

\begin{figure}[!ht]
    \centering
    \includegraphics[width=0.99\linewidth]{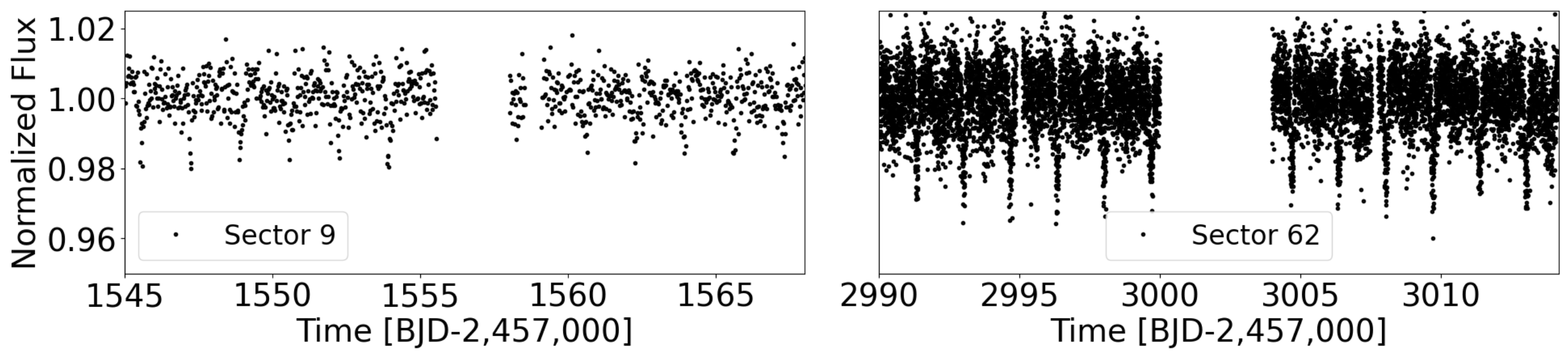}
    \caption{{\em TESS} FFI \textsc{eleanor} lightcurve of TIC 5232381 for Sector 9 (left) and Sector 62 (right). The vertical span is the same for both panels. The eclipse depths are different between the two sectors due to systematic effects cause by contamination from TIC 5232374.}
    \label{fig:dep_changes_systematics}
\end{figure}

It is important to note that the observed eclipse depths often vary from one sector to the next. With a handful of exceptions, mentioned below, these depth variations are due to systematic effects inherent to the lightcurve extraction process. An example is shown in Fig. \ref{fig:dep_changes_systematics} for the case of TIC 5232381, where the eclipses in Sector 9 (left panel) are about half as deep as those in Sector 62 (right panel). This is likely due to the sector-specific background subtraction being affected by TIC 5232374, $\approx13$ arcsec away and about two magnitudes fainter. 

\begin{figure}[!ht]
    \centering
    \includegraphics[width=0.98\linewidth]{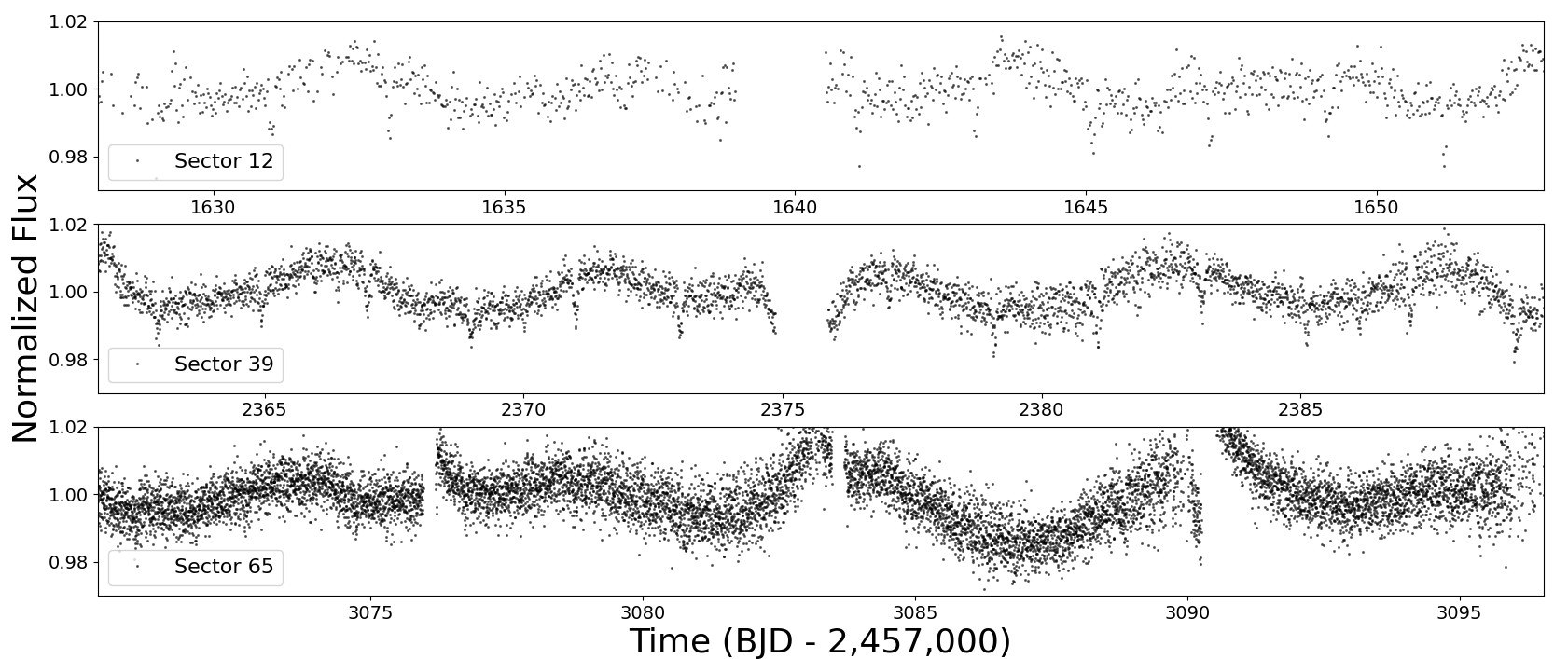}
    \caption{Similar to Fig. \ref{fig:dep_changes_systematics} but for TIC 77392704. The vertical span is the same for all three panels. Here, the eclipses are loud and clear in Sectors 12 and 39, but barely present in Sector 65.}
    \label{fig:extr_dep_var}
\end{figure}

TIC 5232381 is neither an isolated occurrence, nor an outlier. Sometimes, eclipse depths can fluctuate even within a single sector, showing differences before and after {\em TESS} data downlink gaps. In the most extreme cases, the eclipses can be virtually undetectable in certain sectors, as highlighted in Fig. \ref{fig:extr_dep_var} for TIC 77392704 (also, e.g., TIC 63165670). Thus, in these cases it is preferable to exclude such sectors when phase-folding the lightcurve, which we do by visual inspection on a target-by-target and sector-by-sector basis. 

\begin{figure}%[!ht]
    \centering
    \includegraphics[width=0.98\linewidth]{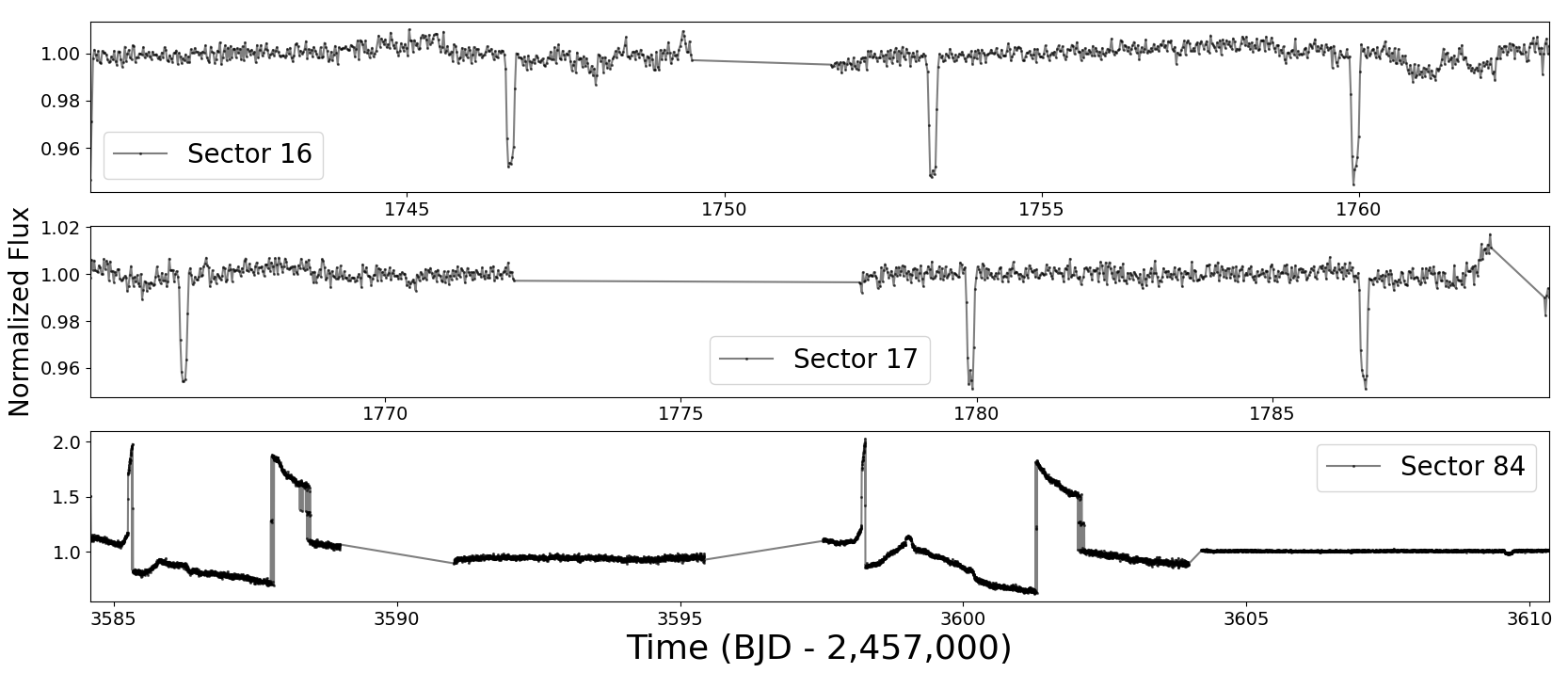}
    \caption{{\em TESS} FFI \textsc{eleanor} lightcurve of TIC 173706211 for all available sectors at the time of writing. Sector 84 is completely dominated by systematics. As a result, the ephemeris measured from Sectors 16 and 17 is not significantly improved with the addition of the latest data.}
    \label{fig:last_sector_bad}
\end{figure}

These complications are often further exacerbated when the number of detected eclipses is small due to relatively long orbital period, data gaps, and systematic effects. Sometimes, even the addition of new sectors does not help improve the measurements. An example of this is shown in Fig. \ref{fig:last_sector_bad} for the case of TIC 173706211, where the Sector 84 lightcurve is completely dominated by systematics and there is a single useful eclipse near the end of the sector. As a result, while obtaining reliable eclipse depths and durations for individual sectors is, in general, relatively straightforward, extending these measurements across multiple sectors is challenging and sector-averaged depths and durations can be misleading. 

Collectively, these factors underscore the complexity of obtaining accurate and precise measurements for many {\em TESS} FFI EBs. Thus, it is important to emphasize that even after thorough scrutiny it is still possible that the ephemerides provided in this catalog are slightly off, especially when the number of {\em TESS} observations is small, the eclipses are few and shallow, the SNR is low, and the lightcurve is dominated by systematics. Unfortunately, resolving these issues by, e.g., cross-checking EBs between different lightcurve pipelines is far from straightforward. For instance, it is not uncommon to see depth differences when comparing \textsc{eleanor} to \textsc{QLP} or \textsc{TESS-SPOC} data, likely due to different treatments of crowding correction\footnote{We note that, at the time of writing, the \textsc{QLP} lightcurves beyond Sector 74 were not well-suited for ephemerides determination due to a timing error of about 3 min. One of us (DS) noticed this issue during our investigations and brought it to the attention of the \textsc{QLP} team.}.

Finally, it is worth pointing out two further considerations regarding eclipse depth variations in particular. First, these do not substantially affect the ephemerides measurements as these are done on an eclipse-by-eclipse basis -- each eclipse is independently modeled with a generalized Gaussian which fits for the depth by design. Neither do the depth variations dramatically impact our photocenter-based vetting as it mostly depends on the eclipse durations\footnote{Naturally, it would be important to keep track of such effects when validating targets based on eclipse depths and contamination from nearby stars.}. 

\subsection{EBs in Multiple Stellar Systems}

Multiple stellar systems are not uncommon. About one in ten binary stars reside in hierarchical (2+1) triples, and thousands of even higher-order systems have already been discovered \citep[e.g.,][and references therein]{Raghavan2010,2020MNRAS.493.5583T}. The higher the multiplicity of the system, the higher its complexity in terms of orbital and physical parameters, formation and evolution pathways, and long-term dynamical stability \citep[e.g.,][and references therein]{2017ApJS..230...15M,Tokovinin2021,2020MNRAS.493.5583T}. In general, wide multiple systems are prime targets for long-term astrometric monitoring, while compact multiples are ideally-suited for observing short-term dynamical interactions between its components such as eclipse timing variations (ETVs).

\begin{figure}[!ht]
    \centering
    \includegraphics[width=0.99\linewidth]{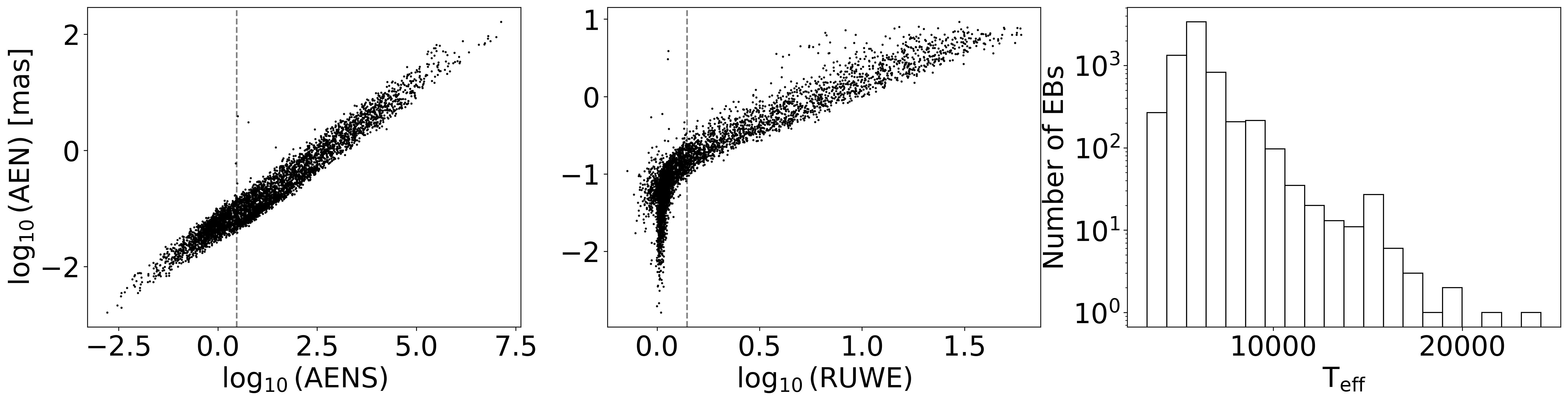}
    \caption{The distributions of Gaia's \texttt{astrometric\_excess\_noise} (AEN), \texttt{astrometric\_excess\_noise\_sig} (AENS), renormalized unit weight error (RUWE), and effective temperatures for the \newEBs new EBs presented here. The vertical dashed lines represent AENS = 3 (left panel) and RUWE = 1.4 (middle panel), potentially suggesting unresolved companions.}
    \label{fig:gaia_data}
\end{figure}

Cross-matching our catalog with the Gaia DR3 astrometric measurements, we extracted the available \verb|astrometric_excess_noise| (AEN), \verb|astrometric_excess_noise_sig| (AENS), and renormalized unit weight error (RUWE). These can be used to test for unseen companions \citep[e.g.,][and references therein]{Belokurov2020,Penoyre2020,Stassun2021,Gandhi2022,Majewski2025} which, if indeed present, would potentially mark the EBs as components in systems of three (or more) stars. The corresponding distributions are shown in Fig. \ref{fig:gaia_data}, highlighting several interesting features. In particular, the AEN is greater than 10 mas for hundreds of targets, and the AENS is greater than 3 for $\sim40\%$ of the EBs, reaching values of tens to even hundreds of thousands for dozens of targets. Similarly, RUWE is greater than 1.4 -- suggesting unresolved companions \citep{Stassun2021} -- for about one in every four targets. Altogether, these considerations indicate that a potentially large fraction of the \newEBs new EBs presented here may reside in multiple stellar systems. 

Another option for finding multiple stellar systems is through the presence of extra events in the lightcurves of EBs. Indeed, {\em TESS} has already enabled the detection of thousands of such events, practically revolutionizing the field by discovering hundreds of new 2+1 triply-eclipsing triple systems \citep[e.g.,][and references therein]{2022Galax..10....9B,2022MNRAS.513.4341R,2024A&A...686A..27R,2024ApJ...974...25K}, 2+2 eclipsing quadruple systems \citep[][]{Kostov2022_quadcat1,2024A&A...687A...6Z,Kostov_quadcat2,2025arXiv250412239P}, as well as unusual (2+1)+1 eclipsing quadruples \citep[e.g.,][]{2022ApJ...938..133P}, several (2+1)+2 quintuple systems \citep{Kostov2022_quadcat1,Kostov_quadcat2}, the first two (2+2)+2 eclipsing sextuple systems \citep{2021AJ....161..162P,2023MNRAS.520.3127Z}, and even two transiting circumbinary planets \citep{2020AJ....159..253K,2021AJ....162..234K}. Volunteers at EBP independently re-discovered many of these and, naturally, a significant number of false positives that mimic 2+2 eclipsing quadruples due to blended light from two unrelated EBs\footnote{See, for example, \url{https://www.zooniverse.org/projects/vbkostov/eclipsing-binary-patrol/talk/tags/multiple-system-candidate}}, and also identified several new eclipsing triple and quadruple candidates (paper in preparation).

\subsection{Known EBs observed by {\em TESS}}

\begin{figure}[!ht]
    \centering
    \includegraphics[width=0.99\linewidth]{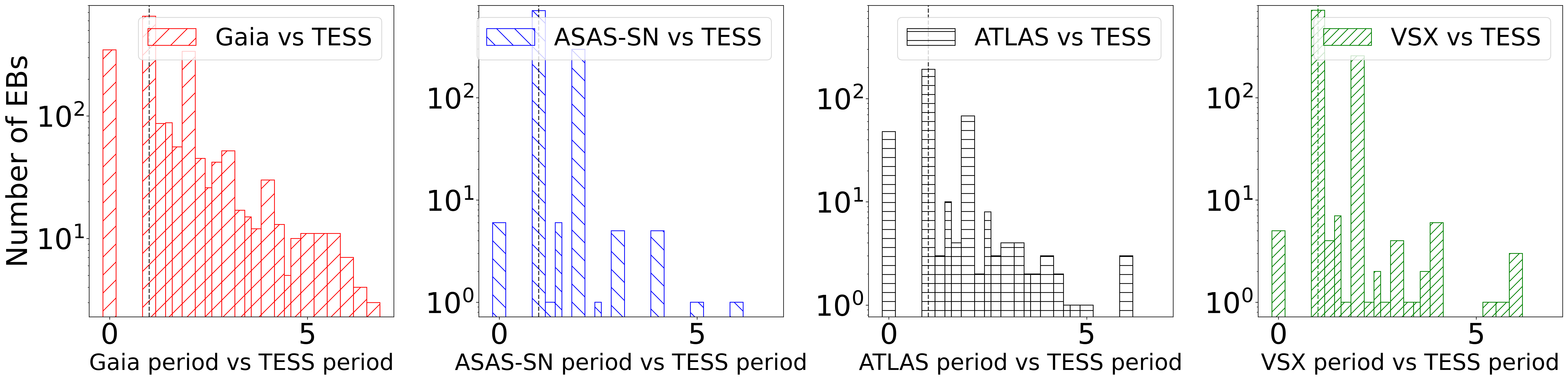}
    \caption{Distributions of period ratios between the correct {\em TESS} periods and the incorrect periods from Gaia (first panel from left), ASAS-SN (second panel from left), ATLAS (third panel from left) and VSX (last panel from left) for \knownEBs known EBs. For simplicity, the reciprocal fractions are combined, i.e., 1/2 with 2/1, 1/3 with 3/1, etc. The zero values represent targets where the Gaia/ASAS-SN/ATLAS/VSX periods are not within 10\% of a corresponding integer fraction of the {\em TESS} periods.}
    \label{fig:tess_vs_gaia_distribution}
\end{figure}

Given our ML search was effectively blind, it was inevitable that it picked up a large number of known EBs. And indeed, as discussed above, about one in four of the identified candidates are within 1 pixel of known EBs. Thus, in order to verify the efficiency and reliability of our automated ephemeris and vetting pipeline, we applied it to $\approx$ 30,000 such targets and tracked its performance. Interestingly, during the early stages of the EBP project, the volunteers noticed that the correctly-measured periods from {\em TESS} are sometimes different from literature values. Altogether, we marked \knownEBs such cases. As an example, the distributions of the period ratios between {\em TESS} on one hand and Gaia, ASAS-SN, ATLAS, and VSX on the other are highlighted in Fig. \ref{fig:tess_vs_gaia_distribution}.

\begin{figure*}[!ht]
    \centering
    \includegraphics[width=0.49\linewidth]{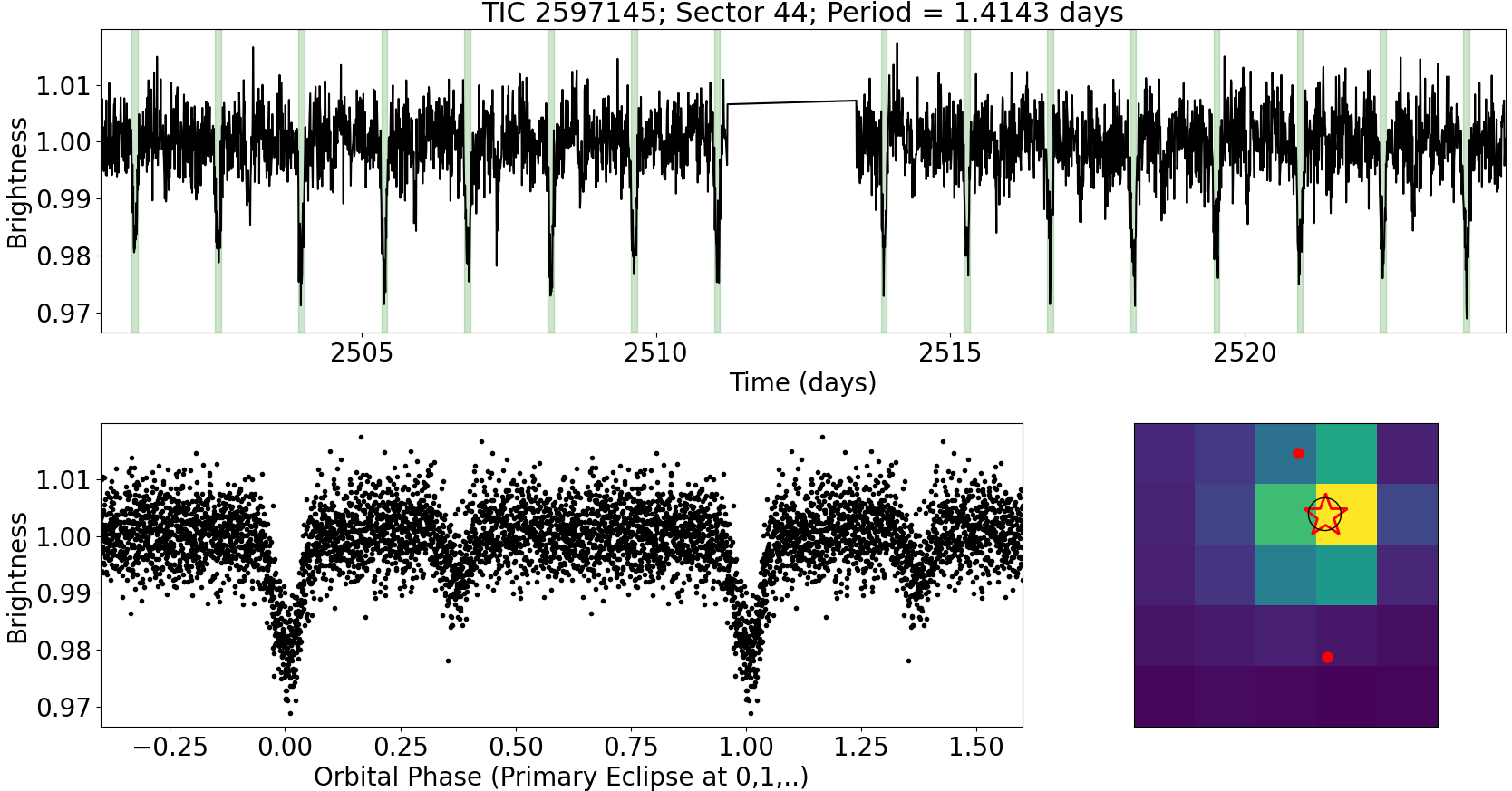}
    \includegraphics[width=0.49\linewidth]{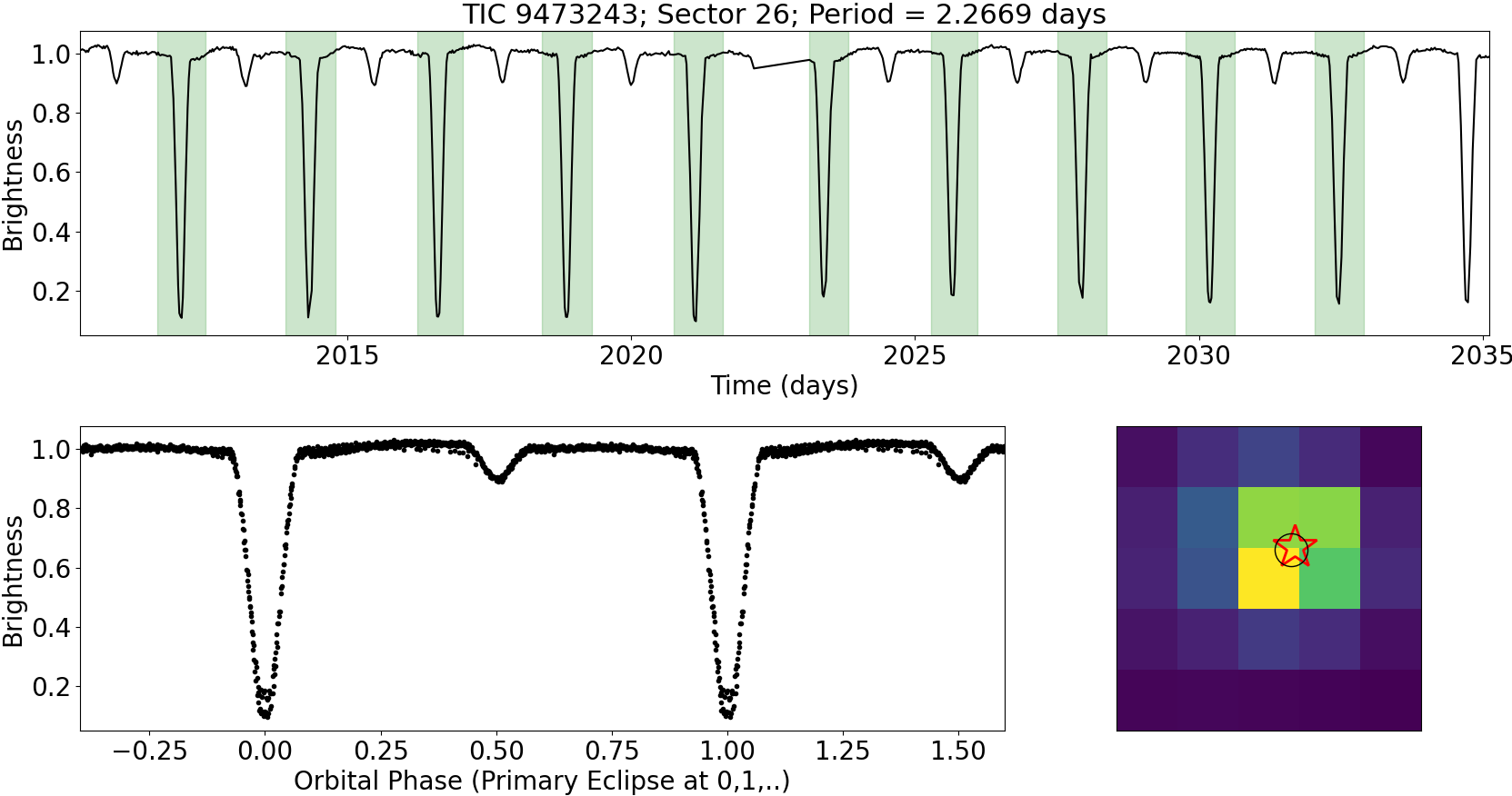}
    \caption{Left panels: Ephemeris and vetting pipeline results for TIC 2597145. The correct period measured from {\em TESS} is 1.4143 days -- twice as long as the period listed in Gaia (0.7072 days); ASAS-SN and VSX provide the correct period. Right panels: Same as left but for TIC 9473243. Here, the Gaia period is four times the correct period, the ASAS-SN period is twice the correct period, and the WISE period is half of the correct period.}
    \label{fig:tess_vs_gaia_ratios}
\end{figure*}

\begin{figure*}
    \centering
    \includegraphics[width=0.48\linewidth]{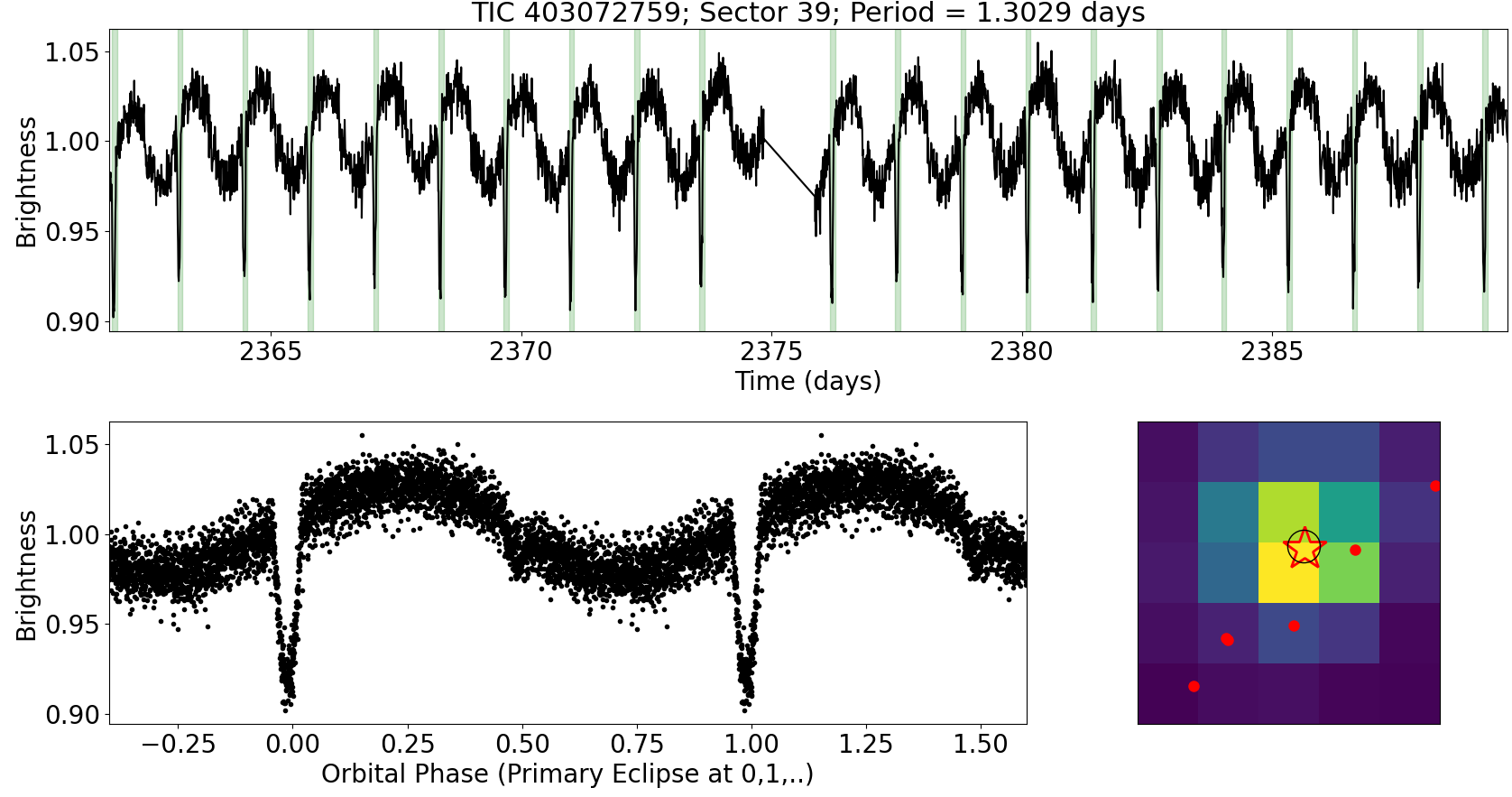}
    \includegraphics[width=0.48\linewidth]{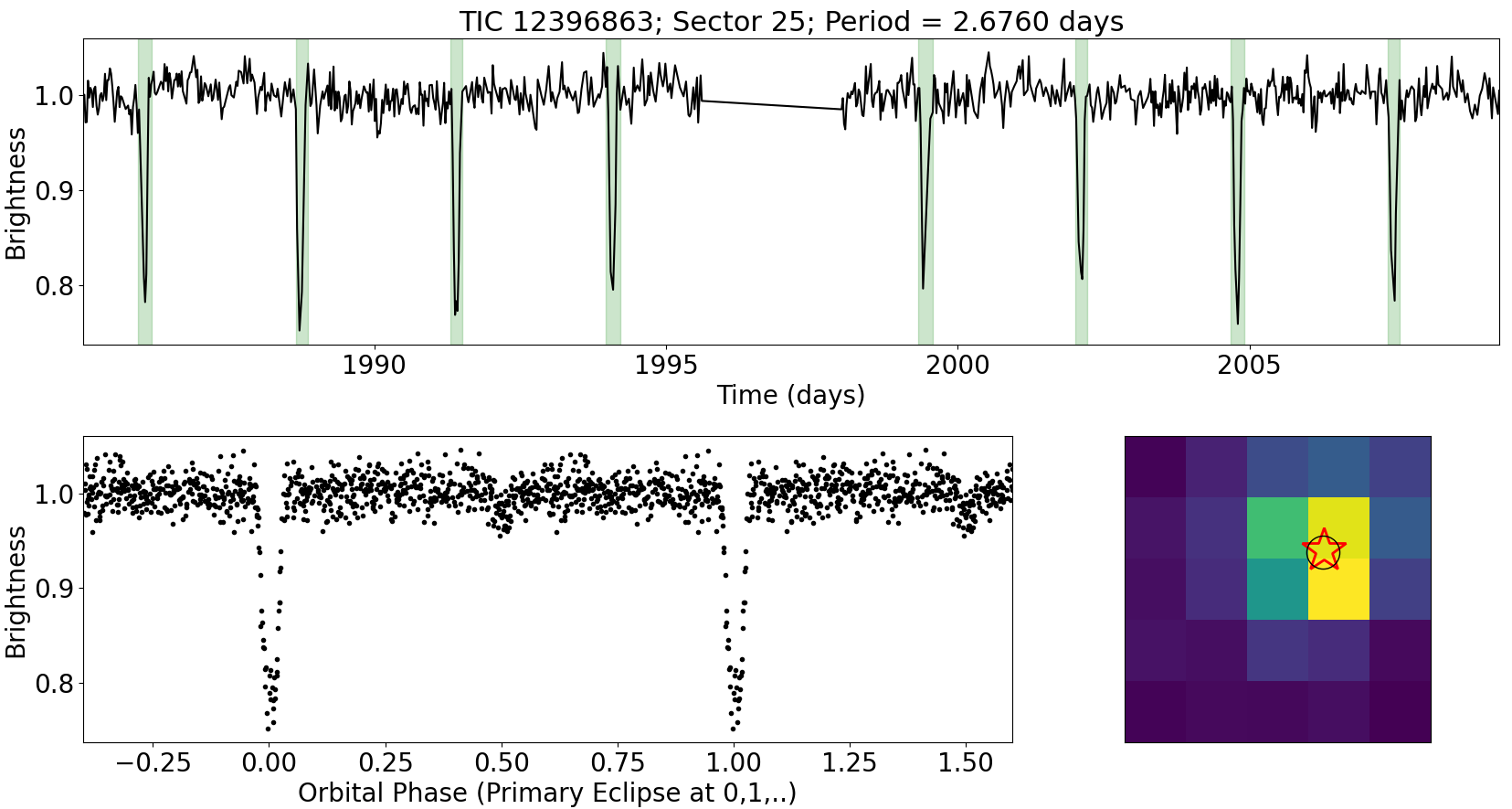}
    \caption{Same as Fig. \ref{fig:tess_vs_gaia_ratios} but for TIC 403072759 (left), and TIC 12396863 (right). Thanks to {\em TESS}, shallow secondary eclipses can be detected near phase 0.5 for both targets, confirming the corresponding periods are 1.3029 days and 2.6760 days. For comparison, the periods listed in Gaia are 0.5210 days for TIC 403072759 (i.e., 2/5 of the true period) and 5.3519 days (i.e., twice the true period).}
    \label{fig:403072759}
\end{figure*}

As seen from the figure, most of the Gaia, ASAS-SN, ATLAS, and VSX periods are close to an integer fraction of the true period, where for simplicity `close' is defined here as within 10\% of integer fractions of 2 (from 1/2 to 10/2) and 3 (from 1/3 to 20/3)\footnote{The reciprocal fractions are combined in the figure, i.e., 1/2 with 2/1 , 1/3 with 3/1, 2/3 with 3/2, etc.}. This is perhaps not too surprising given the much longer continuous baseline coverage and higher cadence of {\em TESS} observations compared to other surveys. As an example, Fig. \ref{fig:tess_vs_gaia_ratios} shows TIC 2597145, where the correct period measured from {\em TESS} is 1.4143 days, twice the period listed in Gaia (0.7072 days). For this target, the periods listed in ASAS-SN and VSX are correct. Another example is TIC 9473243, where the correct period measured from {\em TESS} is 2.2669 days, whereas Gaia gives a period of 9.0680 days (four times too long), ASAS-SN gives a period of 4.5338 days (two times too long) and WISE gives a period of 1.1335 days (two times too short). Additionally, {\em TESS} excels at enabling the detection of shallow secondary eclipses. An example of this is shown in Fig. \ref{fig:403072759} for TIC 403072759, highlighting the shallow but clear secondary eclipse near phase of 0.5. Here, the true period measured from {\em TESS} is 1.3029 days whereas Gaia gives a period of 0.52 days, i.e., a 2/5 fraction of the true period.

\begin{figure*}
    \centering
    \includegraphics[width=0.48\linewidth]{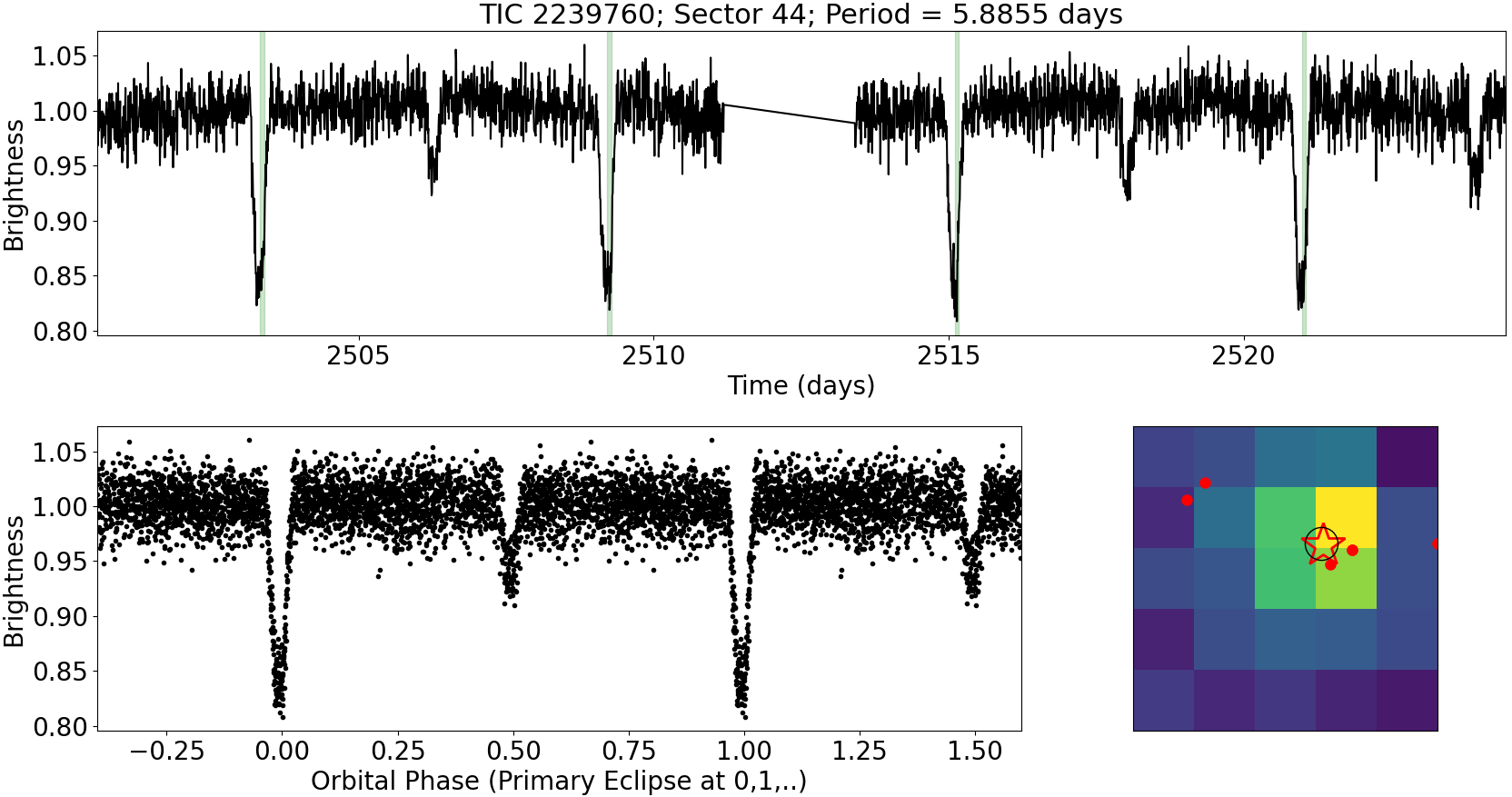}
    \includegraphics[width=0.48\linewidth]{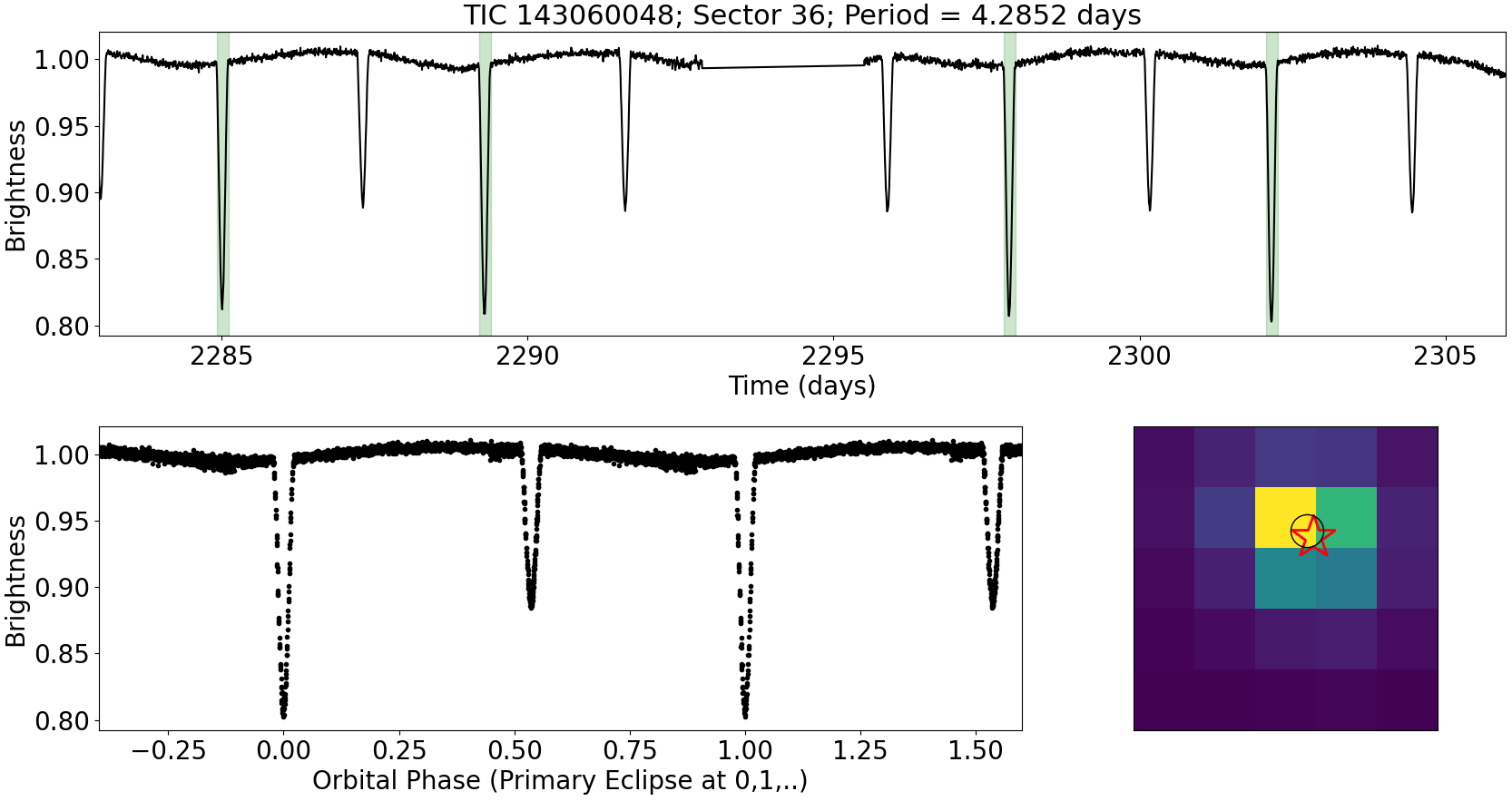}
    \caption{Same as Fig. \ref{fig:tess_vs_gaia_ratios} but for TIC 2239760 (left), and TIC 143060048 (right), where the corresponding ratios between the Gaia and {\em TESS} periods are not close to low-order integer ratios ($\approx0.55$ and $\approx7.11$, respective).}
    \label{fig:tess_vs_gaia_off}
\end{figure*}

Interestingly, about 20\% of the Gaia EB periods seem to be unrelated to the {\em TESS} periods at all. These cases are represented in Fig. \ref{fig:tess_vs_gaia_distribution} by the peak at zero. One example is TIC 2239760 where the correct {\em TESS} period is 5.8855 days, while the Gaia period is 3.2370 days, a ratio of $\approx0.55$ (Fig. \ref{fig:tess_vs_gaia_off}, left panel). Another is TIC 143060048, where the {\em TESS} period is 4.2852 days and the Gaia period is 30.4715 days (ratio of $\approx7.11$) (Fig. \ref{fig:tess_vs_gaia_off}, right panel). Some of the most extreme discrepancies are for TIC 443450339, 152328270, 353628656, 138032974 and 34853800, where the {\em TESS} periods are 2.9032, 9.0706, 2.5816, 1.8713, and 3.2028 days, respectively, whereas the corresponding Gaia periods are orders of magnitude longer, i.e., 406.0884, 381.0741, 355.6257, 189.4914, and 164.4351 days. 

Table \ref{tab:known_ebs} highlights ten random rows of the catalog of \knownEBs known EBs with updated ephemerides produced as part of this work.

\begin{table}[]
    \centering
    \begin{tabular}{c|cccccccccccccc}
        \hline
        \hline
        \rotatebox{90}{TIC} & 
        \rotatebox{90}{RA} & 
        \rotatebox{90}{Dec} & 
        \rotatebox{90}{TESS period (d)} & 
        \rotatebox{90}{TESS T0} & 
        \rotatebox{90}{Gaia period (d)} & 
        \rotatebox{90}{Gaia vs TESS} & 
        \rotatebox{90}{ASAS-SN period (d)} & 
        \rotatebox{90}{ASAS-SN vs TESS} & 
        \rotatebox{90}{ATLAS period (d)} & 
        \rotatebox{90}{ATLAS vs TESS} & 
        \rotatebox{90}{VSX period (d)} & 
        \rotatebox{90}{VSX vs TESS} & 
        \rotatebox{90}{WISE period (d)} & 
        \rotatebox{90}{WISE vs TESS} \\
        761795257 & 111.7000 & 12.5605 & 2.5887 & 1494.1521 & 3.2064 & 1.33 & -- & -- & 5.1740 & 2 & -- & -- & -- & -- \\
        252351823 & 74.9920 & 55.4948 & 4.0213 & 1820.4185 & 7.4545 & 0 & -- & -- & -- & -- & 4.0088 & 1 & -- & -- \\
        436564213 & 70.6023 & 12.9477 & 3.2533 & 1444.2706 & 3.2431 & 1 & 3.2432 & 1 & -- & -- & 3.2431 & 1 & 1.6215 & 2 \\
        386250632 & 139.1054 & -58.3239 & 1.5174 & 1546.1097 & 1.5174 & 1 & 9.1040 & 6 & -- & -- & 1.5172 & 1 & 0.7587 & 2 \\
        369995729 & 20.8105 & 59.8022 & 2.6458 & 1792.0862 & 2.0944 & 1.33 & 0.2571 & 0 & -- & -- & 0.2571 & 0 & -- & -- \\
        68543179 & 103.6037 & 32.4242 & 4.1924 & 1845.8610 & 4.6823 & 0 & 4.1927 & 1 & -- & -- & 4.1925 & 1 & -- & -- \\
        311651226 & 260.8442 & -77.3195 & 1.0494 & 1629.9139 & 1.0494 & 1 & 1.0494 & 1 & -- & -- & 1.0524 & 1 & 0.5247 & 2 \\
        427654873 & 349.4827 & 70.1097 & 4.2869 & 1764.9825 & 8.5734 & 2 & -- & -- & -- & -- & 4.2871 & 1 & 2.1434 & 2 \\
        410498300 & 26.0089 & 48.0453 & 2.2905 & 1792.8908 & 2.2905 & 1 & 2.2903 & 1 & -- & -- & 2.2904 & 1 & 1.1453 & 2 \\
        123135027 & 118.6583 & -5.2361 & 2.1691 & 1494.1874 & 0.2870 & 0 & 2.1689 & 1 & -- & -- & 2.1689 & 1 & -- & -- \\
    \end{tabular}
    \caption{Comparison between the correct period measured from {\em TESS} and the periods from Gaia, ASAS-SN, ATLAS, VSX, and WISE for \knownEBs known EBs. Table available in full as a machine-readable online supplement.}
    \label{tab:known_ebs}
\end{table}

\subsection{Interesting Systems}
\label{sec:interesting}

\begin{figure}[!ht]
    \centering
    \includegraphics[width=0.95\linewidth]{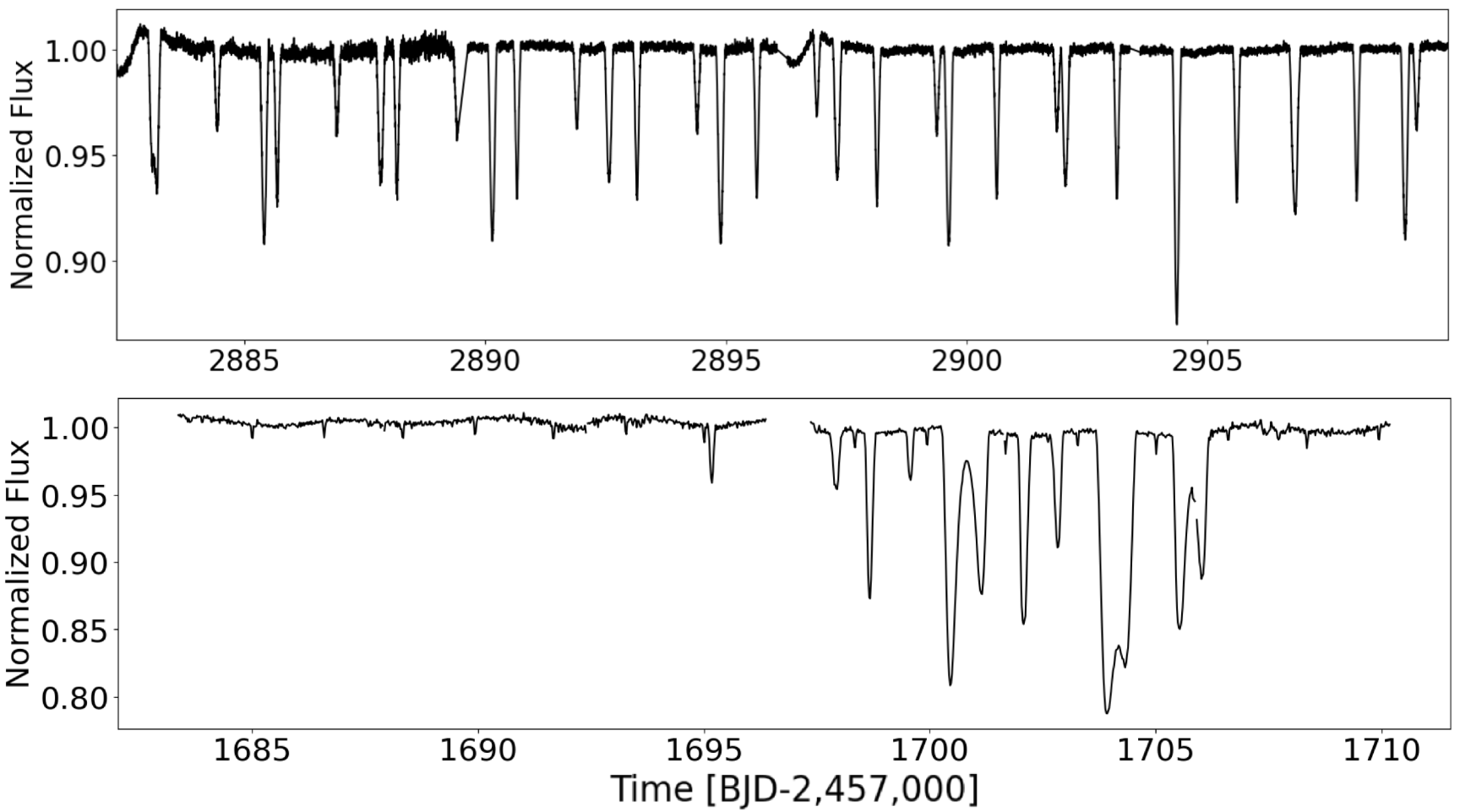}
    \caption{Upper panel: {\em TESS} FFI \textsc{eleanor} lightcurve of TIC 307119043, an eclipsing 2+2 quadruple  exhibiting two sets of primary and secondary eclipses \citep{Kostov2022_quadcat1} Lower panel: eclipsing (2+1)+1 quadruple system TIC 114936199 exhibiting a complex eclipse on the outer orbit \citep{2022ApJ...938..133P}.}
    \label{fig:quads}
\end{figure}

\begin{figure}[!ht]
    \centering
    \includegraphics[width=0.98\linewidth]{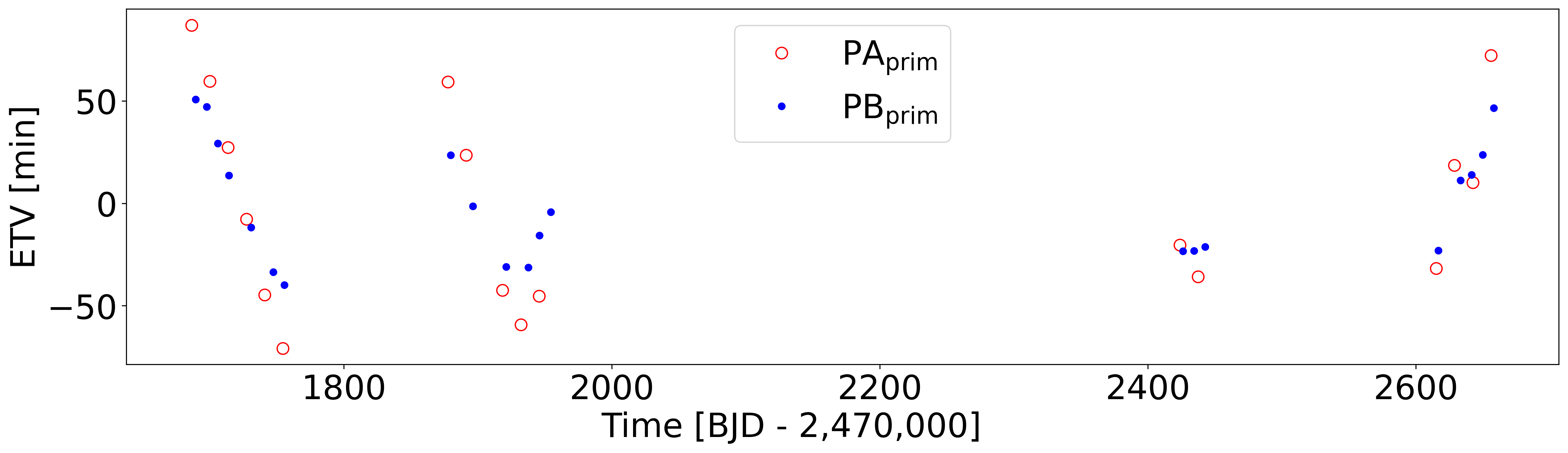}
    \includegraphics[width=0.98\linewidth]{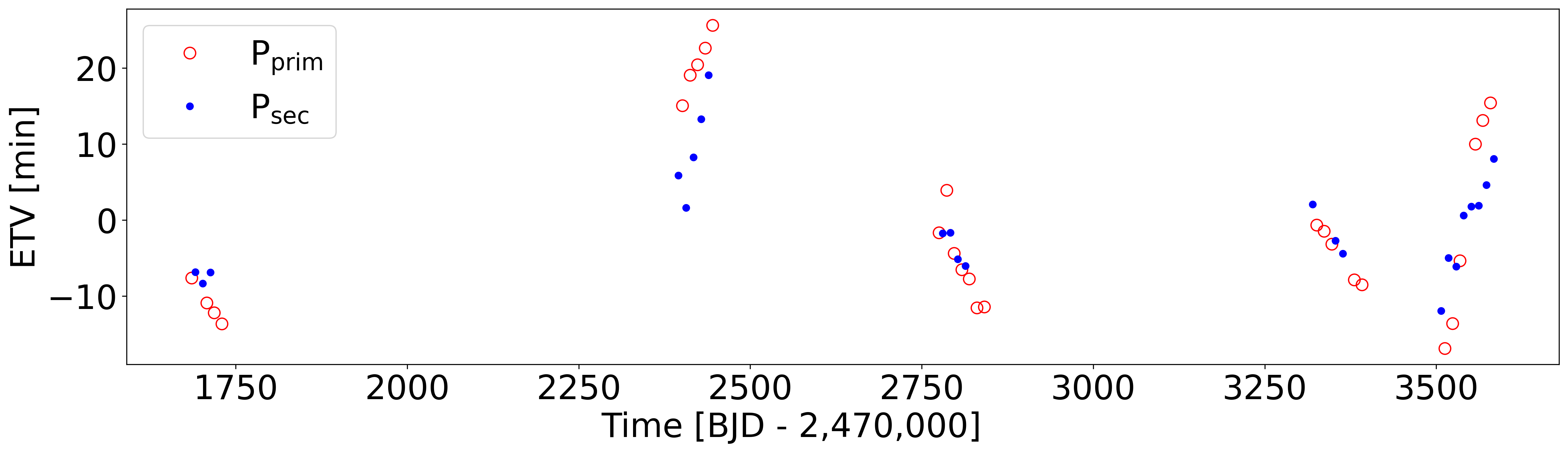}
    \caption{Upper panel: Primary ETVs for the two EBs in the 2+2 quadruple system TIC 219006972, confirming the two sub-systems are gravitationally-bound \citep{2023MNRAS.522...90K} with an outer period of 168 days. Lower panel: Primary (red) and secondary (blue) ETVs of the known EB TIC 26542657, suggesting a 2+1 triple system with an outer period of about 300 days.}
    \label{fig:219006972_and_26542657}
\end{figure}

\begin{figure}[!ht]
    \centering
    \includegraphics[width=0.95\linewidth]{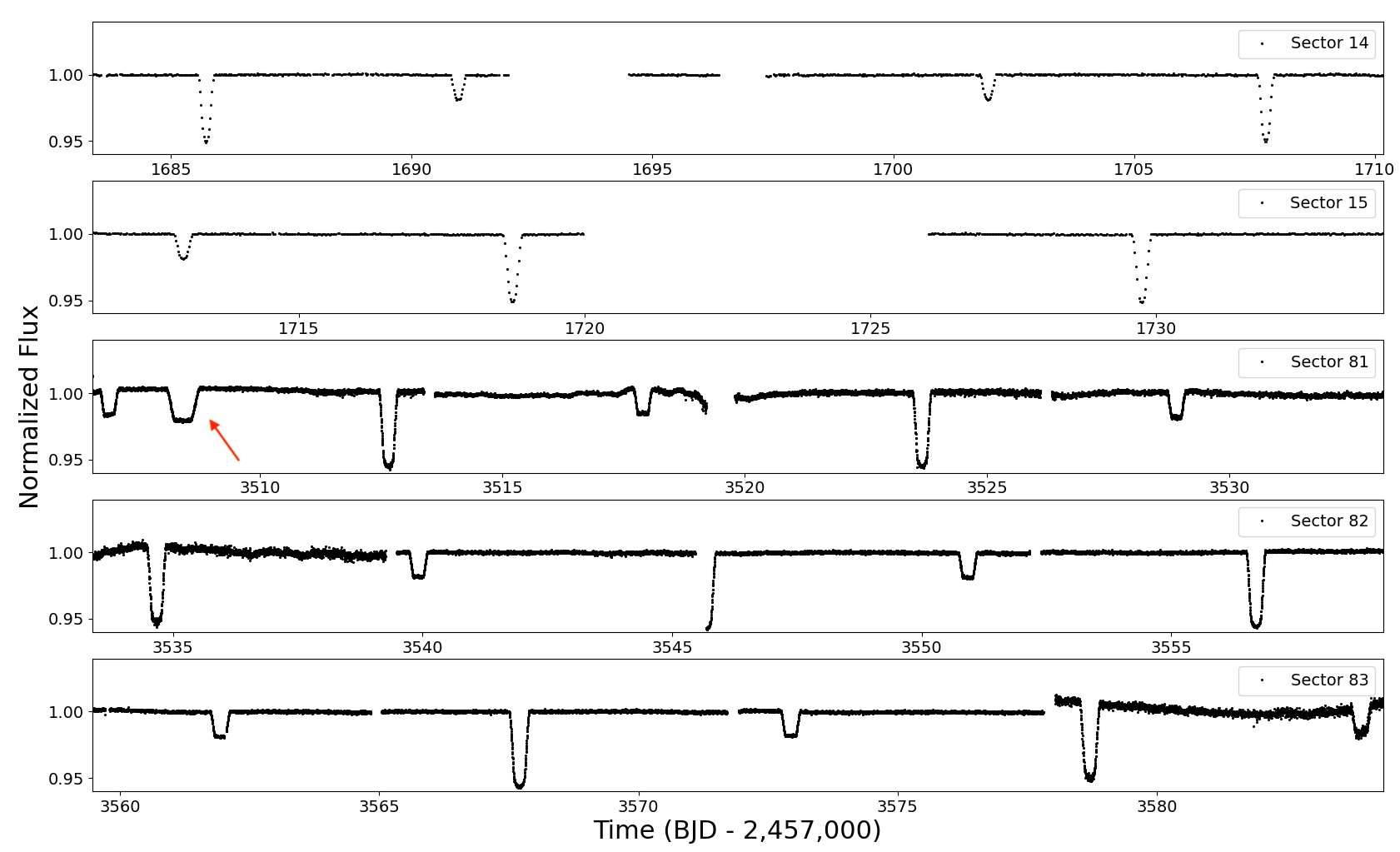}
    \caption{{\em TESS} \textsc{eleanor} lightcurve of the known EB TIC 26542657, highlighting clear changes in the shape of the primary and secondary eclipses between the first two Sectors (14 and 15; sharp primary, rounded secondary) and the last three sectors (81, 82, 83; rounded primary, flat secondary), along with a prominent tertiary eclipse in Sector 81 (red arrow). }
    \label{fig:26542657_lc}
\end{figure}

\begin{figure}[!ht]
    \centering
    \includegraphics[width=0.95\linewidth]{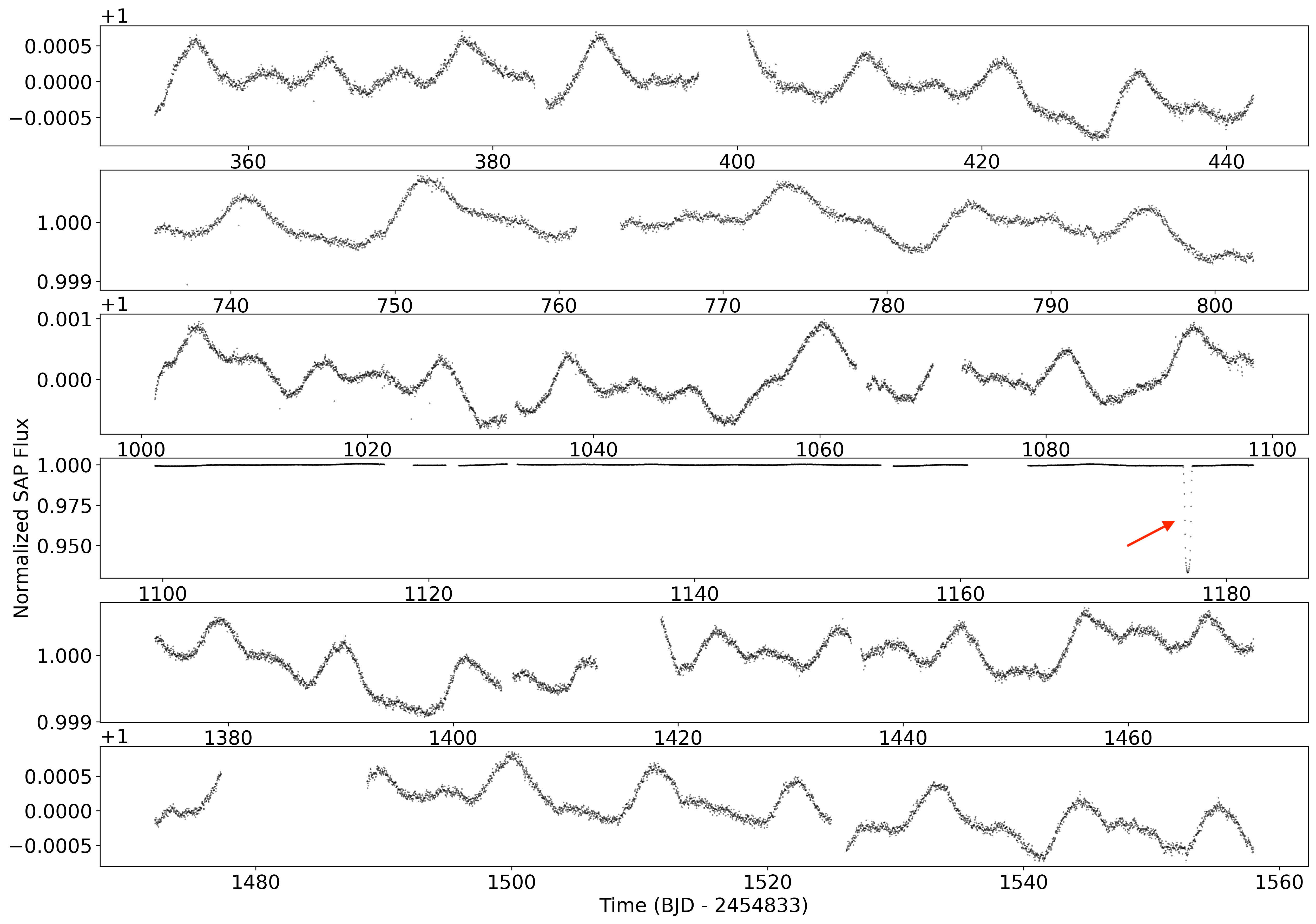}
    \caption{Kepler lightcurve of the known {\em TESS} EB TIC 26542657 showing a single {\it tertiary} eclipse (marked with a red arrow) and no discernible eclipses from the $\sim11$-days EB.}
    \label{fig:26542657_kepler}
\end{figure}

\begin{figure}[!ht]
    \centering
    \includegraphics[width=0.95\linewidth]{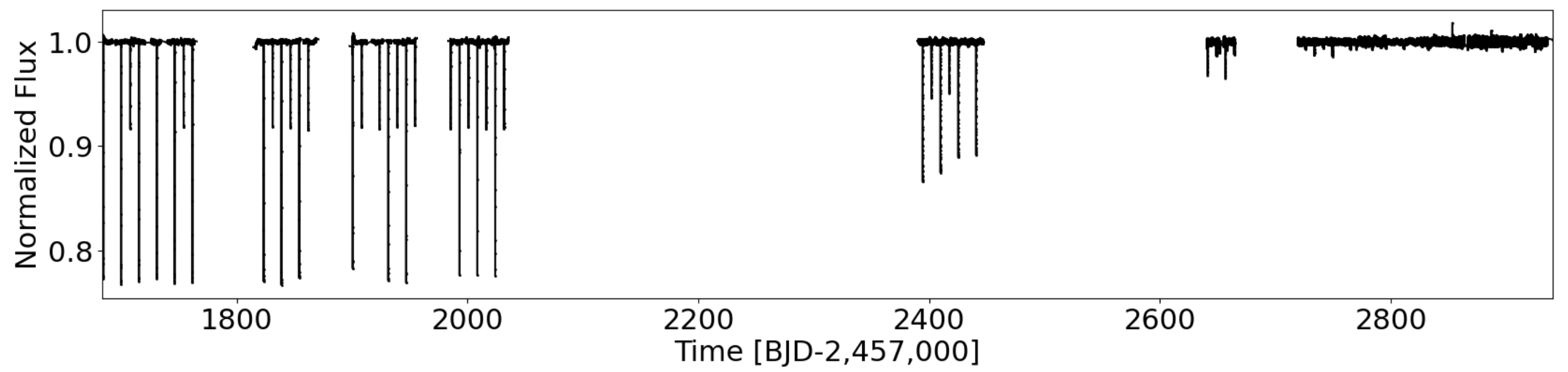}
    \includegraphics[width=0.95\linewidth]{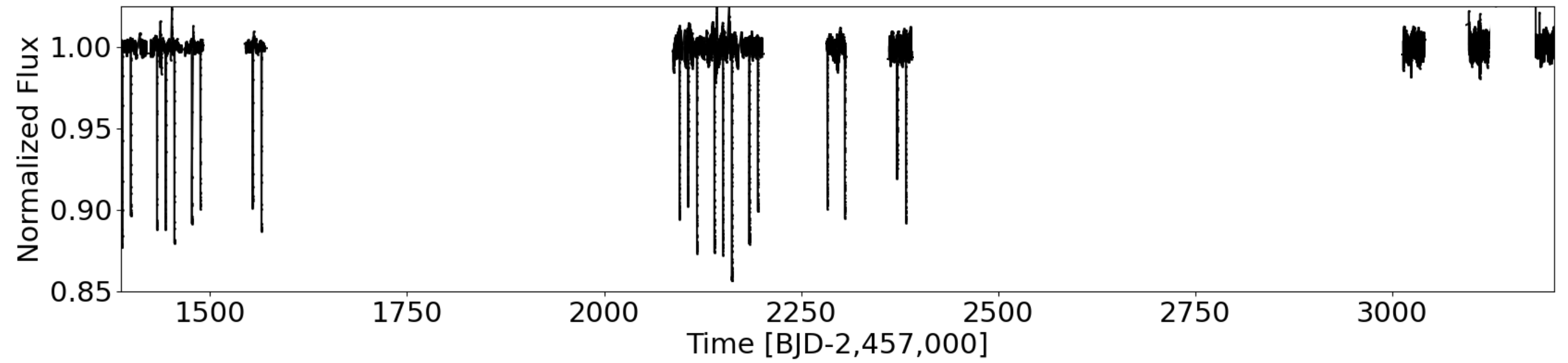}
    \caption{Disappearing eclipses due to dynamical interactions with unseen companions. Upper panel: TIC 236774836 \citep{2024A&A...685A..43M}; lower panel: TIC 220410224.}
    \label{fig:edvs}
\end{figure}

\begin{figure}[!ht]
    \centering
    \includegraphics[width=0.95\linewidth]{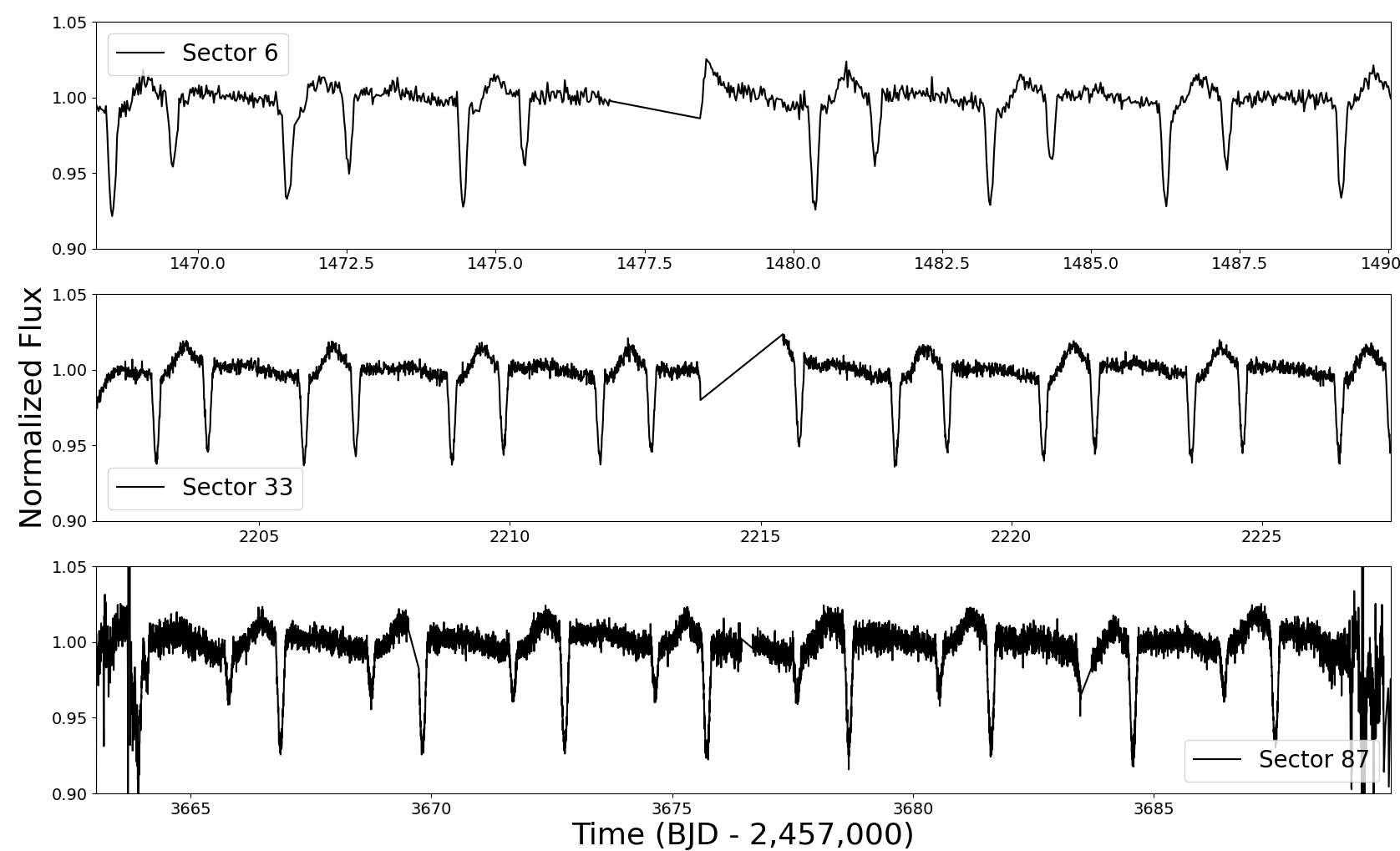}
    \caption{{\em TESS} lightcurve of the known EB TIC 234229841, where the primary and secondary eclipse ``switch'' places. In Sector 6, the deeper eclipses precede the heartbeat-like hump, the primary and secondary eclipses have similar depths in Sector 33, and in Sector 87 the deeper eclipses follow the hump.}
    \label{fig:eclipse_switch}
\end{figure}

\begin{figure}[!ht]
    \centering
    \includegraphics[width=0.95\linewidth]{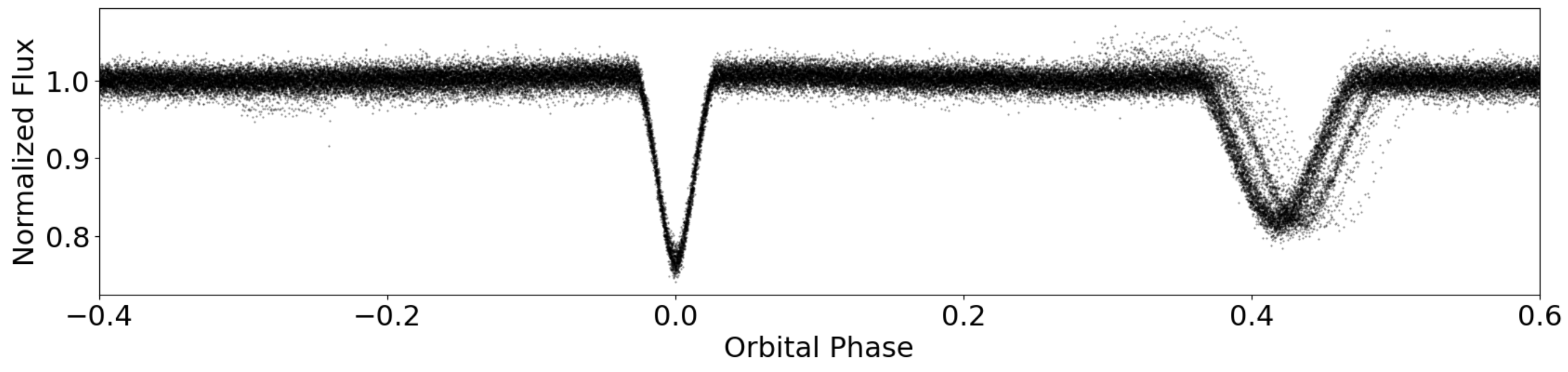}
    \includegraphics[width=0.95\linewidth]{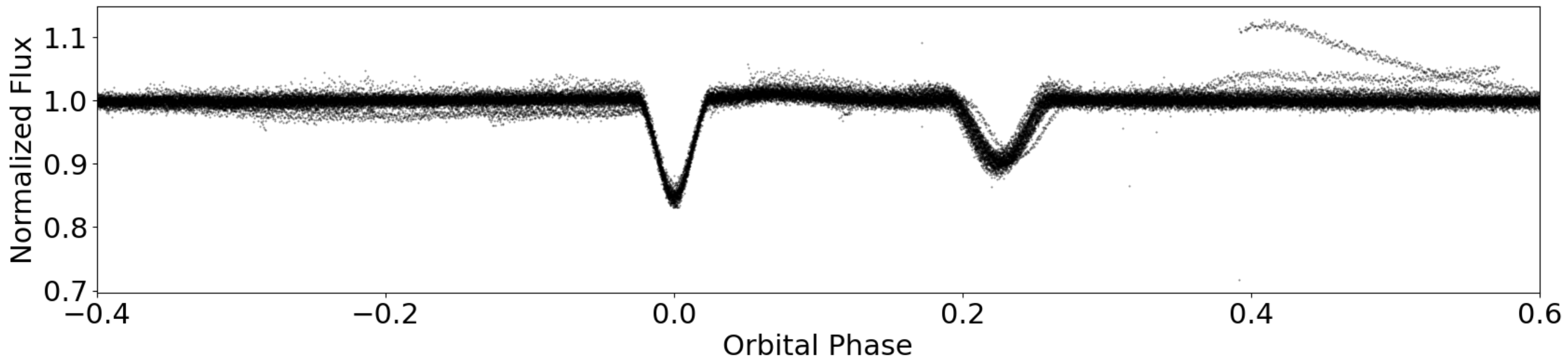}
    \caption{Pronounced apsidal motion exhibited by TIC 189281140 (upper panel) and TIC 470715046 (lower panel);}
    \label{fig:apsidal}
\end{figure}

\begin{figure}
    \centering
    \includegraphics[width=0.98\linewidth]{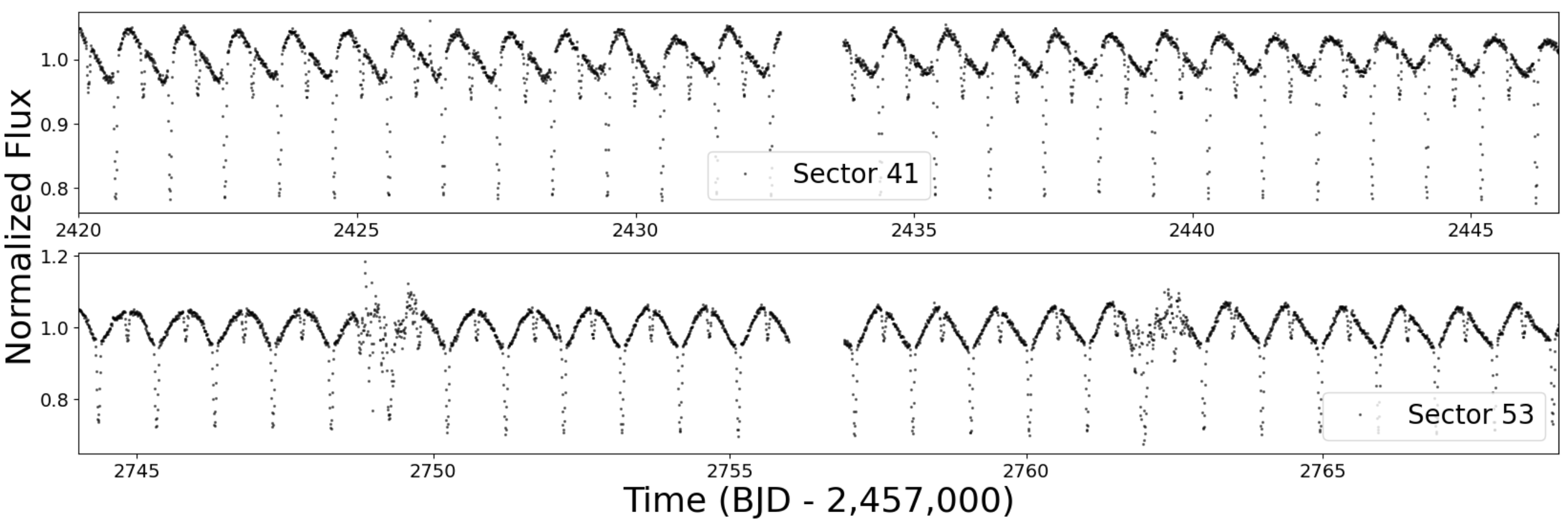}
    \caption{Two sectors of \textsc{eleanor} data for TIC 21159577, an EB showcasing an evolving pattern of starspot-induced rotational modulations.}
    \label{fig:rotator}
\end{figure}

\begin{figure}
    \centering
    \includegraphics[width=0.98\linewidth]{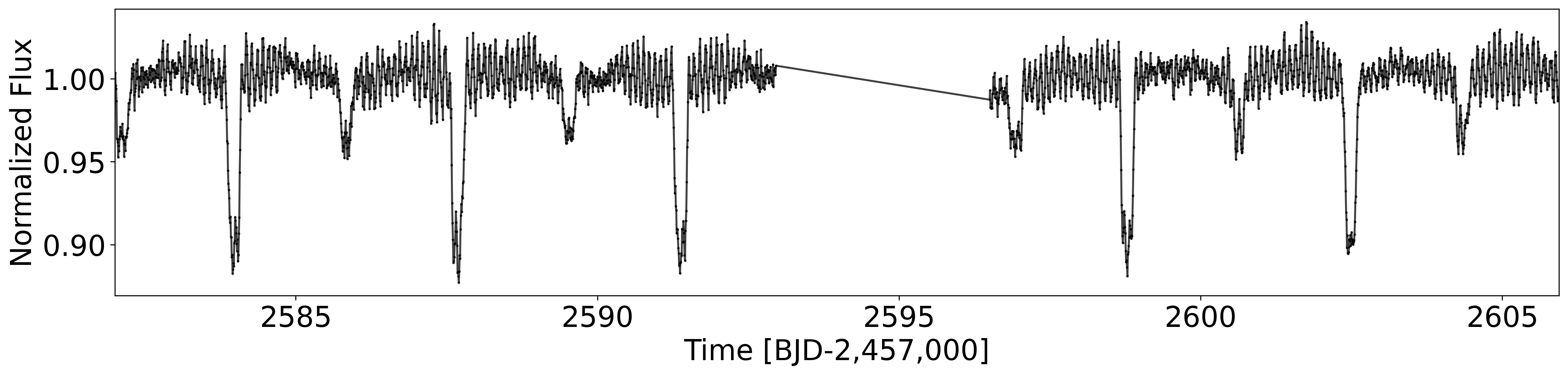}
    \includegraphics[width=0.98\linewidth]{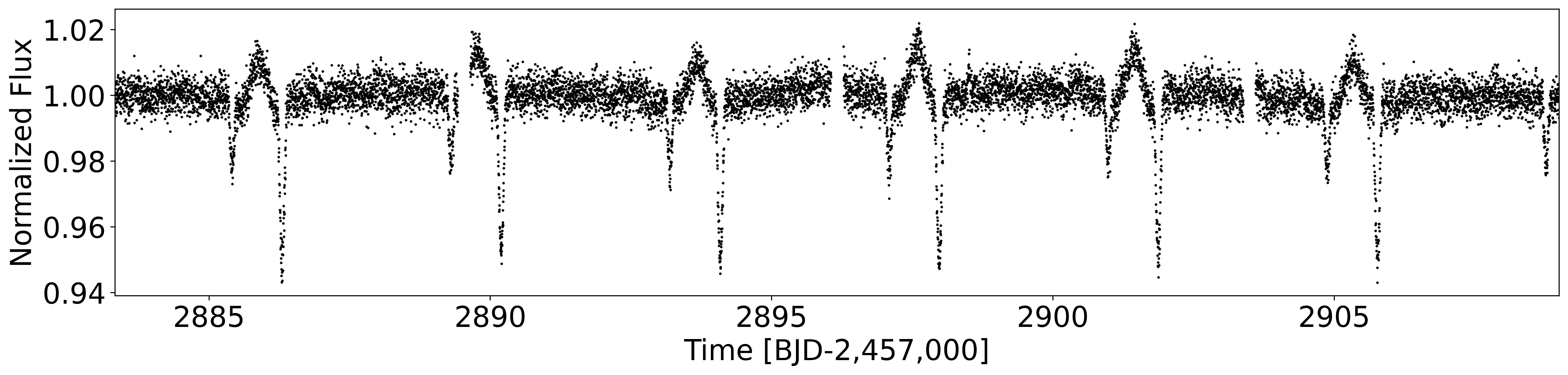}
    \caption{Upper panel: {\em TESS} lightcurve of an EB with a pulsating component (TIC 22621932). Lower panel: {\em TESS} lightcurve of an EB exhibiting a pronounced heartbeat pattern (TIC 336538437, \citep{2025ApJS..276...17S})}
    \label{fig:pulsator}
\end{figure}

\begin{figure}
    \centering
    \includegraphics[width=0.98\linewidth]{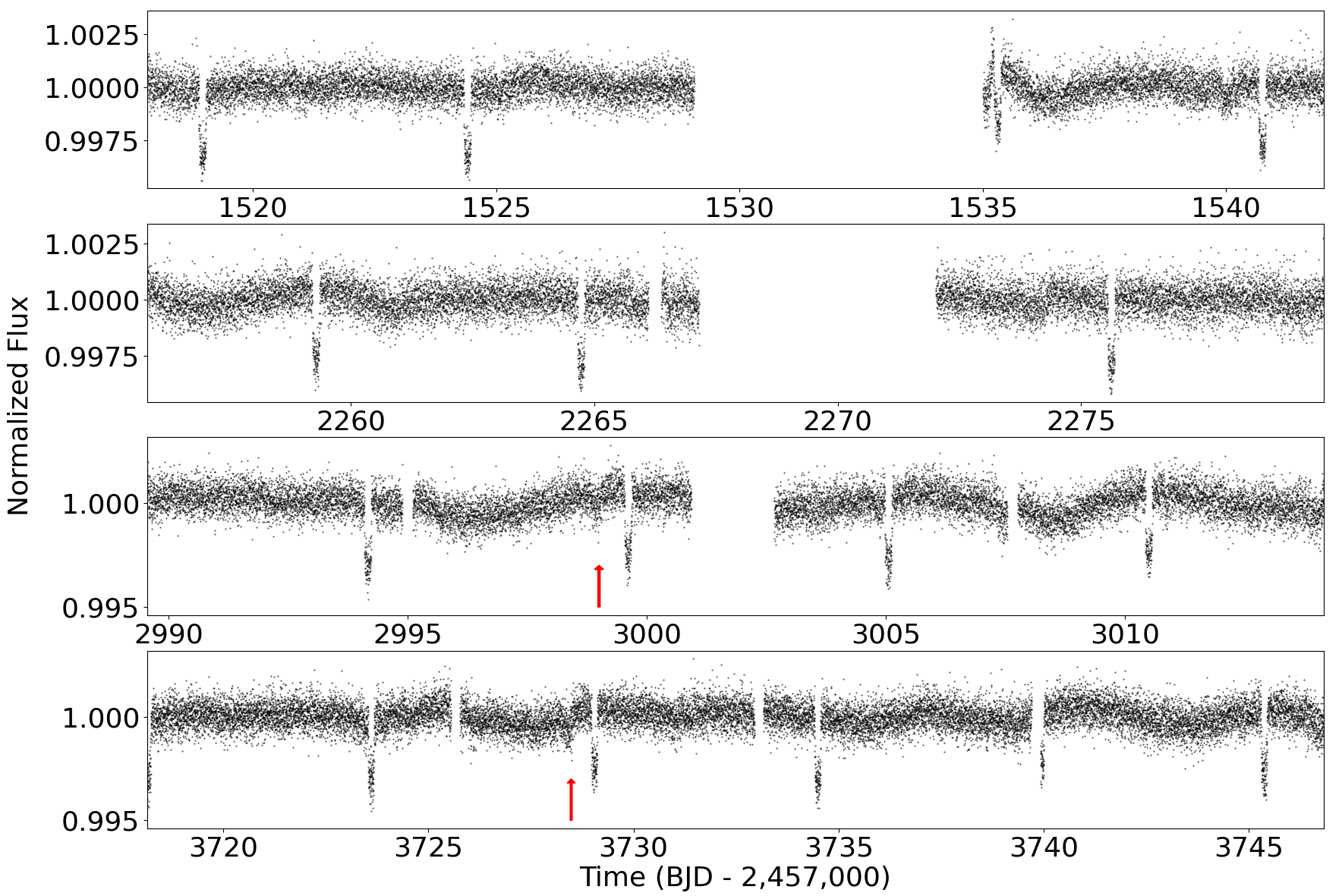}
    \caption{Four sectors of short-cadence {\em TESS} data for WASP-166 (TIC 408310006) showing the prominent transits of the known planet WASP-166 b. Two additional transit-like events can be seen in Sectors 62 and 89 (highlighted with red arrows), suggesting the potential presence of a second planet in the system.}
    \label{fig:wasp_166}
\end{figure}

Here, we highlight some of the more interesting targets independently identified as part of this effort, split into the following categories: 

\begin{itemize}
  \item Additional Eclipses: Targets exhibiting extra events not association with the EB signal, such as a second set of eclipses following a different period (representing a 2+2 quadruple system consisting of two EBs) or complex tertiary events (representing triply-eclipsing 2+1 triple systems or eclipsing (2+1)+1 quadruple systems). Fig. \ref{fig:quads} highlights two such examples.
  \item Eclipse Timing Variations (ETVs): Targets where the eclipse times deviate from linear ephemeris, suggesting potential dynamical interactions with additional bodies. An example is shown in the upper panel of Fig. \ref{fig:219006972_and_26542657} for the case of the 2+2 quadruple system TIC 219006972 where the two EB subsystems are dynamically-interactive on observable timescales \citep{2023MNRAS.522...90K}. Another example is the known EB TIC 26542657 \citep{Prsa2022} which, through our comparison with the {\em Kepler} catalog, we determined does not show the $\sim 11$-day eclipses in {\em Kepler} and must therefore be a higher order system. It is separated by only $\sim 1~{\rm arcsec}$ from TIC 1882992210, which is only $\sim 0.1$ mag fainter in {\em TESS}, making photocenter confirmation of the eclipse source effectively impossible from {\em TESS}. However, the target exhibits clear primary and secondary ETVs (see Fig. \ref{fig:219006972_and_26542657}, lower panel), a tertiary eclipse in Sector 81, as well as prominent changes in the shape of both the primary secondary eclipses between Sectors 14/15 (narrow, sharper primary, more rounded secondary) and later sectors (more rounded primary, flat secondary; see Fig. \ref{fig:26542657_lc}). Taken together, these provide strong evidence that either TIC 26542657 or TIC 1882992210 is a dynamically-interacting, triply-eclipsing triple system with an outer period of about 300 days. Additionally, as seen from Fig \ref{fig:26542657_kepler}, the Kepler lightcurve of the target hows one {\it tertiary} eclipse suggesting that the system was out of the eclipsing window for the $\sim11$-days EB during the Kepler era due to orbital precession\footnote{We note that the tertiary eclipse has non-zero quality flags, which was not uncommon in Kepler data, and could be easily missed if one only investigates the `quality = 0' data.}.
  \item Disappearing Eclipses: Targets where the detected eclipses exhibit prominent depth variations due to precession of the EB orbital plane, to the point of eventually ceasing altogether. Fig. \ref{fig:edvs} shows the {\em TESS} lightcurves of TIC 236774836 \citep{2024A&A...685A..43M} and TIC 220410224, indicating dynamical interaction with unseen companions. 
  \item ``Switching'' Eclipses: Similar to the previous example, but here the depth ratio between the primary and secondary eclipse changes between sectors. Fig. \ref{fig:eclipse_switch} shows an example of this effect for the case of the known EB TIC 234229841. 
  \item Apsidal Motion: Targets exhibiting pronounced `smear' of the secondary eclipses in orbital phase, indicating apsidal motion. Fig. \ref{fig:apsidal} highlights two such targets, TIC 189281140 and TIC 470715046.
  \item Stellar Variability: Targets exhibiting prominent lightcurve modulations due to e.g., rotational variability (spotted stars), pulsating components, heartbeat patterns, etc. Fig. \ref{fig:rotator} and \ref{fig:pulsator} show examples of each category, represented by TIC 21159577, TIC 22621932, and TIC 336538437. 
  \item Transiting Planets: It is only logical that a search for stellar eclipses will result in finding planetary transits as well. Indeed, our ML pipeline picked up the confirmed planet TIC 408310006 (WASP-166 b). Interestingly, an eagle-eyed volunteer on EBP (DI) noticed an additional transit-like event in the lightcurve of the target in Sector 62\footnote{\url{https://www.zooniverse.org/projects/vbkostov/eclipsing-binary-patrol/talk/6324/3432679?comment=5638772&page=1}}. Further investigation showed another event in Sector 89 (see Fig. \ref{fig:wasp_166}), suggesting the potential presence of a second transiting planet in the system. 
\end{itemize}

\section{Summary}
\label{sec:summary}

We have presented the {\em TESS} Ten-Thousand catalog containing 10,001 uniformly-vetted and -validated EBs observed by {\em TESS} in FFI data. \newEBs of these are new EBs while the remaining \knownEBs are known EBs where the period listed in one or more catalogs is incorrect. The targets were detected by a neural network search applied to Sectors 1 through 26 lightcurves. These were produced with a local implementation of the \textsc{eleanor} pipeline, and extracted for all stars brighter than {\em TESS} magnitude T = 15. The EBs passed comprehensive automated analysis and thorough visual scrutiny by citizen scientists, including confirmation of the measured ephemerides and photocenter offsets, and cross-matching against millions of known EBs from multiple catalogs. Most of the \newEBs new EBs are on the fainter end (median magnitude T = 13.8), have short orbital periods (median period of 3.5 days), the eclipses originate within $\sim1-2$ arcsec of the respective TIC, and have been observed in at least three {\em TESS} sectors. For the \knownEBs known EBs, we correct the ephemerides available at the time of writing. Astrometric measurements from Gaia suggest that a significant fraction of the new EBs may have unresolved companions and thus be part of higher-order stellar systems. In addition, some of the new EBs show eclipse timing variations, apsidal motion, and even extra eclipses due to additional stars. These are excellent targets for further in-depth investigation aimed at unraveling the underlying architecture and dynamics. Finally, we provide a list of $\sim900,000$ unvetted and unvalidated TESS targets for which the neural network identified eclipse-like features and scored higher than 0.9, and for which there are no known EBs within a sky-projected separation of 1 {\em TESS} pixel (21 arcsec). 

\acknowledgments
This paper includes data collected by the {\em TESS} mission, which are publicly available from the Mikulski Archive for Space Telescopes (MAST). Funding for the {\em TESS} mission is provided by NASA's Science Mission directorate.

This research has made use of the Exoplanet Follow-up Observation Program website, which is operated by the California Institute of Technology, under contract with the National Aeronautics and Space Administration under the Exoplanet Exploration Program. 

This work has made use of data from the European Space Agency (ESA) mission {\it Gaia} (\url{https://www.cosmos.esa.int/gaia}), processed by the {\it Gaia} Data Processing and Analysis Consortium (DPAC, \url{https://www.cosmos.esa.int/web/gaia/dpac/consortium}). Funding for the DPAC has been provided by national institutions, in particular the institutions participating in the {\it Gaia} Multilateral Agreement.

This publication uses data generated via the Zooniverse.org platform, development of which is funded by generous support, including a Global Impact Award from Google, and by a grant from the Alfred P. Sloan Foundation.

The Eclipsing Binary Patrol project has been made possible by the participation of nearly 2,000 incredible citizen scientists who volunteered their free time for the advancement of our understanding of eclipsing binary stars. The project as a whole is referred to as the Eclipsing Binary Patrol Collaboration in the affiliations listed above.

Resources supporting this work were provided by the NASA High-End Computing (HEC) Program through the NASA Center for Climate Simulation (NCCS) at Goddard Space Flight Center.

Resources supporting this work were provided by the NASA High-End Computing (HEC) Program through the NASA Advanced Supercomputing (NAS) Division at Ames Research Center for the production of the SPOC data products.

V.\,B.\,K. is grateful for financial support from NASA grant 80NSSC21K0631; V.\,B.\,K., J.\,O., and W.\,W. acknowledge support from NSF grant AST-2206814. 

\bibliography{refs}{}

\begin{thebibliography}{}
\expandafter\ifx\csname natexlab\endcsname\relax\def\natexlab#1{#1}\fi
\providecommand{\url}[1]{\href{#1}{#1}}
\providecommand{\dodoi}[1]{doi:~\href{http://doi.org/#1}{\nolinkurl{#1}}}
\providecommand{\doeprint}[1]{\href{http://ascl.net/#1}{\nolinkurl{http://ascl.net/#1}}}
\providecommand{\doarXiv}[1]{\href{https://arxiv.org/abs/#1}{\nolinkurl{https://arxiv.org/abs/#1}}}

\bibitem[{Abadi {et~al.}(2015)Abadi, Agarwal, Barham, Brevdo, Chen, Citro, Corrado, Davis, Dean, Devin, Ghemawat, Goodfellow, Harp, Irving, Isard, Jia, Jozefowicz, Kaiser, Kudlur, Levenberg, Man\'{e}, Monga, Moore, Murray, Olah, Schuster, Shlens, Steiner, Sutskever, Talwar, Tucker, Vanhoucke, Vasudevan, Vi\'{e}gas, Vinyals, Warden, Wattenberg, Wicke, Yu, \& Zheng}]{tensorflow}
Abadi, M., Agarwal, A., Barham, P., {et~al.} 2015, {TensorFlow}: Large-Scale Machine Learning on Heterogeneous Systems.
\newblock \url{https://www.tensorflow.org/}

\bibitem[{{Abdul-Masih} {et~al.}(2016){Abdul-Masih}, {Pr{\v{s}}a}, {Conroy}, {Bloemen}, {Boyajian}, {Doyle}, {Johnston}, {Kostov}, {Latham}, {Matijevi{\v{c}}}, {Shporer}, \& {Southworth}}]{2016AJ....151..101A}
{Abdul-Masih}, M., {Pr{\v{s}}a}, A., {Conroy}, K., {et~al.} 2016, \aj, 151, 101, \dodoi{10.3847/0004-6256/151/4/101}

\bibitem[{{Andersen}(1991)}]{Andersen1991}
{Andersen}, J. 1991, \aapr, 3, 91, \dodoi{10.1007/BF00873538}

\bibitem[{{Armstrong} {et~al.}(2021){Armstrong}, {Gamper}, \& {Damoulas}}]{2021MNRAS.504.5327A}
{Armstrong}, D.~J., {Gamper}, J., \& {Damoulas}, T. 2021, \mnras, 504, 5327, \dodoi{10.1093/mnras/staa2498}

\bibitem[{{Astropy Collaboration} {et~al.}(2013){Astropy Collaboration}, {Robitaille}, {Tollerud}, {Greenfield}, {Droettboom}, {Bray}, {Aldcroft}, {Davis}, {Ginsburg}, {Price-Whelan}, {Kerzendorf}, {Conley}, {Crighton}, {Barbary}, {Muna}, {Ferguson}, {Grollier}, {Parikh}, {Nair}, {Unther}, {Deil}, {Woillez}, {Conseil}, {Kramer}, {Turner}, {Singer}, {Fox}, {Weaver}, {Zabalza}, {Edwards}, {Azalee Bostroem}, {Burke}, {Casey}, {Crawford}, {Dencheva}, {Ely}, {Jenness}, {Labrie}, {Lim}, {Pierfederici}, {Pontzen}, {Ptak}, {Refsdal}, {Servillat}, \& {Streicher}}]{astropy2013}
{Astropy Collaboration}, {Robitaille}, T.~P., {Tollerud}, E.~J., {et~al.} 2013, \aap, 558, A33, \dodoi{10.1051/0004-6361/201322068}

\bibitem[{{Astropy Collaboration} {et~al.}(2018){Astropy Collaboration}, {Price-Whelan}, {Sip{\H{o}}cz}, {G{\"u}nther}, {Lim}, {Crawford}, {Conseil}, {Shupe}, {Craig}, {Dencheva}, {Ginsburg}, {Vand erPlas}, {Bradley}, {P{\'e}rez-Su{\'a}rez}, {de Val-Borro}, {Aldcroft}, {Cruz}, {Robitaille}, {Tollerud}, {Ardelean}, {Babej}, {Bach}, {Bachetti}, {Bakanov}, {Bamford}, {Barentsen}, {Barmby}, {Baumbach}, {Berry}, {Biscani}, {Boquien}, {Bostroem}, {Bouma}, {Brammer}, {Bray}, {Breytenbach}, {Buddelmeijer}, {Burke}, {Calderone}, {Cano Rodr{\'\i}guez}, {Cara}, {Cardoso}, {Cheedella}, {Copin}, {Corrales}, {Crichton}, {D'Avella}, {Deil}, {Depagne}, {Dietrich}, {Donath}, {Droettboom}, {Earl}, {Erben}, {Fabbro}, {Ferreira}, {Finethy}, {Fox}, {Garrison}, {Gibbons}, {Goldstein}, {Gommers}, {Greco}, {Greenfield}, {Groener}, {Grollier}, {Hagen}, {Hirst}, {Homeier}, {Horton}, {Hosseinzadeh}, {Hu}, {Hunkeler}, {Ivezi{\'c}}, {Jain}, {Jenness}, {Kanarek}, {Kendrew}, {Kern}, {Kerzendorf}, {Khvalko}, {King}, {Kirkby}, {Kulkarni},
  {Kumar}, {Lee}, {Lenz}, {Littlefair}, {Ma}, {Macleod}, {Mastropietro}, {McCully}, {Montagnac}, {Morris}, {Mueller}, {Mumford}, {Muna}, {Murphy}, {Nelson}, {Nguyen}, {Ninan}, {N{\"o}the}, {Ogaz}, {Oh}, {Parejko}, {Parley}, {Pascual}, {Patil}, {Patil}, {Plunkett}, {Prochaska}, {Rastogi}, {Reddy Janga}, {Sabater}, {Sakurikar}, {Seifert}, {Sherbert}, {Sherwood-Taylor}, {Shih}, {Sick}, {Silbiger}, {Singanamalla}, {Singer}, {Sladen}, {Sooley}, {Sornarajah}, {Streicher}, {Teuben}, {Thomas}, {Tremblay}, {Turner}, {Terr{\'o}n}, {van Kerkwijk}, {de la Vega}, {Watkins}, {Weaver}, {Whitmore}, {Woillez}, {Zabalza}, \& {Astropy Contributors}}]{astropy2018}
{Astropy Collaboration}, {Price-Whelan}, A.~M., {Sip{\H{o}}cz}, B.~M., {et~al.} 2018, \aj, 156, 123, \dodoi{10.3847/1538-3881/aabc4f}

\bibitem[{{Astropy Collaboration} {et~al.}(2022){Astropy Collaboration}, {Price-Whelan}, {Lim}, {Earl}, {Starkman}, {Bradley}, {Shupe}, {Patil}, {Corrales}, {Brasseur}, {N{\"o}the}, {Donath}, {Tollerud}, {Morris}, {Ginsburg}, {Vaher}, {Weaver}, {Tocknell}, {Jamieson}, {van Kerkwijk}, {Robitaille}, {Merry}, {Bachetti}, {G{\"u}nther}, {Aldcroft}, {Alvarado-Montes}, {Archibald}, {B{\'o}di}, {Bapat}, {Barentsen}, {Baz{\'a}n}, {Biswas}, {Boquien}, {Burke}, {Cara}, {Cara}, {Conroy}, {Conseil}, {Craig}, {Cross}, {Cruz}, {D'Eugenio}, {Dencheva}, {Devillepoix}, {Dietrich}, {Eigenbrot}, {Erben}, {Ferreira}, {Foreman-Mackey}, {Fox}, {Freij}, {Garg}, {Geda}, {Glattly}, {Gondhalekar}, {Gordon}, {Grant}, {Greenfield}, {Groener}, {Guest}, {Gurovich}, {Handberg}, {Hart}, {Hatfield-Dodds}, {Homeier}, {Hosseinzadeh}, {Jenness}, {Jones}, {Joseph}, {Kalmbach}, {Karamehmetoglu}, {Ka{\l}uszy{\'n}ski}, {Kelley}, {Kern}, {Kerzendorf}, {Koch}, {Kulumani}, {Lee}, {Ly}, {Ma}, {MacBride}, {Maljaars}, {Muna}, {Murphy}, {Norman},
  {O'Steen}, {Oman}, {Pacifici}, {Pascual}, {Pascual-Granado}, {Patil}, {Perren}, {Pickering}, {Rastogi}, {Roulston}, {Ryan}, {Rykoff}, {Sabater}, {Sakurikar}, {Salgado}, {Sanghi}, {Saunders}, {Savchenko}, {Schwardt}, {Seifert-Eckert}, {Shih}, {Jain}, {Shukla}, {Sick}, {Simpson}, {Singanamalla}, {Singer}, {Singhal}, {Sinha}, {Sip{\H{o}}cz}, {Spitler}, {Stansby}, {Streicher}, {{\v{S}}umak}, {Swinbank}, {Taranu}, {Tewary}, {Tremblay}, {de Val-Borro}, {Van Kooten}, {Vasovi{\'c}}, {Verma}, {de Miranda Cardoso}, {Williams}, {Wilson}, {Winkel}, {Wood-Vasey}, {Xue}, {Yoachim}, {Zhang}, {Zonca}, \& {Astropy Project Contributors}}]{astropy2022}
{Astropy Collaboration}, {Price-Whelan}, A.~M., {Lim}, P.~L., {et~al.} 2022, \apj, 935, 167, \dodoi{10.3847/1538-4357/ac7c74}

\bibitem[{Bahdanau {et~al.}(2014)Bahdanau, Cho, \& Bengio}]{Bahdanau2014}
Bahdanau, D., Cho, K., \& Bengio, Y. 2014, arXiv preprint arXiv:1409.0473, \dodoi{10.48550/arXiv.1409.0473}

\bibitem[{{Belokurov} {et~al.}(2020){Belokurov}, {Penoyre}, {Oh}, {Iorio}, {Hodgkin}, {Evans}, {Everall}, {Koposov}, {Tout}, {Izzard}, {Clarke}, \& {Brown}}]{Belokurov2020}
{Belokurov}, V., {Penoyre}, Z., {Oh}, S., {et~al.} 2020, \mnras, 496, 1922, \dodoi{10.1093/mnras/staa1522}

\bibitem[{{Borkovits}(2022)}]{2022Galax..10....9B}
{Borkovits}, T. 2022, Galaxies, 10, 9, \dodoi{10.3390/galaxies10010009}

\bibitem[{{Borkovits} {et~al.}(2022){Borkovits}, {Mitnyan}, {Rappaport}, {Pribulla}, {Powell}, {Kostov}, {B{\'\i}r{\'o}}, {Cs{\'a}nyi}, {Garai}, {Gary}, {Kaye}, {Kom{\v{z}}{\'\i}k}, {Terentev}, {Omohundro}, {Gagliano}, {Jacobs}, {Kristiansen}, {LaCourse}, {Schwengeler}, {Czavalinga}, {Seli}, {Huang}, {P{\'a}l}, {Vanderburg}, {Rodriguez}, \& {Stevens}}]{2022MNRAS.510.1352B}
{Borkovits}, T., {Mitnyan}, T., {Rappaport}, S.~A., {et~al.} 2022, \mnras, 510, 1352, \dodoi{10.1093/mnras/stab3397}

\bibitem[{{Borucki} {et~al.}(2010){Borucki}, {Koch}, {Basri}, {Batalha}, {Brown}, {Caldwell}, {Caldwell}, {Christensen-Dalsgaard}, {Cochran}, {DeVore}, {Dunham}, {Dupree}, {Gautier}, {Geary}, {Gilliland}, {Gould}, {Howell}, {Jenkins}, {Kondo}, {Latham}, {Marcy}, {Meibom}, {Kjeldsen}, {Lissauer}, {Monet}, {Morrison}, {Sasselov}, {Tarter}, {Boss}, {Brownlee}, {Owen}, {Buzasi}, {Charbonneau}, {Doyle}, {Fortney}, {Ford}, {Holman}, {Seager}, {Steffen}, {Welsh}, {Rowe}, {Anderson}, {Buchhave}, {Ciardi}, {Walkowicz}, {Sherry}, {Horch}, {Isaacson}, {Everett}, {Fischer}, {Torres}, {Johnson}, {Endl}, {MacQueen}, {Bryson}, {Dotson}, {Haas}, {Kolodziejczak}, {Van Cleve}, {Chandrasekaran}, {Twicken}, {Quintana}, {Clarke}, {Allen}, {Li}, {Wu}, {Tenenbaum}, {Verner}, {Bruhweiler}, {Barnes}, \& {Prsa}}]{2010Sci...327..977B}
{Borucki}, W.~J., {Koch}, D., {Basri}, G., {et~al.} 2010, Science, 327, 977, \dodoi{10.1126/science.1185402}

\bibitem[{{Boyajian} {et~al.}(2016){Boyajian}, {LaCourse}, {Rappaport}, {Fabrycky}, {Fischer}, {Gandolfi}, {Kennedy}, {Korhonen}, {Liu}, {Moor}, {Olah}, {Vida}, {Wyatt}, {Best}, {Brewer}, {Ciesla}, {Cs{\'a}k}, {Deeg}, {Dupuy}, {Handler}, {Heng}, {Howell}, {Ishikawa}, {Kov{\'a}cs}, {Kozakis}, {Kriskovics}, {Lehtinen}, {Lintott}, {Lynn}, {Nespral}, {Nikbakhsh}, {Schawinski}, {Schmitt}, {Smith}, {Szabo}, {Szabo}, {Viuho}, {Wang}, {Weiksnar}, {Bosch}, {Connors}, {Goodman}, {Green}, {Hoekstra}, {Jebson}, {Jek}, {Omohundro}, {Schwengeler}, \& {Szewczyk}}]{Boyajian2016}
{Boyajian}, T.~S., {LaCourse}, D.~M., {Rappaport}, S.~A., {et~al.} 2016, \mnras, 457, 3988, \dodoi{10.1093/mnras/stw218}

\bibitem[{{Burke} {et~al.}(2020){Burke}, {Levine}, {Fausnaugh}, {Vanderspek}, {Barclay}, {Libby-Roberts}, {Morris}, {Sipocz}, {Owens}, {Feinstein}, \& {Camacho}}]{2020ascl.soft03001B}
{Burke}, C.~J., {Levine}, A., {Fausnaugh}, M., {et~al.} 2020, {TESS-Point: High precision TESS pointing tool}, Astrophysics Source Code Library.
\newblock \doeprint{2003.001}

\bibitem[{{Cacciapuoti} {et~al.}(2022){Cacciapuoti}, {Kostov}, {Kuchner}, {Quintana}, {Col{\'o}n}, {Brande}, {Mullally}, {Chance}, {Christiansen}, {Ahlers}, {Di Fraia}, {Durantini Luca}, {Ienco}, {Gallo}, {de Lima}, {Hyogo}, {Andr{\'e}s-Carcasona}, {Fornear}, {de Lambilly}, {Salik}, {Yablonsky}, {Wallace}, \& {Acharya}}]{2022MNRAS.513..102C}
{Cacciapuoti}, L., {Kostov}, V.~B., {Kuchner}, M., {et~al.} 2022, \mnras, 513, 102, \dodoi{10.1093/mnras/stac652}

\bibitem[{{Caldwell} {et~al.}(2020){Caldwell}, {Tenenbaum}, {Twicken}, {Jenkins}, {Ting}, {Smith}, {Hedges}, {Fausnaugh}, {Rose}, \& {Burke}}]{2020RNAAS...4..201C}
{Caldwell}, D.~A., {Tenenbaum}, P., {Twicken}, J.~D., {et~al.} 2020, Research Notes of the American Astronomical Society, 4, 201, \dodoi{10.3847/2515-5172/abc9b3}

\bibitem[{{Capistrant} {et~al.}(2022){Capistrant}, {Soares-Furtado}, {Vanderburg}, {Kounkel}, {Rappaport}, {Omohundro}, {Powell}, {Gagliano}, {Jacobs}, {Kostov}, {Kristiansen}, {LaCourse}, {Schmitt}, {Schwengeler}, \& {Terentev}}]{2022ApJS..263...14C}
{Capistrant}, B.~K., {Soares-Furtado}, M., {Vanderburg}, A., {et~al.} 2022, \apjs, 263, 14, \dodoi{10.3847/1538-4365/ac9125}

\bibitem[{Chollet {et~al.}(2015)}]{keras}
Chollet, F., {et~al.} 2015, Keras, \url{https://keras.io}

\bibitem[{{Conroy} {et~al.}(2014{\natexlab{a}}){Conroy}, {Pr{\v{s}}a}, {Stassun}, {Orosz}, {Fabrycky}, \& {Welsh}}]{2014AJ....147...45C}
{Conroy}, K.~E., {Pr{\v{s}}a}, A., {Stassun}, K.~G., {et~al.} 2014{\natexlab{a}}, \aj, 147, 45, \dodoi{10.1088/0004-6256/147/2/45}

\bibitem[{{Conroy} {et~al.}(2014{\natexlab{b}}){Conroy}, {Pr{\v{s}}a}, {Stassun}, {Bloemen}, {Parvizi}, {Quarles}, {Boyajian}, {Barclay}, {Shporer}, {Latham}, \& {Abdul-Masih}}]{2014PASP..126..914C}
---. 2014{\natexlab{b}}, \pasp, 126, 914, \dodoi{10.1086/678953}

\bibitem[{{Coughlin} {et~al.}(2014){Coughlin}, {Thompson}, {Bryson}, {Burke}, {Caldwell}, {Christiansen}, {Haas}, {Howell}, {Jenkins}, {Kolodziejczak}, {Mullally}, \& {Rowe}}]{2014AJ....147..119C}
{Coughlin}, J.~L., {Thompson}, S.~E., {Bryson}, S.~T., {et~al.} 2014, \aj, 147, 119, \dodoi{10.1088/0004-6256/147/5/119}

\bibitem[{{Eggen}(1957)}]{Eggen1957}
{Eggen}, O.~J. 1957, The Observatory, 77, 191

\bibitem[{{Eisner} {et~al.}(2021){Eisner}, {Barrag{\'a}n}, {Lintott}, {Aigrain}, {Nicholson}, {Boyajian}, {Howell}, {Johnston}, {Lakeland}, {Miller}, {McMaster}, {Parviainen}, {Safron}, {Schwamb}, {Trouille}, {Vaughan}, {Zicher}, {Allen}, {Allen}, {Bouslog}, {Johnson}, {Simon}, {Wolfenbarger}, {Baeten}, {Bundy}, \& {Hoffman}}]{Eisner2021}
{Eisner}, N.~L., {Barrag{\'a}n}, O., {Lintott}, C., {et~al.} 2021, \mnras, 501, 4669, \dodoi{10.1093/mnras/staa3739}

\bibitem[{{Fang} {et~al.}(2018){Fang}, {Thompson}, \& {Hirata}}]{2018MNRAS.476.4234F}
{Fang}, X., {Thompson}, T.~A., \& {Hirata}, C.~M. 2018, \mnras, 476, 4234, \dodoi{10.1093/mnras/sty472}

\bibitem[{{Feinstein} {et~al.}(2019{\natexlab{a}}){Feinstein}, {Montet}, {Foreman-Mackey}, {Bedell}, {Saunders}, {Bean}, {Christiansen}, {Hedges}, {Luger}, {Scolnic}, \& {Cardoso}}]{eleanor2019}
{Feinstein}, A.~D., {Montet}, B.~T., {Foreman-Mackey}, D., {et~al.} 2019{\natexlab{a}}, \pasp, 131, 094502, \dodoi{10.1088/1538-3873/ab291c}

\bibitem[{{Feinstein} {et~al.}(2019{\natexlab{b}}){Feinstein}, {Montet}, {Foreman-Mackey}, {Bedell}, {Saunders}, {Bean}, {Christiansen}, {Hedges}, {Luger}, {Scolnic}, \& {Cardoso}}]{2019PASP..131i4502F}
---. 2019{\natexlab{b}}, \pasp, 131, 094502, \dodoi{10.1088/1538-3873/ab291c}

\bibitem[{{Fischer} {et~al.}(2012){Fischer}, {Schwamb}, {Schawinski}, {Lintott}, {Brewer}, {Giguere}, {Lynn}, {Parrish}, {Sartori}, {Simpson}, {Smith}, {Spronck}, {Batalha}, {Rowe}, {Jenkins}, {Bryson}, {Prsa}, {Tenenbaum}, {Crepp}, {Morton}, {Howard}, {Beleu}, {Kaplan}, {Vannispen}, {Sharzer}, {Defouw}, {Hajduk}, {Neal}, {Nemec}, {Schuepbach}, \& {Zimmermann}}]{Fischer2012}
{Fischer}, D.~A., {Schwamb}, M.~E., {Schawinski}, K., {et~al.} 2012, \mnras, 419, 2900, \dodoi{10.1111/j.1365-2966.2011.19932.x}

\bibitem[{{Fragione} \& {Kocsis}(2019)}]{2019MNRAS.486.4781F}
{Fragione}, G., \& {Kocsis}, B. 2019, \mnras, 486, 4781, \dodoi{10.1093/mnras/stz1175}

\bibitem[{{Gandhi} {et~al.}(2022){Gandhi}, {Buckley}, {Charles}, {Hodgkin}, {Scaringi}, {Knigge}, {Rao}, {Paice}, \& {Zhao}}]{Gandhi2022}
{Gandhi}, P., {Buckley}, D.~A.~H., {Charles}, P.~A., {et~al.} 2022, \mnras, 510, 3885, \dodoi{10.1093/mnras/stab3771}

\bibitem[{{Gao} {et~al.}(2025){Gao}, {Chen}, {Wang}, \& {Liu}}]{2025ApJS..276...57G}
{Gao}, X., {Chen}, X., {Wang}, S., \& {Liu}, J. 2025, \apjs, 276, 57, \dodoi{10.3847/1538-4365/ad9dd6}

\bibitem[{Goodfellow {et~al.}(2016)Goodfellow, Bengio, \& Courville}]{Goodfellow2016}
Goodfellow, I., Bengio, Y., \& Courville, A. 2016, Deep Learning (Cambridge, MA: MIT Press).
\newblock \url{http://www.deeplearningbook.org}

\bibitem[{{Goodricke}(1783)}]{Goodricke1783}
{Goodricke}, J. 1783, Philosophical Transactions of the Royal Society of London Series I, 73, 474

\bibitem[{{Green} {et~al.}(2023){Green}, {Maoz}, {Mazeh}, {Faigler}, {Shahaf}, {Gomel}, {El-Badry}, \& {Rix}}]{2023MNRAS.522...29G}
{Green}, M.~J., {Maoz}, D., {Mazeh}, T., {et~al.} 2023, \mnras, 522, 29, \dodoi{10.1093/mnras/stad915}

\bibitem[{{Hamers} {et~al.}(2021){Hamers}, {Rantala}, {Neunteufel}, {Preece}, \& {Vynatheya}}]{2021MNRAS.502.4479H}
{Hamers}, A.~S., {Rantala}, A., {Neunteufel}, P., {Preece}, H., \& {Vynatheya}, P. 2021, \mnras, 502, 4479, \dodoi{10.1093/mnras/stab287}

\bibitem[{{Han} \& {Brandt}(2023)}]{2023AJ....165...71H}
{Han}, T., \& {Brandt}, T.~D. 2023, \aj, 165, 71, \dodoi{10.3847/1538-3881/acaaa7}

\bibitem[{{Handler} {et~al.}(2020){Handler}, {Kurtz}, {Rappaport}, {Saio}, {Fuller}, {Jones}, {Guo}, {Chowdhury}, {Sowicka}, {Kahraman Ali{\c{c}}avu{\c{s}}}, {Streamer}, {Murphy}, {Gagliano}, {Jacobs}, \& {Vanderburg}}]{Handler2020}
{Handler}, G., {Kurtz}, D.~W., {Rappaport}, S.~A., {et~al.} 2020, Nature Astronomy, 4, 684, \dodoi{10.1038/s41550-020-1035-1}

\bibitem[{Harris {et~al.}(2020)Harris, Millman, van~der Walt, Gommers, Virtanen, Cournapeau, Wieser, Taylor, Berg, Smith, Kern, Picus, Hoyer, van Kerkwijk, Brett, Haldane, del R{\'{i}}o, Wiebe, Peterson, G{\'{e}}rard-Marchant, Sheppard, Reddy, Weckesser, Abbasi, Gohlke, \& Oliphant}]{numpy}
Harris, C.~R., Millman, K.~J., van~der Walt, S.~J., {et~al.} 2020, Nature, 585, 357, \dodoi{10.1038/s41586-020-2649-2}

\bibitem[{{Hartman} {et~al.}(2025){Hartman}, {Bakos}, {Bouma}, \& {Csubry}}]{2025PASP..137b4501H}
{Hartman}, J.~D., {Bakos}, G.~{\'A}., {Bouma}, L.~G., \& {Csubry}, Z. 2025, \pasp, 137, 024501, \dodoi{10.1088/1538-3873/adad42}

\bibitem[{He {et~al.}(2016)He, Zhang, Ren, \& Sun}]{He2016}
He, K., Zhang, X., Ren, S., \& Sun, J. 2016, in Proceedings of the IEEE Conference on Computer Vision and Pattern Recognition (CVPR), 770--778, \dodoi{10.1109/CVPR.2016.90}

\bibitem[{{Heinze} {et~al.}(2018){Heinze}, {Tonry}, {Denneau}, {Flewelling}, {Stalder}, {Rest}, {Smith}, {Smartt}, \& {Weiland}}]{Heinze2018}
{Heinze}, A.~N., {Tonry}, J.~L., {Denneau}, L., {et~al.} 2018, \aj, 156, 241, \dodoi{10.3847/1538-3881/aae47f}

\bibitem[{Hochreiter \& Schmidhuber(1997)}]{Hochreiter1997}
Hochreiter, S., \& Schmidhuber, J. 1997, Neural Computation, 9, 1735, \dodoi{10.1162/neco.1997.9.8.1735}

\bibitem[{{Howard} {et~al.}(2022){Howard}, {Davenport}, \& {Covey}}]{2022RNAAS...6...96H}
{Howard}, E.~L., {Davenport}, J. R.~A., \& {Covey}, K.~R. 2022, Research Notes of the American Astronomical Society, 6, 96, \dodoi{10.3847/2515-5172/ac6e42}

\bibitem[{{Huang} {et~al.}(2020{\natexlab{a}}){Huang}, {Vanderburg}, {P{\'a}l}, {Sha}, {Yu}, {Fong}, {Fausnaugh}, {Shporer}, {Guerrero}, {Vanderspek}, \& {Ricker}}]{2020RNAAS...4..204H}
{Huang}, C.~X., {Vanderburg}, A., {P{\'a}l}, A., {et~al.} 2020{\natexlab{a}}, Research Notes of the American Astronomical Society, 4, 204, \dodoi{10.3847/2515-5172/abca2e}

\bibitem[{{Huang} {et~al.}(2020{\natexlab{b}}){Huang}, {Vanderburg}, {P{\'a}l}, {Sha}, {Yu}, {Fong}, {Fausnaugh}, {Shporer}, {Guerrero}, {Vanderspek}, \& {Ricker}}]{2020RNAAS...4..206H}
---. 2020{\natexlab{b}}, Research Notes of the American Astronomical Society, 4, 206, \dodoi{10.3847/2515-5172/abca2d}

\bibitem[{Hunter(2007)}]{matplotlib}
Hunter, J.~D. 2007, Computing in science \& engineering, 9, 90

\bibitem[{{IJspeert} {et~al.}(2024){IJspeert}, {Tkachenko}, {Johnston}, \& {Aerts}}]{2024A&A...691A.242I}
{IJspeert}, L.~W., {Tkachenko}, A., {Johnston}, C., \& {Aerts}, C. 2024, \aap, 691, A242, \dodoi{10.1051/0004-6361/202450507}

\bibitem[{{Jayaraman} {et~al.}(2024){Jayaraman}, {Rappaport}, {Powell}, {Handler}, {Omohundro}, {Gagliano}, {Kostov}, {Fuller}, {Kurtz}, {Zhang}, \& {Ricker}}]{2024ApJ...975..121J}
{Jayaraman}, R., {Rappaport}, S.~A., {Powell}, B., {et~al.} 2024, \apj, 975, 121, \dodoi{10.3847/1538-4357/ad77c3}

\bibitem[{{Kirk} {et~al.}(2016){Kirk}, {Conroy}, {Pr{\v{s}}a}, {Abdul-Masih}, {Kochoska}, {Matijevi{\v{c}}}, {Hambleton}, {Barclay}, {Bloemen}, {Boyajian}, {Doyle}, {Fulton}, {Hoekstra}, {Jek}, {Kane}, {Kostov}, {Latham}, {Mazeh}, {Orosz}, {Pepper}, {Quarles}, {Ragozzine}, {Shporer}, {Southworth}, {Stassun}, {Thompson}, {Welsh}, {Agol}, {Derekas}, {Devor}, {Fischer}, {Green}, {Gropp}, {Jacobs}, {Johnston}, {LaCourse}, {Saetre}, {Schwengeler}, {Toczyski}, {Werner}, {Garrett}, {Gore}, {Martinez}, {Spitzer}, {Stevick}, {Thomadis}, {Vrijmoet}, {Yenawine}, {Batalha}, \& {Borucki}}]{2016AJ....151...68K}
{Kirk}, B., {Conroy}, K., {Pr{\v{s}}a}, A., {et~al.} 2016, \aj, 151, 68, \dodoi{10.3847/0004-6256/151/3/68}

\bibitem[{{Kochanek}(2021)}]{2021MNRAS.507.5832K}
{Kochanek}, C.~S. 2021, \mnras, 507, 5832, \dodoi{10.1093/mnras/stab2483}

\bibitem[{{Kopal}(1956)}]{Kopal1956}
{Kopal}, Z. 1956, Annales d'Astrophysique, 19, 298

\bibitem[{{Kostov}(2023)}]{2023Univ....9..455K}
{Kostov}, V.~B. 2023, Universe, 9, 455, \dodoi{10.3390/universe9100455}

\bibitem[{{Kostov} {et~al.}(2019){Kostov}, {Mullally}, {Quintana}, {Coughlin}, {Mullally}, {Barclay}, {Col{\'o}n}, {Schlieder}, {Barentsen}, \& {Burke}}]{2019AJ....157..124K}
{Kostov}, V.~B., {Mullally}, S.~E., {Quintana}, E.~V., {et~al.} 2019, \aj, 157, 124, \dodoi{10.3847/1538-3881/ab0110}

\bibitem[{{Kostov} {et~al.}(2020){Kostov}, {Orosz}, {Feinstein}, {Welsh}, {Cukier}, {Haghighipour}, {Quarles}, {Martin}, {Montet}, {Torres}, {Triaud}, {Barclay}, {Boyd}, {Briceno}, {Cameron}, {Correia}, {Gilbert}, {Gill}, {Gillon}, {Haqq-Misra}, {Hellier}, {Dressing}, {Fabrycky}, {Furesz}, {Jenkins}, {Kane}, {Kopparapu}, {Hod{\v{z}}i{\'c}}, {Latham}, {Law}, {Levine}, {Li}, {Lintott}, {Lissauer}, {Mann}, {Mazeh}, {Mardling}, {Maxted}, {Eisner}, {Pepe}, {Pepper}, {Pollacco}, {Quinn}, {Quintana}, {Rowe}, {Ricker}, {Rose}, {Seager}, {Santerne}, {S{\'e}gransan}, {Short}, {Smith}, {Standing}, {Tokovinin}, {Trifonov}, {Turner}, {Twicken}, {Udry}, {Vanderspek}, {Winn}, {Wolf}, {Ziegler}, {Ansorge}, {Barnet}, {Bergeron}, {Huten}, {Pappa}, \& {van der Straeten}}]{2020AJ....159..253K}
{Kostov}, V.~B., {Orosz}, J.~A., {Feinstein}, A.~D., {et~al.} 2020, \aj, 159, 253, \dodoi{10.3847/1538-3881/ab8a48}

\bibitem[{{Kostov} {et~al.}(2021{\natexlab{a}}){Kostov}, {Powell}, {Torres}, {Borkovits}, {Rappaport}, {Tokovinin}, {Zasche}, {Anderson}, {Barclay}, {Berlind}, {Brown}, {Calkins}, {Collins}, {Collins}, {Conti}, {Esquerdo}, {Hellier}, {Jensen}, {Kamler}, {Kruse}, {Latham}, {Ma{\v{s}}ek}, {Murgas}, {Olmschenk}, {Orosz}, {P{\'a}l}, {Palle}, {Schwarz}, {Stockdale}, {Tamayo}, {Uhla{\v{r}}}, {Welsh}, \& {West}}]{2021ApJ...917...93K}
{Kostov}, V.~B., {Powell}, B.~P., {Torres}, G., {et~al.} 2021{\natexlab{a}}, \apj, 917, 93, \dodoi{10.3847/1538-4357/ac04ad}

\bibitem[{{Kostov} {et~al.}(2021{\natexlab{b}}){Kostov}, {Powell}, {Orosz}, {Welsh}, {Cochran}, {Collins}, {Endl}, {Hellier}, {Latham}, {MacQueen}, {Pepper}, {Quarles}, {Sairam}, {Torres}, {Wilson}, {Bergeron}, {Boyce}, {Bieryla}, {Buchheim}, {Ben Christiansen}, {Ciardi}, {Collins}, {Conti}, {Dixon}, {Guerra}, {Haghighipour}, {Herman}, {Hintz}, {Howard}, {Jensen}, {Kielkopf}, {Kruse}, {Law}, {Martin}, {Maxted}, {Montet}, {Murgas}, {Nelson}, {Olmschenk}, {Otero}, {Quimby}, {Richmond}, {Schwarz}, {Shporer}, {Stassun}, {Stephens}, {Triaud}, {Ulowetz}, {Walter}, {Wiley}, {Wood}, {Yenawine}, {Agol}, {Barclay}, {Beatty}, {Boisse}, {Caldwell}, {Christiansen}, {Col{\'o}n}, {Deleuil}, {Doyle}, {Fausnaugh}, {F{\H{u}}r{\'e}sz}, {Gilbert}, {H{\'e}brard}, {James}, {Jenkins}, {Kane}, {Kidwell}, {Kopparapu}, {Li}, {Lissauer}, {Lund}, {Majewski}, {Mazeh}, {Quinn}, {Quintana}, {Ricker}, {Rodriguez}, {Rowe}, {Santerne}, {Schlieder}, {Seager}, {Standing}, {Stevens}, {Ting}, {Vanderspek}, \& {Winn}}]{2021AJ....162..234K}
{Kostov}, V.~B., {Powell}, B.~P., {Orosz}, J.~A., {et~al.} 2021{\natexlab{b}}, \aj, 162, 234, \dodoi{10.3847/1538-3881/ac223a}

\bibitem[{{Kostov} {et~al.}(2022{\natexlab{a}}){Kostov}, {Powell}, {Rappaport}, {Borkovits}, {Gagliano}, {Jacobs}, {Kristiansen}, {LaCourse}, {Omohundro}, {Orosz}, {Schmitt}, {Schwengeler}, {Terentev}, {Torres}, {Barclay}, {Friedman}, {Kruse}, {Olmschenk}, {Vanderburg}, \& {Welsh}}]{Kostov2022_quadcat1}
{Kostov}, V.~B., {Powell}, B.~P., {Rappaport}, S.~A., {et~al.} 2022{\natexlab{a}}, \apjs, 259, 66, \dodoi{10.3847/1538-4365/ac5458}

\bibitem[{{Kostov} {et~al.}(2022{\natexlab{b}}){Kostov}, {Kuchner}, {Cacciapuoti}, {Acharya}, {Ahlers}, {Andr{\'e}s-Carcasona}, {Brande}, {de Lima}, {Di Fraia}, {Fornear}, {Gallo}, {Hyogo}, {Ienco}, {de Lambilly}, {Luca}, {Quintana}, {Salik}, \& {Yablonsky}}]{2022PASP..134d4401K}
{Kostov}, V.~B., {Kuchner}, M.~J., {Cacciapuoti}, L., {et~al.} 2022{\natexlab{b}}, \pasp, 134, 044401, \dodoi{10.1088/1538-3873/ac5de0}

\bibitem[{{Kostov} {et~al.}(2023){Kostov}, {Borkovits}, {Rappaport}, {Powell}, {P{\'a}l}, {Jacobs}, {Gagliano}, {Kristiansen}, {LaCourse}, {Moe}, {Omohundro}, {Schmitt}, {Schwengeler}, {Terentev}, \& {Vanderburg}}]{2023MNRAS.522...90K}
{Kostov}, V.~B., {Borkovits}, T., {Rappaport}, S.~A., {et~al.} 2023, \mnras, 522, 90, \dodoi{10.1093/mnras/stad941}

\bibitem[{{Kostov} {et~al.}(2024{\natexlab{a}}){Kostov}, {Powell}, {Rappaport}, {Borkovits}, {Gagliano}, {Jacobsy}, {Jayaraman}, {Kristiansen}, {LaCourse}, {Mitnyan}, {Omohundro}, {Orosz}, {P{\'a}l}, {Schmitt}, {Schwengeler}, {Terentev}, {Torres}, {Barclay}, {Vanderburg}, \& {Welsh}}]{Kostov_quadcat2}
{Kostov}, V.~B., {Powell}, B.~P., {Rappaport}, S.~A., {et~al.} 2024{\natexlab{a}}, \mnras, 527, 3995, \dodoi{10.1093/mnras/stad2947}

\bibitem[{{Kostov} {et~al.}(2024{\natexlab{b}}){Kostov}, {Rappaport}, {Borkovits}, {Powell}, {Gagliano}, {Omohundro}, {B{\'\i}r{\'o}}, {Moe}, {Howell}, {Mitnyan}, {Clark}, {Kristiansen}, {Terentev}, {Schwengeler}, {P{\'a}l}, \& {Vanderburg}}]{2024ApJ...974...25K}
{Kostov}, V.~B., {Rappaport}, S.~A., {Borkovits}, T., {et~al.} 2024{\natexlab{b}}, \apj, 974, 25, \dodoi{10.3847/1538-4357/ad7368}

\bibitem[{Kotikalapudi \& contributors(2017)}]{raghakotkerasvis}
Kotikalapudi, R., \& contributors. 2017, keras-vis, \url{https://github.com/raghakot/keras-vis},  GitHub

\bibitem[{{Kov{\'a}cs} {et~al.}(2002){Kov{\'a}cs}, {Zucker}, \& {Mazeh}}]{BLS}
{Kov{\'a}cs}, G., {Zucker}, S., \& {Mazeh}, T. 2002, \aap, 391, 369, \dodoi{10.1051/0004-6361:20020802}

\bibitem[{{Kozai}(1962)}]{1962AJ.....67..591K}
{Kozai}, Y. 1962, \aj, 67, 591, \dodoi{10.1086/108790}

\bibitem[{{Kristiansen} {et~al.}(2022){Kristiansen}, {Rappaport}, {Vanderburg}, {Jacobs}, {Martin Schwengeler}, {Gagliano}, {Terentev}, {LaCourse}, {Omohundro}, {Schmitt}, {Powell}, \& {Kostov}}]{2022PASP..134g4401K}
{Kristiansen}, M. H.~K., {Rappaport}, S.~A., {Vanderburg}, A.~M., {et~al.} 2022, \pasp, 134, 074401, \dodoi{10.1088/1538-3873/ac6e06}

\bibitem[{{Kunimoto} {et~al.}(2024){Kunimoto}, {Bryson}, {Daylan}, {Lissauer}, {Matesic}, {Mullally}, \& {Rowe}}]{2024AAS...24323105K}
{Kunimoto}, M., {Bryson}, S., {Daylan}, T., {et~al.} 2024, in American Astronomical Society Meeting Abstracts, Vol. 243, American Astronomical Society Meeting Abstracts, 231.05

\bibitem[{{Kunimoto} {et~al.}(2022){Kunimoto}, {Tey}, {Fong}, {Hesse}, {Shporer}, {Fausnaugh}, {Vanderspek}, \& {Ricker}}]{qlp2022}
{Kunimoto}, M., {Tey}, E., {Fong}, W., {et~al.} 2022, Research Notes of the American Astronomical Society, 6, 236, \dodoi{10.3847/2515-5172/aca158}

\bibitem[{{LaCourse} {et~al.}(2015){LaCourse}, {Jek}, {Jacobs}, {Winarski}, {Boyajian}, {Rappaport}, {Sanchis-Ojeda}, {Conroy}, {Nelson}, {Barclay}, {Fischer}, {Schmitt}, {Wang}, {Stassun}, {Pepper}, {Coughlin}, {Shporer}, \& {Pr{\v{s}}a}}]{2015MNRAS.452.3561L}
{LaCourse}, D.~M., {Jek}, K.~J., {Jacobs}, T.~L., {et~al.} 2015, \mnras, 452, 3561, \dodoi{10.1093/mnras/stv1475}

\bibitem[{Lea {et~al.}(2017)Lea, Flynn, Vidal, Reiter, \& Hager}]{lea2017temporal}
Lea, C., Flynn, M.~D., Vidal, R., Reiter, A., \& Hager, G.~D. 2017, in Proceedings of the IEEE Conference on Computer Vision and Pattern Recognition, 156--165, \dodoi{10.1109/CVPR.2017.113}

\bibitem[{LeCun {et~al.}(1989)LeCun, Boser, Denker, Henderson, Howard, Hubbard, \& Jackel}]{LeCun1989}
LeCun, Y., Boser, B., Denker, J.~S., {et~al.} 1989, Neural Computation, 1, 541, \dodoi{10.1162/neco.1989.1.4.541}

\bibitem[{{Lidov}(1962)}]{1962P&SS....9..719L}
{Lidov}, M.~L. 1962, \planss, 9, 719, \dodoi{10.1016/0032-0633(62)90129-0}

\bibitem[{{Lightkurve Collaboration} {et~al.}(2018){Lightkurve Collaboration}, {Cardoso}, {Hedges}, {Gully-Santiago}, {Saunders}, {Cody}, {Barclay}, {Hall}, {Sagear}, {Turtelboom}, {Zhang}, {Tzanidakis}, {Mighell}, {Coughlin}, {Bell}, {Berta-Thompson}, {Williams}, {Dotson}, \& {Barentsen}}]{lightkurve}
{Lightkurve Collaboration}, {Cardoso}, J.~V.~d.~M., {Hedges}, C., {et~al.} 2018, {Lightkurve: Kepler and TESS time series analysis in Python}, Astrophysics Source Code Library.
\newblock \doeprint{1812.013}

\bibitem[{{Liu} \& {Lai}(2019)}]{2019MNRAS.483.4060L}
{Liu}, B., \& {Lai}, D. 2019, \mnras, 483, 4060, \dodoi{10.1093/mnras/sty3432}

\bibitem[{Luong {et~al.}(2015)Luong, Pham, \& Manning}]{Luong2015}
Luong, M.-T., Pham, H., \& Manning, C.~D. 2015, in Proceedings of the 2015 Conference on Empirical Methods in Natural Language Processing (Association for Computational Linguistics), 1412--1421, \dodoi{10.18653/v1/D15-1166}

\bibitem[{{Magliano} {et~al.}(2023){Magliano}, {Kostov}, {Cacciapuoti}, {Covone}, {Inno}, {Fiscale}, {Kuchner}, {Quintana}, {Salik}, {Saggese}, {Yablonsky}, {Fornear}, {Hyogo}, {Di Fraia}, {Luca}, {de Lambilly}, {Oliva}, {Pagano}, {Ienco}, {de Lima}, {Andr{\'e}s-Carcasona}, {Gallo}, \& {Acharya}}]{2023MNRAS.521.3749M}
{Magliano}, C., {Kostov}, V., {Cacciapuoti}, L., {et~al.} 2023, \mnras, 521, 3749, \dodoi{10.1093/mnras/stad683}

\bibitem[{Majewski(2025)}]{Majewski2025}
Majewski, S. R. e.~a. 2025

\bibitem[{{Matijevi{\v{c}}} {et~al.}(2012){Matijevi{\v{c}}}, {Pr{\v{s}}a}, {Orosz}, {Welsh}, {Bloemen}, \& {Barclay}}]{2012AJ....143..123M}
{Matijevi{\v{c}}}, G., {Pr{\v{s}}a}, A., {Orosz}, J.~A., {et~al.} 2012, \aj, 143, 123, \dodoi{10.1088/0004-6256/143/5/123}

\bibitem[{McKinney(2010)}]{pandas}
McKinney, W. 2010, in Proceedings of the 9th Python in Science Conference, ed. S.~van~der Walt \& J.~Millman, 51 -- 56

\bibitem[{{Melton} {et~al.}(2024){Melton}, {Feigelson}, {Montalto}, {Caceres}, {Rosenswie}, \& {Abelson}}]{2024AJ....167..203M}
{Melton}, E.~J., {Feigelson}, E.~D., {Montalto}, M., {et~al.} 2024, \aj, 167, 203, \dodoi{10.3847/1538-3881/ad29f1}

\bibitem[{{Mitnyan} {et~al.}(2024{\natexlab{a}}){Mitnyan}, {Borkovits}, {Czavalinga}, {Rappaport}, {P{\'a}l}, {Powell}, \& {Hajdu}}]{2024A&A...686C...3M}
{Mitnyan}, T., {Borkovits}, T., {Czavalinga}, D.~R., {et~al.} 2024{\natexlab{a}}, \aap, 686, C3, \dodoi{10.1051/0004-6361/202450750e}

\bibitem[{{Mitnyan} {et~al.}(2024{\natexlab{b}}){Mitnyan}, {Borkovits}, {Czavalinga}, {Rappaport}, {P{\'a}l}, {Powell}, \& {Hajdu}}]{2024A&A...685A..43M}
---. 2024{\natexlab{b}}, \aap, 685, A43, \dodoi{10.1051/0004-6361/202348909}

\bibitem[{{Moe} \& {Di Stefano}(2017)}]{2017ApJS..230...15M}
{Moe}, M., \& {Di Stefano}, R. 2017, \apjs, 230, 15, \dodoi{10.3847/1538-4365/aa6fb6}

\bibitem[{{Montalto}(2023)}]{Montalto2023}
{Montalto}, M. 2023, \mnras, 518, L31, \dodoi{10.1093/mnrasl/slac131}

\bibitem[{{Mowlavi} {et~al.}(2023){Mowlavi}, {Holl}, {Lecoeur-Ta{\"\i}bi}, {Barblan}, {Kochoska}, {Pr{\v{s}}a}, {Mazeh}, {Rimoldini}, {Gavras}, {Audard}, {Jevardat de Fombelle}, {Nienartowicz}, {Garc{\'\i}a-Lario}, \& {Eyer}}]{2023A&A...674A..16M}
{Mowlavi}, N., {Holl}, B., {Lecoeur-Ta{\"\i}bi}, I., {et~al.} 2023, \aap, 674, A16, \dodoi{10.1051/0004-6361/202245330}

\bibitem[{Nair \& Hinton(2010)}]{Nair2010}
Nair, V., \& Hinton, G.~E. 2010, in Proceedings of the 27th International Conference on Machine Learning (ICML-10) (Omnipress), 807--814

\bibitem[{Nelson(2008)}]{Nelson2008}
Nelson, S. 2008, Nature, 455, 36, \dodoi{10.1038/455036a}

\bibitem[{{Niemela}(2001)}]{2001RMxAC..11...23N}
{Niemela}, V. 2001, in Revista Mexicana de Astronomia y Astrofisica Conference Series, Vol.~11, Revista Mexicana de Astronomia y Astrofisica Conference Series, 23--26

\bibitem[{{Oelkers} \& {Stassun}(2018)}]{2018AJ....156..132O}
{Oelkers}, R.~J., \& {Stassun}, K.~G. 2018, \aj, 156, 132, \dodoi{10.3847/1538-3881/aad68e}

\bibitem[{{Offner} {et~al.}(2023){Offner}, {Moe}, {Kratter}, {Sadavoy}, {Jensen}, \& {Tobin}}]{Offner2023}
{Offner}, S.~S.~R., {Moe}, M., {Kratter}, K.~M., {et~al.} 2023, in Astronomical Society of the Pacific Conference Series, Vol. 534, Protostars and Planets VII, ed. S.~{Inutsuka}, Y.~{Aikawa}, T.~{Muto}, K.~{Tomida}, \& M.~{Tamura}, 275, \dodoi{10.48550/arXiv.2203.10066}

\bibitem[{{Ol{\'a}h} {et~al.}(2025){Ol{\'a}h}, {Seli}, {Haris}, {Rappaport}, {Tuomi}, {Gagliano}, {Jacobs}, {Kristiansen}, {Schwengeler}, {Omohundro}, {Terentev}, {Vanderburg}, {Powell}, {Kostov}, \& {K{\H{o}}v{\'a}ri}}]{2025arXiv250415389O}
{Ol{\'a}h}, K., {Seli}, B., {Haris}, A., {et~al.} 2025, arXiv e-prints, arXiv:2504.15389, \dodoi{10.48550/arXiv.2504.15389}

\bibitem[{{Osterbrock}(1953)}]{Osterbrock1953}
{Osterbrock}, D.~E. 1953, \apj, 118, 529, \dodoi{10.1086/145781}

\bibitem[{Pedregosa {et~al.}(2011)Pedregosa, Varoquaux, Gramfort, Michel, Thirion, Grisel, Blondel, Prettenhofer, Weiss, Dubourg, Vanderplas, Passos, Cournapeau, Brucher, Perrot, \& Duchesnay}]{scikit-learn}
Pedregosa, F., Varoquaux, G., Gramfort, A., {et~al.} 2011, Journal of Machine Learning Research, 12, 2825

\bibitem[{{Pejcha} {et~al.}(2013){Pejcha}, {Antognini}, {Shappee}, \& {Thompson}}]{2013MNRAS.435..943P}
{Pejcha}, O., {Antognini}, J.~M., {Shappee}, B.~J., \& {Thompson}, T.~A. 2013, \mnras, 435, 943, \dodoi{10.1093/mnras/stt1281}

\bibitem[{{Penoyre} {et~al.}(2020){Penoyre}, {Belokurov}, {Wyn Evans}, {Everall}, \& {Koposov}}]{Penoyre2020}
{Penoyre}, Z., {Belokurov}, V., {Wyn Evans}, N., {Everall}, A., \& {Koposov}, S.~E. 2020, \mnras, 495, 321, \dodoi{10.1093/mnras/staa1148}

\bibitem[{{Petrosky} {et~al.}(2021){Petrosky}, {Hwang}, {Zakamska}, {Chandra}, \& {Hill}}]{2021MNRAS.503.3975P}
{Petrosky}, E., {Hwang}, H.-C., {Zakamska}, N.~L., {Chandra}, V., \& {Hill}, M.~J. 2021, \mnras, 503, 3975, \dodoi{10.1093/mnras/stab592}

\bibitem[{{Powell} {et~al.}(2023){Powell}, {Kostov}, \& {Tokovinin}}]{2023MNRAS.524.4296P}
{Powell}, B.~P., {Kostov}, V.~B., \& {Tokovinin}, A. 2023, \mnras, 524, 4296, \dodoi{10.1093/mnras/stad2065}

\bibitem[{{Powell} {et~al.}(2021{\natexlab{a}}){Powell}, {Kostov}, {Rappaport}, {Borkovits}, {Zasche}, {Tokovinin}, {Kruse}, {Latham}, {Montet}, {Jensen}, {Jayaraman}, {Collins}, {Ma{\v{s}}ek}, {Hellier}, {Evans}, {Tan}, {Schlieder}, {Torres}, {Smale}, {Friedman}, {Barclay}, {Gagliano}, {Quintana}, {Jacobs}, {Gilbert}, {Kristiansen}, {Col{\'o}n}, {LaCourse}, {Olmschenk}, {Omohundro}, {Schnittman}, {Schwengeler}, {Barry}, {Terentev}, {Boyd}, {Schmitt}, {Quinn}, {Vanderburg}, {Palle}, {Armstrong}, {Ricker}, {Vanderspek}, {Seager}, {Winn}, {Jenkins}, {Caldwell}, {Wohler}, {Shiao}, {Burke}, {Daylan}, \& {Villase{\~n}or}}]{2021AJ....161..162P}
{Powell}, B.~P., {Kostov}, V.~B., {Rappaport}, S.~A., {et~al.} 2021{\natexlab{a}}, \aj, 161, 162, \dodoi{10.3847/1538-3881/abddb5}

\bibitem[{{Powell} {et~al.}(2021{\natexlab{b}}){Powell}, {Kostov}, {Rappaport}, {Tokovinin}, {Shporer}, {Collins}, {Corbett}, {Borkovits}, {Gary}, {Chiang}, {Rodriguez}, {Law}, {Barclay}, {Gagliano}, {Vanderburg}, {Olmschenk}, {Kruse}, {Schlieder}, {Soto}, {Goeke}, {Jacobs}, {Kristiansen}, {LaCourse}, {Omohundro}, {Schwengeler}, {Terentev}, \& {Schmitt}}]{2021AJ....162..299P}
---. 2021{\natexlab{b}}, \aj, 162, 299, \dodoi{10.3847/1538-3881/ac2c81}

\bibitem[{{Powell} {et~al.}(2022{\natexlab{a}}){Powell}, {Kruse}, {Montet}, {Feinstein}, {Lewis}, {Foreman-Mackey}, {Barclay}, {Quintana}, {Col{\'o}n}, {Kostov}, {Boyd}, {Smale}, {Mullally}, {Schlieder}, {Schnittman}, {Carroll}, {Carriere}, {Salmon}, {Strong}, {Acks}, {Pfaff}, {Gerner}, \& {Burch}}]{2022RNAAS...6..111P}
{Powell}, B.~P., {Kruse}, E., {Montet}, B.~T., {et~al.} 2022{\natexlab{a}}, Research Notes of the American Astronomical Society, 6, 111, \dodoi{10.3847/2515-5172/ac74c4}

\bibitem[{{Powell} {et~al.}(2022{\natexlab{b}}){Powell}, {Rappaport}, {Borkovits}, {Kostov}, {Torres}, {Jayaraman}, {Latham}, {Ku{\v{c}}{\'a}kov{\'a}}, {Garai}, {Pribulla}, {Vanderburg}, {Kruse}, {Barclay}, {Olmschenk}, {Kristiansen}, {Gagliano}, {Jacobs}, {LaCourse}, {Omohundro}, {Schwengeler}, {Terentev}, \& {Schmitt}}]{2022ApJ...938..133P}
{Powell}, B.~P., {Rappaport}, S.~A., {Borkovits}, T., {et~al.} 2022{\natexlab{b}}, \apj, 938, 133, \dodoi{10.3847/1538-4357/ac8934}

\bibitem[{{Powell} {et~al.}(2025{\natexlab{a}}){Powell}, {Torres}, {Kostov}, {Borkovits}, {Rappaport}, {Moe}, {Latham}, {Jacobs}, {Gagliano}, {Kristiansen}, {Omohundro}, {Schwengeler}, {LaCourse}, {Terentev}, \& {Schmitt}}]{2025ApJ...985..213P}
{Powell}, B.~P., {Torres}, G., {Kostov}, V.~B., {et~al.} 2025{\natexlab{a}}, \apj, 985, 213, \dodoi{10.3847/1538-4357/adcece}

\bibitem[{{Powell} {et~al.}(2025{\natexlab{b}}){Powell}, {Torres}, {Kostov}, {Borkovits}, {Rappaport}, {Moe}, {Latham}, {Jacobs}, {Gagliano}, {Kristiansen}, {Omohundro}, {Schwengeler}, {LaCourse}, {Terentev}, \& {Schmitt}}]{2025arXiv250412239P}
---. 2025{\natexlab{b}}, arXiv e-prints, arXiv:2504.12239, \dodoi{10.48550/arXiv.2504.12239}

\bibitem[{{Pr{\v{s}}a} {et~al.}(2011){Pr{\v{s}}a}, {Batalha}, {Slawson}, {Doyle}, {Welsh}, {Orosz}, {Seager}, {Rucker}, {Mjaseth}, {Engle}, {Conroy}, {Jenkins}, {Caldwell}, {Koch}, \& {Borucki}}]{2011AJ....141...83P}
{Pr{\v{s}}a}, A., {Batalha}, N., {Slawson}, R.~W., {et~al.} 2011, \aj, 141, 83, \dodoi{10.1088/0004-6256/141/3/83}

\bibitem[{{Pr{\v{s}}a} {et~al.}(2022{\natexlab{a}}){Pr{\v{s}}a}, {Kochoska}, {Conroy}, {Eisner}, {Hey}, {IJspeert}, {Kruse}, {Fleming}, {Johnston}, {Kristiansen}, {LaCourse}, {Mortensen}, {Pepper}, {Stassun}, {Torres}, {Abdul-Masih}, {Chakraborty}, {Gagliano}, {Guo}, {Hambleton}, {Hong}, {Jacobs}, {Jones}, {Kostov}, {Lee}, {Omohundro}, {Orosz}, {Page}, {Powell}, {Rappaport}, {Reed}, {Schnittman}, {Schwengeler}, {Shporer}, {Terentev}, {Vanderburg}, {Welsh}, {Caldwell}, {Doty}, {Jenkins}, {Latham}, {Ricker}, {Seager}, {Schlieder}, {Shiao}, {Vanderspek}, \& {Winn}}]{Prsa2022}
{Pr{\v{s}}a}, A., {Kochoska}, A., {Conroy}, K.~E., {et~al.} 2022{\natexlab{a}}, \apjs, 258, 16, \dodoi{10.3847/1538-4365/ac324a}

\bibitem[{{Pr{\v{s}}a} {et~al.}(2022{\natexlab{b}}){Pr{\v{s}}a}, {Kochoska}, {Conroy}, {Eisner}, {Hey}, {IJspeert}, {Kruse}, {Fleming}, {Johnston}, {Kristiansen}, {LaCourse}, {Mortensen}, {Pepper}, {Stassun}, {Torres}, {Abdul-Masih}, {Chakraborty}, {Gagliano}, {Guo}, {Hambleton}, {Hong}, {Jacobs}, {Jones}, {Kostov}, {Lee}, {Omohundro}, {Orosz}, {Page}, {Powell}, {Rappaport}, {Reed}, {Schnittman}, {Schwengeler}, {Shporer}, {Terentev}, {Vanderburg}, {Welsh}, {Caldwell}, {Doty}, {Jenkins}, {Latham}, {Ricker}, {Seager}, {Schlieder}, {Shiao}, {Vanderspek}, \& {Winn}}]{2022ApJS..258...16P}
---. 2022{\natexlab{b}}, \apjs, 258, 16, \dodoi{10.3847/1538-4365/ac324a}

\bibitem[{{Raghavan} {et~al.}(2010){Raghavan}, {McAlister}, {Henry}, {Latham}, {Marcy}, {Mason}, {Gies}, {White}, \& {ten Brummelaar}}]{Raghavan2010}
{Raghavan}, D., {McAlister}, H.~A., {Henry}, T.~J., {et~al.} 2010, \apjs, 190, 1, \dodoi{10.1088/0067-0049/190/1/1}

\bibitem[{{Rappaport} {et~al.}(2018){Rappaport}, {Vanderburg}, {Jacobs}, {LaCourse}, {Jenkins}, {Kraus}, {Rizzuto}, {Latham}, {Bieryla}, {Lazarevic}, \& {Schmitt}}]{Rappaport2018}
{Rappaport}, S., {Vanderburg}, A., {Jacobs}, T., {et~al.} 2018, \mnras, 474, 1453, \dodoi{10.1093/mnras/stx2735}

\bibitem[{{Rappaport} {et~al.}(2022){Rappaport}, {Borkovits}, {Gagliano}, {Jacobs}, {Kostov}, {Powell}, {Terentev}, {Omohundro}, {Torres}, {Vanderburg}, {Mitnyan}, {Kristiansen}, {LaCourse}, {Schwengeler}, {Kaye}, {P{\'a}l}, {Pribulla}, {B{\'\i}r{\'o}}, {Cs{\'a}nyi}, {Garai}, {Zasche}, {Maxted}, {Rodriguez}, \& {Stevens}}]{2022MNRAS.513.4341R}
{Rappaport}, S.~A., {Borkovits}, T., {Gagliano}, R., {et~al.} 2022, \mnras, 513, 4341, \dodoi{10.1093/mnras/stac957}

\bibitem[{{Rappaport} {et~al.}(2023){Rappaport}, {Borkovits}, {Gagliano}, {Jacobs}, {Tokovinin}, {Mitnyan}, {Kom{\v{z}}{\'\i}k}, {Kostov}, {Powell}, {Torres}, {Terentev}, {Omohundro}, {Pribulla}, {Vanderburg}, {Kristiansen}, {Latham}, {Schwengeler}, {LaCourse}, {B{\'\i}r{\'o}}, {Cs{\'a}nyi}, {Czavalinga}, {Garai}, {P{\'a}l}, {Rodriguez}, \& {Stevens}}]{2023MNRAS.521..558R}
---. 2023, \mnras, 521, 558, \dodoi{10.1093/mnras/stad367}

\bibitem[{{Rappaport} {et~al.}(2024){Rappaport}, {Borkovits}, {Mitnyan}, {Gagliano}, {Eisner}, {Jacobs}, {Tokovinin}, {Powell}, {Kostov}, {Omohundro}, {Kristiansen}, {Jayaraman}, {Terentev}, {Schwengeler}, {LaCourse}, {Garai}, {Pribulla}, {Maxted}, {B{\'\i}r{\'o}}, {Cs{\'a}nyi}, {P{\'a}l}, \& {Vanderburg}}]{2024A&A...686A..27R}
{Rappaport}, S.~A., {Borkovits}, T., {Mitnyan}, T., {et~al.} 2024, \aap, 686, A27, \dodoi{10.1051/0004-6361/202449273}

\bibitem[{{Ricker} {et~al.}(2015){Ricker}, {Winn}, {Vanderspek}, {Latham}, {Bakos}, {Bean}, {Berta-Thompson}, {Brown}, {Buchhave}, {Butler}, {Butler}, {Chaplin}, {Charbonneau}, {Christensen-Dalsgaard}, {Clampin}, {Deming}, {Doty}, {De Lee}, {Dressing}, {Dunham}, {Endl}, {Fressin}, {Ge}, {Henning}, {Holman}, {Howard}, {Ida}, {Jenkins}, {Jernigan}, {Johnson}, {Kaltenegger}, {Kawai}, {Kjeldsen}, {Laughlin}, {Levine}, {Lin}, {Lissauer}, {MacQueen}, {Marcy}, {McCullough}, {Morton}, {Narita}, {Paegert}, {Palle}, {Pepe}, {Pepper}, {Quirrenbach}, {Rinehart}, {Sasselov}, {Sato}, {Seager}, {Sozzetti}, {Stassun}, {Sullivan}, {Szentgyorgyi}, {Torres}, {Udry}, \& {Villasenor}}]{Ricker2015}
{Ricker}, G.~R., {Winn}, J.~N., {Vanderspek}, R., {et~al.} 2015, Journal of Astronomical Telescopes, Instruments, and Systems, 1, 014003, \dodoi{10.1117/1.JATIS.1.1.014003}

\bibitem[{{Rowan} {et~al.}(2022){Rowan}, {Jayasinghe}, {Stanek}, {Kochanek}, {Thompson}, {Shappee}, {Holoien}, {Prieto}, \& {Giles}}]{2022MNRAS.517.2190R}
{Rowan}, D.~M., {Jayasinghe}, T., {Stanek}, K.~Z., {et~al.} 2022, \mnras, 517, 2190, \dodoi{10.1093/mnras/stac2520}

\bibitem[{Rumelhart {et~al.}(1986)Rumelhart, Hinton, \& Williams}]{Rumelhart1986}
Rumelhart, D.~E., Hinton, G.~E., \& Williams, R.~J. 1986, Nature, 323, 533, \dodoi{10.1038/323533a0}

\bibitem[{{Russell}(1948)}]{1948HarMo...7..181R}
{Russell}, H.~N. 1948, in Harvard Observatory Monographs, Vol.~7, 181

\bibitem[{{Shan} {et~al.}(2025){Shan}, {Chen}, {Zhang}, {Wang}, {Zou}, \& {Li}}]{2025arXiv250415875S}
{Shan}, Y., {Chen}, J., {Zhang}, Z., {et~al.} 2025, arXiv e-prints, arXiv:2504.15875.
\newblock \doarXiv{2504.15875}

\bibitem[{{Shara} {et~al.}(2021){Shara}, {Howell}, {Furlan}, {Gnilka}, {Moffat}, {Scott}, \& {Zurek}}]{2021MNRAS.507..560S}
{Shara}, M.~M., {Howell}, S.~B., {Furlan}, E., {et~al.} 2021, \mnras, 507, 560, \dodoi{10.1093/mnras/stab2212}

\bibitem[{Shi {et~al.}(2015)Shi, Chen, Wang, Yeung, Wong, \& Woo}]{Shi2015}
Shi, X., Chen, Z., Wang, H., {et~al.} 2015, in Advances in Neural Information Processing Systems, Vol.~28 (Curran Associates, Inc.), 802--810, \dodoi{10.48550/arXiv.1506.04214}

\bibitem[{Simonyan \& Zisserman(2014)}]{Simonyan2014}
Simonyan, K., \& Zisserman, A. 2014, arXiv preprint arXiv:1409.1556, \dodoi{10.48550/arXiv.1409.1556}

\bibitem[{{Slawson} {et~al.}(2011){Slawson}, {Pr{\v{s}}a}, {Welsh}, {Orosz}, {Rucker}, {Batalha}, {Doyle}, {Engle}, {Conroy}, {Coughlin}, {Gregg}, {Fetherolf}, {Short}, {Windmiller}, {Fabrycky}, {Howell}, {Jenkins}, {Uddin}, {Mullally}, {Seader}, {Thompson}, {Sanderfer}, {Borucki}, \& {Koch}}]{2011AJ....142..160S}
{Slawson}, R.~W., {Pr{\v{s}}a}, A., {Welsh}, W.~F., {et~al.} 2011, \aj, 142, 160, \dodoi{10.1088/0004-6256/142/5/160}

\bibitem[{{Solanki} {et~al.}(2025){Solanki}, {Cieplak}, {Schnittman}, {Baker}, {Barclay}, {Barry}, {Kostov}, {Kruse}, {Olmschenk}, {Powell}, {Ishitani Silva}, \& {Torres}}]{2025ApJS..276...17S}
{Solanki}, S., {Cieplak}, A.~M., {Schnittman}, J., {et~al.} 2025, \apjs, 276, 17, \dodoi{10.3847/1538-4365/ad8a62}

\bibitem[{{Soszy{\'n}ski} {et~al.}(2017){Soszy{\'n}ski}, {Udalski}, {Szyma{\'n}ski}, {Wyrzykowski}, {Ulaczyk}, {Poleski}, {Pietrukowicz}, {Koz{\l}owski}, {Skowron}, {Skowron}, {Mr{\'o}z}, {Pawlak}, {Rybicki}, \& {Jacyszyn-Dobrzeniecka}}]{2017AcA....67..297S}
{Soszy{\'n}ski}, I., {Udalski}, A., {Szyma{\'n}ski}, M.~K., {et~al.} 2017, \actaa, 67, 297, \dodoi{10.32023/0001-5237/67.4.1}

\bibitem[{{Stassun} \& {Torres}(2021)}]{Stassun2021}
{Stassun}, K.~G., \& {Torres}, G. 2021, \apjl, 907, L33, \dodoi{10.3847/2041-8213/abdaad}

\bibitem[{{Stassun} {et~al.}(2019){Stassun}, {Oelkers}, {Paegert}, {Torres}, {Pepper}, {De Lee}, {Collins}, {Latham}, {Muirhead}, {Chittidi}, {Rojas-Ayala}, {Fleming}, {Rose}, {Tenenbaum}, {Ting}, {Kane}, {Barclay}, {Bean}, {Brassuer}, {Charbonneau}, {Ge}, {Lissauer}, {Mann}, {McLean}, {Mullally}, {Narita}, {Plavchan}, {Ricker}, {Sasselov}, {Seager}, {Sharma}, {Shiao}, {Sozzetti}, {Stello}, {Vanderspek}, {Wallace}, \& {Winn}}]{2019AJ....158..138S}
{Stassun}, K.~G., {Oelkers}, R.~J., {Paegert}, M., {et~al.} 2019, \aj, 158, 138, \dodoi{10.3847/1538-3881/ab3467}

\bibitem[{{Sullivan} {et~al.}(2015){Sullivan}, {Winn}, {Berta-Thompson}, {Charbonneau}, {Deming}, {Dressing}, {Latham}, {Levine}, {McCullough}, {Morton}, {Ricker}, {Vanderspek}, \& {Woods}}]{2015ApJ...809...77S}
{Sullivan}, P.~W., {Winn}, J.~N., {Berta-Thompson}, Z.~K., {et~al.} 2015, \apj, 809, 77, \dodoi{10.1088/0004-637X/809/1/77}

\bibitem[{Sun {et~al.}(2017)Sun, Shrivastava, Singh, \& Gupta}]{Sun2017}
Sun, C., Shrivastava, A., Singh, S., \& Gupta, A. 2017, in Proceedings of the IEEE International Conference on Computer Vision, 843--852, \dodoi{10.1109/ICCV.2017.97}

\bibitem[{{Thiemann} {et~al.}(2021){Thiemann}, {Norton}, {Dickinson}, {McMaster}, \& {Kolb}}]{2021MNRAS.502.1299T}
{Thiemann}, H.~B., {Norton}, A.~J., {Dickinson}, H.~J., {McMaster}, A., \& {Kolb}, U.~C. 2021, \mnras, 502, 1299, \dodoi{10.1093/mnras/stab140}

\bibitem[{{Thompson} {et~al.}(2015){Thompson}, {Mullally}, {Coughlin}, {Christiansen}, {Henze}, {Haas}, \& {Burke}}]{2015ApJ...812...46T}
{Thompson}, S.~E., {Mullally}, F., {Coughlin}, J., {et~al.} 2015, \apj, 812, 46, \dodoi{10.1088/0004-637X/812/1/46}

\bibitem[{{Thompson} {et~al.}(2018){Thompson}, {Coughlin}, {Hoffman}, {Mullally}, {Christiansen}, {Burke}, {Bryson}, {Batalha}, {Haas}, {Catanzarite}, {Rowe}, {Barentsen}, {Caldwell}, {Clarke}, {Jenkins}, {Li}, {Latham}, {Lissauer}, {Mathur}, {Morris}, {Seader}, {Smith}, {Klaus}, {Twicken}, {Van Cleve}, {Wohler}, {Akeson}, {Ciardi}, {Cochran}, {Henze}, {Howell}, {Huber}, {Pr{\v{s}}a}, {Ram{\'\i}rez}, {Morton}, {Barclay}, {Campbell}, {Chaplin}, {Charbonneau}, {Christensen-Dalsgaard}, {Dotson}, {Doyle}, {Dunham}, {Dupree}, {Ford}, {Geary}, {Girouard}, {Isaacson}, {Kjeldsen}, {Quintana}, {Ragozzine}, {Shabram}, {Shporer}, {Silva Aguirre}, {Steffen}, {Still}, {Tenenbaum}, {Welsh}, {Wolfgang}, {Zamudio}, {Koch}, \& {Borucki}}]{2018ApJS..235...38T}
{Thompson}, S.~E., {Coughlin}, J.~L., {Hoffman}, K., {et~al.} 2018, \apjs, 235, 38, \dodoi{10.3847/1538-4365/aab4f9}

\bibitem[{Tieleman \& Hinton(2012)}]{Tieleman2012}
Tieleman, T., \& Hinton, G. 2012, COURSERA: Neural networks for machine learning, 4, 26

\bibitem[{{Tokovinin}(2021)}]{Tokovinin2021}
{Tokovinin}, A. 2021, Universe, 7, 352, \dodoi{10.3390/universe7090352}

\bibitem[{{Torres} {et~al.}(2010){Torres}, {Andersen}, \& {Gim{\'e}nez}}]{Torres2010}
{Torres}, G., {Andersen}, J., \& {Gim{\'e}nez}, A. 2010, \aapr, 18, 67, \dodoi{10.1007/s00159-009-0025-1}

\bibitem[{{Trani} {et~al.}(2022){Trani}, {Rastello}, {Di Carlo}, {Santoliquido}, {Tanikawa}, \& {Mapelli}}]{2022MNRAS.511.1362T}
{Trani}, A.~A., {Rastello}, S., {Di Carlo}, U.~N., {et~al.} 2022, \mnras, 511, 1362, \dodoi{10.1093/mnras/stac122}

\bibitem[{{Tremaine}(2020)}]{2020MNRAS.493.5583T}
{Tremaine}, S. 2020, \mnras, 493, 5583, \dodoi{10.1093/mnras/staa643}

\bibitem[{{Twicken} {et~al.}(2018){Twicken}, {Catanzarite}, {Clarke}, {Girouard}, {Jenkins}, {Klaus}, {Li}, {McCauliff}, {Seader}, {Tenenbaum}, {Wohler}, {Bryson}, {Burke}, {Caldwell}, {Haas}, {Henze}, \& {Sanderfer}}]{2018PASP..130f4502T}
{Twicken}, J.~D., {Catanzarite}, J.~H., {Clarke}, B.~D., {et~al.} 2018, \pasp, 130, 064502, \dodoi{10.1088/1538-3873/aab694}

\bibitem[{{Valizadegan} {et~al.}(2025){Valizadegan}, {Martinho}, {Jenkins}, {Twicken}, {Caldwell}, {Maynard}, {Wei}, {Zhong}, {Yates}, {Donald}, {Collins}, {Latham}, {Barkaoui}, {Berlind}, {Calkins}, {Carden}, {Chazov}, {Esquerdo}, {Guillot}, {Krushinsky}, {Nowak}, {Rackham}, {Triaud}, {Schwarz}, {Stephens}, {Stockdale}, {Wang}, {Watkins}, \& {Wilkin}}]{2025arXiv250209790V}
{Valizadegan}, H., {Martinho}, M. J.~S., {Jenkins}, J.~M., {et~al.} 2025, arXiv e-prints, arXiv:2502.09790, \dodoi{10.48550/arXiv.2502.09790}

\bibitem[{{Virtanen} {et~al.}(2020){Virtanen}, {Gommers}, {Oliphant}, {Haberland}, {Reddy}, {Cournapeau}, {Burovski}, {Peterson}, {Weckesser}, {Bright}, {van der Walt}, {Brett}, {Wilson}, {Jarrod Millman}, {Mayorov}, {Nelson}, {Jones}, {Kern}, {Larson}, {Carey}, {Polat}, {Feng}, {Moore}, {VanderPlas}, {Laxalde}, {Perktold}, {Cimrman}, {Henriksen}, {Quintero}, {Harris}, {Archibald}, {Ribeiro}, {Pedregosa}, {van Mulbregt}, \& {Contributors}}]{scipy}
{Virtanen}, P., {Gommers}, R., {Oliphant}, T.~E., {et~al.} 2020, Nature Methods, \dodoi{https://doi.org/10.1038/s41592-019-0686-2}

\bibitem[{{von Zeipel}(1910)}]{1910AN....183..345V}
{von Zeipel}, H. 1910, Astronomische Nachrichten, 183, 345, \dodoi{10.1002/asna.19091832202}

\bibitem[{{Vynatheya} \& {Hamers}(2022)}]{2022ApJ...926..195V}
{Vynatheya}, P., \& {Hamers}, A.~S. 2022, \apj, 926, 195, \dodoi{10.3847/1538-4357/ac4892}

\bibitem[{{Wang} {et~al.}(2015){Wang}, {Fischer}, {Barclay}, {Picard}, {Ma}, {Bowler}, {Schmitt}, {Boyajian}, {Jek}, {LaCourse}, {Baranec}, {Riddle}, {Law}, {Lintott}, {Schawinski}, {Simister}, {Gr{\'e}goire}, {Babin}, {Poile}, {Jacobs}, {Jebson}, {Omohundro}, {Schwengeler}, {Sejpka}, {Terentev}, {Gagliano}, {Paakkonen}, {Otnes Berge}, {Winarski}, {Green}, {Schmitt}, {Kristiansen}, \& {Hoekstra}}]{Wang2015}
{Wang}, J., {Fischer}, D.~A., {Barclay}, T., {et~al.} 2015, \apj, 815, 127, \dodoi{10.1088/0004-637X/815/2/127}

\bibitem[{{Welsh} \& {Orosz}(2018)}]{2018haex.bookE..34W}
{Welsh}, W.~F., \& {Orosz}, J.~A. 2018, in Handbook of Exoplanets, ed. H.~J. {Deeg} \& J.~A. {Belmonte}, 34, \dodoi{10.1007/978-3-319-55333-7_34}

\bibitem[{{Zasche} {et~al.}(2024){Zasche}, {Henzl}, {Merc}, {K{\'a}ra}, \& {Ku{\v{c}}{\'a}kov{\'a}}}]{2024A&A...687A...6Z}
{Zasche}, P., {Henzl}, Z., {Merc}, J., {K{\'a}ra}, J., \& {Ku{\v{c}}{\'a}kov{\'a}}, H. 2024, \aap, 687, A6, \dodoi{10.1051/0004-6361/202450400}

\bibitem[{{Zasche} {et~al.}(2023){Zasche}, {Borkovits}, {Jayaraman}, {Rappaport}, {Bro{\v{z}}}, {Vokrouhlick{\'y}}, {B{\'\i}r{\'o}}, {Heged{\"u}s}, {Kiss}, {Uhla{\v{r}}}, {Schwengeler}, {P{\'a}l}, {Ma{\v{s}}ek}, {Howell}, {Dallaporta}, {Munari}, {Gagliano}, {Jacobs}, {Kristiansen}, {LaCourse}, {Omohundro}, {Terentev}, {Vanderburg}, {Henzl}, {Powell}, \& {Kostov}}]{2023MNRAS.520.3127Z}
{Zasche}, P., {Borkovits}, T., {Jayaraman}, R., {et~al.} 2023, \mnras, 520, 3127, \dodoi{10.1093/mnras/stad328}

\end{thebibliography}
\bibliographystyle{aasjournal}

\facilities{
\emph{Gaia},
MAST,
TESS,}

\software{
Exogram (\url{https://exogram.vercel.app/}),
Fast Lighcurve Inspector (\url{https://fast-lightcurve-inspector.osc-fr1.scalingo.io/},
astropy \citep{astropy2013, astropy2018, astropy2022}, 
Keras \citep{keras},
Lightkurve \citep{lightkurve}, 
% M\_-M\_K- \citep{mann19}, 
Matplotlib \citep{matplotlib},
NumPy \citep{numpy}, 
Pandas \citep{pandas}, 
% PyMC3 \citep{exoplanet:pymc3}, 
SciPy \citep{scipy}, 
Tensorflow \citep{tensorflow},
}

\end{document}